\documentclass[twoside]{report}
\usepackage[a4paper]{geometry}
\usepackage[latin1]{inputenc} 

\usepackage{tensor}
\usepackage{color}
\newcommand{\blue}{\color{blue}}

\usepackage{graphicx}
\usepackage{ upgreek }
\usepackage{graphicx, amsmath,  amssymb}
\usepackage{accents}

\newcommand*{\dt}[1]{%
  \accentset{\mbox{\large\bfseries .}}{#1}}
\newcommand*{\dtv}[1]{%
  \accentset{\mbox{\large\bfseries {\scriptsize$v$}}}{#1}}

\usepackage{amsthm}

\usepackage{algorithm}
\usepackage{algorithmicx}
\usepackage[noend]{algpseudocode}

\newtheorem{definition}{Definition}[chapter]
\newtheorem{rem}{Remark}[chapter]
\newtheorem{theorem}{Theorem}[chapter]

\newcommand{\giat}{\widehat{g}}

\newcommand{\E}{\mathcal{E}}

\newcommand{\Mn}{\mathbb{R}^n}
\newcommand{\uideg}{\textnormal{deg}_w}

\begin{document}

\newcommand{\revtex}{REV\TeX\ }
\newcommand{\classoption}[1]{\texttt{#1}}
\newcommand{\macro}[1]{\texttt{\textbackslash#1}}
\newcommand{\m}[1]{\macro{#1}}
\newcommand{\env}[1]{\texttt{#1}}

\newcommand{\C}{\mathcal{C}}

\newcommand{\SUIO}{$\mathcal{SUIO}$}
\newcommand{\B}{$\mathcal{B}$}
\newcommand{\VS}{$\mathcal{V}$}
\newcommand{\IMU}{$\mathcal{I}$}
\newcommand{\tIMU}{\textnormal{\IMU}}
\newcommand{\tVS}{\textnormal{\VS}}
\newcommand{\tih}{\widetilde{h}}
\newcommand{\tobs}{\widetilde{\OBS}}
\newcommand{\NO}{s}
\newcommand{\ND}{r}

\newcommand{\M}{\mathcal{M}}
\newcommand{\Li}{\mathcal{L}}
\newcommand{\RM}{\mathcal{RM}}
\newcommand{\Hf}{\mathcal{H}}

\newcommand{\oplusn}{+}
\newcommand{\bigoplusn}{\sum}

\setlength{\textheight}{9.5in}

\newcommand{\Obs}{$\mathcal{O}$}
\newcommand{\OBS}{\mathcal{O}}




\title{Identifiability of nonlinear ODE Models with Time-Varying Parameters: the General Analytical Solution  and Applications in Viral Dynamics} 

\author{Agostino Martinelli
\thanks{A. Martinelli is with INRIA Rhone Alpes,
Montbonnot, France e-mail: {\tt agostino.martinelli@inria.fr}} }

\maketitle



\tableofcontents


\begin{abstract}
Identifiability is a structural property of any ODE model characterized by a set of unknown parameters. It describes the possibility of determining the values of these parameters from fusing the observations of the system inputs and outputs.
This paper finds the general analytical solution of this fundamental problem and, based on this, provides a general and automated analytical method to determine the identifiability of the unknown parameters. In particular,
the method can handle any model, regardless of its complexity and type of non-linearity, and provides the identifiability of the parameters even when they are time-varying.
In addition, it is automatic as it simply needs to follow the steps of a systematic procedure that 
only requires to perform the calculation of derivatives and matrix ranks.
Time-varying parameters are treated as unknown inputs and their identification is based on the very recent analytical solution of the unknown input observability problem \cite{SIAMbook,IF22}.
The method is used to determine the identifiability of the unknown time-varying parameters that characterize two non-linear models in the field of viral dynamics (HIV and Covid-19) and a non-linear model that characterizes the genetic toggle switch. 
New fundamental properties that characterize these models are determined and
discussed in detail through a comparison with the state-of-the-art results. In particular, regarding the very popular HIV ODE model and the genetic toggle switch model, the method automatically finds new important results that are in contrast with the results in the current literature.
\end{abstract}




\chapter*{Highlights}

\begin{enumerate}

\item Algorithm that solves the most general Unknown Input Observability problem usable by a non specialist user.


\item Algorithm that determines the time-varying parameter identifiability of any ODE model

\item Determination of the continuous transformations (one parameter Lie groups) which allow us to construct the set of indistinguishable states and indistinguishable unknown inputs 
in the presence of unobservability and unidentifiability.



\item Observability and identifiability of a very popular HIV model, a Covid-19 model, and a model that characterizes the genetic toggle switch.

\item Detection of errors in the state of the art about the HIV ODE model identifiability and the genetic toggle switch identifiability. In particular, detection of a serious error in Section 6.2 of \cite{Miao11}\footnote{The interested reader can find a detailed explanation of the serious error made by the authors of \cite{Miao11} in \cite{arXivErratum}.
Note that Section 6.2 of \cite{Miao11} contains two distinct errors. The former, which is a simple typo, was recently acknowledged by the authors (see \cite{MiaoErratum}). However, in \cite{MiaoErratum}, the authors still concluded that the system is observable/identifiable.
The very serious error committed by the authors is highlighted in Section 3.2 of \cite{arXivErratum} and lies in an incorrect use of the Inverse Function Theorem. In particular, when using this theorem, the authors did not take into account the constraints due to the dynamics of the system, as shown in Section 3.2 of \cite{arXivErratum}.}.


\end{enumerate}

\chapter{Introduction}\label{ChapterIntroduction}\label{ChapterIntroduction}

Ordinary Differential Equations (ODEs) are used to model a plethora of phenomena in many scientific domains, ranging from the natural and applied sciences up to the social sciences.

An ODE model is characterized by a state that consists of several scalar quantities. Its time evolution is precisely described by a set of ordinary differential equations (one per each state component).
In addition to the state components, an ODE model can also include a set of parameters which may be known or unknown.
Finally, an ODE model is characterized by a set of outputs and, very often, by a set of inputs. The inputs are functions of time that act on the state dynamics and that can be assigned (with some restrictions that depend on the specific case).
The outputs are available quantities and can be expressed in terms of the state and, in some cases, also in terms of the above inputs and the above model parameters.

{\it State observability} and {\it parameter identifiability} are two structural properties of an ODE model. The former characterizes the possibility of inferring the values that the state components take, starting from the knowledge of the system inputs and outputs
(e.g., \cite{Kalman61,Kalman63,Her77,Casti82,Isi95}). The latter characterizes the possibility of inferring the values of the model parameters, again, from the knowledge of the system inputs and outputs
(e.g., \cite{Grew76,Tuna87,Wal96}).

Performing an observability and an identifiability analysis is fundamental in order to set up an ODE model, i.e., in order to correctly define the state and the parameters that the information contained in the sensor measurements allows us to estimate. Characterizing a given system with a state that is unobservable and/or with parameters that are unidentifiable can have dramatic consequences during the estimation and/or the identification process.
Note that, for many complex sensor fusion problems studied in the past, this task (observability and identifiability analysis) was accomplished and played a crucial role to set up a suitable ODE model. For example, this has been the case for the following sensor fusion problems: visual and inertial (e.g., see the works \cite{[*1],[*2],[*3],[*4],[*5]}), visual and pressure (e.g., \cite{[*6]}),
GPS and odometry (e.g., \cite{[*7]}),
wheel encoders and ultra-wideband (e.g., \cite{[*8]}), radio beacons and inertial (e.g., \cite{[**1]}). Additionally, this has also been the case in the framework of 
distributed estimation over a low-cost sensor network (e.g., see \cite{[**2]} and references therein).
In most previous works (and in many others), the observability analysis of the considered sensor fusion problem was essential to set up a proper ODE model (note that in the above works, the concept of identifiability was not distinguished from observability and, with observability analysis, it was actually intended observability and identifiability analysis).
More in general, the key role of observability in information fusion has recently been investigated in \cite{[*9]}.

An ODE model may also include the presence of {\it unknown} inputs, i.e., inputs that act on the system dynamics precisely as the aforementioned inputs. However, they differ from them because they are unknown. From a mathematical point of view, an unknown input plays exactly the same role of an unknown time-varying parameter. In the literature, the possibility of inferring their values has been called in different manners:
{\it input observability}, {\it input reconstruction}, and sometimes {\it system invertibility}
(e.g., \cite{Basile69,Guido71,Hou98,Krya93,Basile73,Hirs79,Sing81,Fli86}).

Throughout this paper, we adopt the following terminology. {\it Observability} refers to the components of the state. {\it Unknown Input Observability} (abbreviated UIO) 
still refers only to the state components, but when the ODE model also includes the presence of unknown inputs (or time-varying parameters).
{\it Identifiability} refers to all the unknown parameters. When a parameter is time-varying we also use the term {\it unknown input reconstruction} to mean its identifiability.

We emphasize that the identifiability of constant parameters can be addressed using methods to study observability (e.g., through the use of the observability rank condition, \cite{Her77,Casti82}, when the system does not contain unknown time-varying parameters). Indeed, the time evolution of a constant parameter is trivially known (its time derivative vanishes, by definition). As a result, by including it in the state, we get a new ODE model in which the presence of this parameter is eliminated as it has become a component of the state (see the last paragraph of Section \ref{SectionProblemStatement}). 
In the presence of time-varying unknown parameters, the same approach does not provide the same immediate result.
~If a time-varying parameter is included in the state, the resulting ODE model will be characterized by a new unknown parameter, which is its time derivative (in general time-varying).

%

ODE identifiability analysis is present in a large variety of scientific domains and several interesting methods have been proposed. An exhaustive review of the methods proposed up to 2011 can be found in \cite{Miao11}. 
Some of these methods are very interesting and, in many cases, they can successfully be used to detect all the identifiability properties of a given ODE model. However,
they have the following fundamental limitations:

\begin{itemize}


\item They are not general. They are based on several properties that sometimes cannot be exploited in the presence of a given type of system nonlinearity (i.e., a given function $f$ and/or $h$ in Equation (\ref{EquationSystemDefinitionUIOGeneral})).

\item They cannot be executed automatically. In most of cases, they must be adapted to the specific case under investigation and this adjustment requires inventiveness by the user. In particular, this process cannot be carried out by simply following the steps of a systematic procedure (e.g., by running a code without human intervention).

\end{itemize}

The second limitation becomes particularly relevant in the case of time varying parameters where, in some cases, an erroneous application of these methods provided wrong results. Specifically, in our study about the identifiability of the ODE HIV model given in Section \ref{SectionHIVIdentifiability}, we obtain a fundamental new result that contradicts the result obtained in Section 6.2 of \cite{Miao11}, which was obtained by using one of these methods (a detailed explanation of the serious error made by the authors of \cite{Miao11} is available in \cite{arXivErratum}).

In the last decade, 
thanks to the progress in UIO, several new interesting approaches have been proposed.
Specifically,
the nonlinear UIO problem was
approached by introducing an extended state that includes the original state together with the unknown inputs and their time derivatives up to a given order \cite{Belo10,MED15,Villa19b,Maes19,Villa16b}. 
In \cite{Villa19b,Maes19,Villa16b}, this was used to study the identifiability of the unknown time-varying parameters. 
In particular, in \cite{Villa19b,Maes19,Villa16b}, several
automatic iterative algorithms were introduced. These algorithms work automatically and are able to check the identifiability of time-varying parameters. 
On the other hand, they suffer from the following  limitations:

\begin{itemize}

\item They do not converge, automatically. 
In particular, at each iterative step, the state is extended by including new time derivatives of the unknown inputs. Consequently, if the extended state is observable at a given step, convergence is achieved. However, if this were not the case, we can never exclude that, at a later step, the extended state becomes observable. Therefore, in the presence of unobservability, all these algorithms remain inconclusive.

\item Due to the previous state augmentation, the computational burden can easily become prohibitive after a few steps.


\end{itemize}

Very recently, we introduced the analytical solution of the nonlinear UIO \cite{IF22}, starting from the results/derivations presented in \cite{SIAMbook}\footnote{This analytical solution is also based on the results obtained in \cite{TAC19} and \cite{SARAFRAZI}, which deal with special systems, namely characterized by a single unknown input, and dynamics linear with respect to this unknown input.}.
In particular, the algorithm introduced in \cite{IF22} does not encounter the aforementioned limitations\footnote{Note that, the solution introduced in \cite{IF22} could need the inclusion of some of the UIs in the state. However, this inclusion is targeted and terminates in a finite number of steps, as soon as the extended system achieves its {\it highest unknown input degree of reconstructability} (see Section 5 in \cite{IF22}).} and provides a full answer to the problem of the state observability for any nonlinear ODE model, in the presence of unknown inputs (or time-varying parameters). On the other hand, the algorithm introduced in \cite{IF22} deals with the observability of the state and not with the identifiability of the time-varying parameters.
In \cite{IF22}, we exploited this analytical solution to provide a preliminary result also about the identifiability of the time-varying parameters (unknown input reconstruction).
One of the goals of this paper is the extension of this preliminary and partial result, and the introduction of a complete systematic procedure to study the identifiability of the time-varying parameters. Then, we adopt this systematic procedure to study the identifiability of a very popular HIV model (e.g., \cite{Miao11,Villa19b,Per99,Per02,Chen08}) and a Covid-19 model
\cite{VillaCovid,SEIARPribylova}.


The contributions of this paper are the following four:

\begin{enumerate}

\item Compact presentation of the analytical solution introduced in \cite{IF22} (Chapter \ref{ChapterSolutionUIO}) together with its basic mathematical foundation introduced in \cite{SIAMbook} (Chapter \ref{ChapterProblemDefinition}) in order to make this solution usable by a non specialist user.

\item Introduction of the systematic procedure to determine the identifiability of all the time-varying parameters of any ODE model (Chapter \ref{ChapterUIRec}).

\item Determination of the continuous transformations (one parameter Lie groups) that allow us to build the set of indistinguishable states and indistinguishable unknown inputs 
in the presence of unobservability and unidentifiability (Chapter \ref{ChapterIndistinguishableStatesAndUIsAndMI} and, in particular, the two systems of differential equations in (\ref{EquationDiffEqUICanonic}) and (\ref{EquationDiffEqStatesUI})).

\item 
Determination of the observability and the identifiability properties of a very popular HIV model, a Covid-19 model, and a genetic toggle switch model (Chapters \ref{ChapterHIV}, \ref{ChapterCovid}, and \ref{ChapterTS}, respectively), obtained by using the three above contributions.

\end{enumerate}

Note that the first two contributions consist of two systematic procedures that fully allow us to perform the observability analysis and the identifiability analysis, respectively. Specifically, they are Algorithm \ref{AlgoFull} in Chapter \ref{ChapterSolutionUIO} and Algorithm \ref{AlgoFullIDE} in Chapter 
\ref{ChapterUIRec}. In other words, given any nonlinear system, we can obtain the observability of the state and the identifiability of the parameters by simply following the steps of these systematic procedures.

In contrast, the third contribution does not consist of a systematic procedure. In particular, all that can be done automatically is the determination of the set of differential equations in (\ref{EquationDiffEqUICanonic}) and (\ref{EquationDiffEqStatesUI}) and the determination of the analytical solution of (\ref{EquationDiffEqUICanonic}), which is trivial. However, the method to solve (\ref{EquationDiffEqStatesUI}) depends on the specific case and it may also be not possible to obtain an analytical solution. 
Note that this third contribution is unnecessary for the observability and the identifiability analyses (which are fully achieved by running Algorithm \ref{AlgoFull} and Algorithm \ref{AlgoFullIDE}, respectively). On the other hand, the determination of the analytical solution of (\ref{EquationDiffEqStatesUI}), when possible, provides very useful insights and additional very interesting properties of the ODE model under investigation.

This third contribution adopts precisely the same definition of Lie symmetry introduced in \cite{Shi20,Shi21}. However, in \cite{Shi20,Shi21} this definition was then adopted by using the aforementioned extended state, which includes the original state and the unknown inputs with their time derivatives up to a given order (i.e., the augmented state introduced in
\cite{Belo10,MED15,Villa19b,Maes19,Villa16b} to investigate the state observability in the presence of unknown inputs).

Regarding the fourth contribution, 
our results improve previous results in the state of the art. In particular, in all the three cases, Algorithm \ref{AlgoFullIDE} automatically provides the unidentifiability of the time-varying parameters (for the HIV model and the genetic toggle switch model this result is in contrast with the state of the art).
In other words, in all the three cases, the time-varying parameters cannot be uniquely identified.
~In addition, for these ODE models, it is possible to obtain an analytical solution of the differential equation in (\ref{EquationDiffEqStatesUI}).
This allows us to determine indistinguishable unknown inputs that agree with the same outputs of the model (these models do not contain known inputs and, consequently, the only source of information about the state and the model parameters comes from the outputs). 
Regarding the HIV ODE model, in Section \ref{SectionHIVComparisonSOTA} we use a data set available in the literature and, by using the solution of (\ref{EquationDiffEqStatesUI}), we generate infinite different unknown inputs that produce the same outputs, starting from the true unknown input and the true states. This unequivocally shows the unidentifiability of the system, in contrast with the results available in the state of the art (e.g., see Section 6.2 of \cite{Miao11} and a detailed explanation of the serious error made by its authors, available in \cite{arXivErratum}).
Finally, we determine the minimal external information (external to the knowledge of the outputs) requested to uniquely determine the system parameters.
In Section \ref{SectionCovidNumerical}, we perform a similar analysis for the Covid-19 ODE model here investigated.

The paper is organized as follows.
Chapter \ref{ChapterProblemDefinition} provides a basic mathematical characterization of the problem together with basic mathematical concepts and definitions. It also introduces an elementary and illustrative example from robotics (from now on, the {\it case study}). We refer to the case study throughout all the paper to better illustrate all the theoretical concepts and the analytical procedures here introduced. In particular, each section ends
with a subsection where the theoretical concepts provided by the section are applied to the
case study. 
Chapter \ref{ChapterSolutionUIO} provides a summary of our recent solution of the nonlinear UIO introduced in \cite{IF22}.
Chapter  \ref{ChapterUIRec} introduces the systematic procedure to obtain the identifiability of any ODE model (i.e., Algorithm \ref{AlgoFullIDE}).
Chapter \ref{ChapterIndistinguishableStatesAndUIsAndMI} introduces the set of differential equations to generate indistinguishable unknown inputs and initial states, simultaneously (i.e., Equations (\ref{EquationDiffEqUICanonic}) and (\ref{EquationDiffEqStatesUI})). 
Chapters \ref{ChapterHIV}, \ref{ChapterCovid}, and \ref{ChapterTS} investigate the HIV model discussed in \cite{Miao11,Villa19b,Per99,Per02,Chen08}, 
the Covid-19 model introduced in \cite{VillaCovid,SEIARPribylova}, and the genetic toggle switch model discussed in \cite{Villa19b}. 
Finally, Chapter \ref{ChapterConclusion} provides our conclusion.


A short version of this paper is also available in \cite{arXivTAC}.

\chapter{Basic definitions and operations}\label{ChapterProblemDefinition}

\section{System characterization}\label{SectionProblem}

We start  from the following very general ODE model:

\begin{equation}\label{EquationSystemDefinitionUIOGeneral}
\left\{\begin{array}{ll}
 \dot{X} &=  f(X(t), ~t, ~U(t), ~Q, ~W(t))\\
  y &= h(X(t), ~t, ~U(t), ~Q, ~W(t)), \\
\end{array}\right.
\end{equation}

 where:
$X(t)\in\mathbb{R}^m$ is the state\footnote{In general, instead of $\mathbb{R}^m$, the state belongs to a differential manifold of dimension $m$. This also holds for the other quantities that appear in (\ref{EquationSystemDefinitionUIOGeneral}) (and also in (\ref{EquationSystemDefinitionUIO})), with the appropriate dimension. In this paper, for the sake of simplicity, we avoid the concept of differential manifold (the reader familiar with this concept can simply replace the Euclidean space of a given dimension with the differential manifold of the same dimension).},
$y(t)\in\mathbb{R}^p$ the output vector,
$U(t)\in\mathbb{R}^{m_u}$ the known system input vector, $Q\in\mathbb{R}^q$ the set of the unknown constant parameters, and
$W(t)\in\mathbb{R}^{m_w}$ the set of the unknown parameters that depend on time. Note that $W(t)$ can be regarded as a system input vector that, precisely as $U(t)$, acts on the system dynamics. However, it differs from $U(t)$ in two fundamental respects: (i) it cannot be assigned, and (ii) it is unknown.

The problem that we solve in this paper is the introduction of the analytic condition that fully characterizes the identifiability of the system parameters, both constant and time-varying. 
This is obtained by exploiting recent results on the unknown input observability problem.
To exploit these results, instead of (\ref{EquationSystemDefinitionUIOGeneral}) we adopt the following system characterization:

\begin{equation}\label{EquationSystemDefinitionUIO}
\left\{\begin{array}{ll}
  \dot{x} &=   g^0(x, t)+\sum_{k=1}^{m_u}f^k (x, t) u_k(t) +  \sum_{j=1}^{m_w}g^j (x, t) w_j(t)  \\
  y &= [h_1(x, t),\ldots,h_p(x, t)], \\
\end{array}\right.
\end{equation}

where:

\begin{itemize}

\item $x\in\mathbb{R}^n$ is the state. 

\item $y\in\mathbb{R}^p$ is the output vector.

\item $u_1(t), \ldots,u_{m_u}(t)$ are the known inputs.

\item $w_1(t), \ldots,w_{m_w}(t)$ are the unknown inputs or the unknown time-varying parameters.

\item $f^1,\ldots,f^{m_u},g^0,g^1,\ldots,g^{m_w}$ are $m_u+m_w+1$ vector fields, which are assumed to be smooth functions of $x$ and $t$.

\end{itemize}

From now on, we denote the system in (\ref{EquationSystemDefinitionUIO}) by $\Sigma$.
Note that the above characterization can easily account for the presence of constant parameters ($Q$) by including all of them in the state $x$ and by suitably setting $f^1,\ldots,f^{m_u},g^0,g^1,\ldots,g^{m_w}$, namely, by setting to zero all their components that yield 
 $\dt{Q}$ in (\ref{EquationSystemDefinitionUIO}).
Note that many nonlinear systems have the structure in (\ref{EquationSystemDefinitionUIO}). When this is not directly the case, it is possible to easily convert (\ref{EquationSystemDefinitionUIOGeneral}) to (\ref{EquationSystemDefinitionUIO}).
This is obtained by setting $u(t)=\dt{U}$, $w(t)=\dt{W}$ (with $u=[u_1,\ldots,u_{m_u}]$ and $w=[w_1,\ldots,w_{m_w}]$) and by including $U(t)$, $W(t)$ and $Q$ in the state (i.e., $x=[X^T, ~U^T, ~W^T, ~Q^T]^T$). 
In the rest of this paper, we directly refer to the characterization given in (\ref{EquationSystemDefinitionUIO}).

\section{Problem statement}\label{SectionProblemStatement}

Given the system characterized by (\ref{EquationSystemDefinitionUIO}) (which is equivalent to (\ref{EquationSystemDefinitionUIOGeneral})), the problem we solve in this paper is the introduction of 
the systematic procedures to automatically perform the observability analysis and the identifiability analysis.

\subsection{Observability analysis}

As we mentioned in the introduction, the paper summarizes (Chapter \ref{ChapterSolutionUIO}) the solution introduced in our previous paper \cite{IF22}.
The solution provides the analytical condition to easily check whether the state that characterizes the system in (\ref{EquationSystemDefinitionUIO}) is observable or not. It actually tells us more than this. It builds the entire observation space, i.e., the set of all the observable functions. An observable function is a scalar function of the state $x$ whose value, at a given time $t_0$, can be expressed in terms of the values that the known inputs and the system outputs take on a given time interval that includes $t_0$\footnote{The definition of observation space is provided in \cite{TAC19}, where Definition 2 defines this concept starting from the concept of indistinguishability. An observable function is precisely an element of this function space.}. In practice, even when the state is unobservable, the solution here summarized provides all the functions of the state that are observable. For instance, let us suppose to have a vehicle that moves on a plane and let us suppose that, with respect to a reference frame, we cannot obtain from the inputs and the outputs its Cartesian coordinates ($x_1$, $x_2$) but we can obtain its distance from the origin. If the state $x$ that characterizes this system contains $x_1$ and $x_2$, the state is unobservable. However, the function of the state $\theta(x):=\sqrt{x_1^2+x_2^2}$ is observable. Note that this is precisely the case of our illustrative example introduced in Section \ref{SectionCaseStudy} (the case study), where, however, we adopt polar coordinates ($\rho$ and $\phi$ instead of $x_1$ and $x_2$).

\subsection{Identifiability analysis}

Chapter \ref{ChapterUIRec} provides the systematic procedure to check if a given time-varying parameter (or unknown input) can be identified. 
If this is not the case, we analytically compute all the set of continuous transformations that transform the value taken by a given unidentifiable parameter in other indistinguishable values for that parameter. This is carried out in Chapter \ref{ChapterIndistinguishableStatesAndUIsAndMI} and the set of continuous transformations are the differential equations in (\ref{EquationDiffEqUICanonic}) and (\ref{EquationDiffEqStatesUI}).

\vskip .2cm

Note that, the identifiability of constant parameters is actually obtained by performing the above observability analysis. Indeed, the constant parameters are included in the state. To check if a given constant parameter $q$ is identifiable, we need to trivially check if the function $\theta(x):=q$ belongs to the observation space (i.e., we need to check if its gradient ($\nabla\theta$) belongs to the observability codistribution ($\OBS$), which is automatically computed by Algorithm \ref{AlgoFull}).

\section{Basic algebraic operations and notions}\label{SectionAlgebraicOperations}

In accordance with the control theory literature, we use the term {\it distribution} to denote the span of a set of $d\le n$ vector fields\footnote{Simply speaking, a vector field is a column vector of dimension $n$ whose components are functions of $x$. Actually, the correct definition of vector field must take into account its tensor nature, i.e., how its components change under a generic change of coordinates. In particular, a vector is a tensor of type $(0,~1)$ (see Section 2.2 of \cite{SIAMbook}).}, $\Delta=$span$\{
\tau^1,~\tau^2,\ldots,\tau^d\}$. It is basically a vector space 
that depends on $x\in\Mn$.
Similarly, we also use the term {\it codistribution} to denote the span of a set of $s\le n$ covector fields\footnote{A covector field is the dual of a vector field. Simply speaking, it is a row vector of dimension $n$ whose components are functions of $x$. Actually, the correct definition of covector field must take into account its tensor nature, i.e., how its components change under a generic change of coordinates. In particular, a covector is a tensor of type $(1,~0)$ (see Section 2.2 of \cite{SIAMbook}).}, $\Omega=$span$\{
\omega_1,~\omega_2,\ldots,\omega_s\}$. Again, it is a vector space that depends on $x\in\Mn$.
In this paper, we always consider non singular distributions and codistributions. In other words, we always refer to an open set of $\Mn$ where the dimension of the considered distribution (or codistribution) takes the same value.

We remind the reader of the definitions of the Lie derivative along a vector field $f$ of a scalar field $\lambda$, of a vector field $\tau$, and of a covector field $\omega$. We have, respectively:
\begin{equation}\label{EquationLieSca}
\Li_f\lambda=\frac{\partial\lambda}{\partial x}\cdot f
\end{equation}
\begin{equation}\label{EquationLieVec}
\Li_f\tau=[f,~\tau]=\frac{\partial\tau}{\partial x}\cdot f-\frac{\partial f}{\partial x}\cdot\tau
\end{equation}
\begin{equation}\label{EquationLieCov}
\Li_f\omega=f^T\cdot\left(
\frac{\partial\omega^T}{\partial x}
\right)^T+
\omega\cdot\left(
\frac{\partial f}{\partial x}
\right)
\end{equation}

The square brackets in (\ref{EquationLieVec}) are called {\it Lie brackets}. In the special case when the covector field $\omega$ is the gradient of a scalar field (i.e., $\omega=\nabla\lambda$), Equations (\ref{EquationLieCov}) and (\ref{EquationLieSca}) yield:

\[
\Li_f\nabla\lambda=\nabla\Li_f\lambda.
\]

We use the following definitions/notation:

\begin{itemize}

\item Given a distribution $\Delta$ and a vector field $f$, we set $\mathcal{L}_f \Delta$ the distribution that is the span of all the vectors $\mathcal{L}_f\tau=[f,\tau]$, for any $\tau\in\Delta$. Similarly,
given a codistribution $\Omega$ and a vector field $f$, we set $\mathcal{L}_f \Omega$ the codistribution that is the span of all the covectors $\mathcal{L}_f\omega$ for any $\omega\in\Omega$.

\item We use the term {\it integrable codistribution} to denote the special codistribution that can be generated by the gradients of a set of scalar fields (i.e.,
$\Omega=$span$\{\nabla\lambda_1,~\nabla\lambda_2,\ldots,\nabla\lambda_s\}$). In this case, we often use the term {\it generators} to denote the scalar functions $\lambda_1,\ldots,\lambda_s$ (instead of their gradients).

\item Given two vector spaces $V_1$ and $V_2$,  $V_1{+}V_2$ is their sum, i.e., the span of all the generators of $V_1$ and $V_2$. When we have $k(>2)$ vector spaces $V_1,\ldots, V_k$, we denote their sum in the compact notation $\sum_{i=1}^kV_i$.

\item Greek indices take non negative values (e.g., $\alpha=0,1,2,\ldots$). Latin indices take positive values (e.g., $i=1,2,\ldots$).

\end{itemize}

In addition, we remind the reader of a further algebraic operation introduced in \cite{IF22} and called the {\it autobracket}. Given a set of $L+1$ vector fields, $\tau^0,\tau^1,\ldots,\tau^L$, and a non singular two-index tensor $\sigma^\alpha_\beta$, $\alpha,\beta=0,1,\ldots,L$, the autobracket of a vector field $f$ along $\gamma$ ($\gamma=0,1,\ldots,L$), denoted by $^\sigma_\tau[f]^\gamma$, is defined as follows:

\begin{equation}\label{EquationAutobracket}
 ^\sigma_\tau[f]^\gamma = 
 \sum_{\beta=0}^L \sigma^\gamma_\beta~ [\tau^{\beta},~f]~+~\delta^\gamma_0\frac{\partial f}{\partial  t},~~~\gamma=0,1,\ldots,L,
\end{equation}
\noindent where the square brackets on the right hand side are the Lie brackets and $\delta^\gamma_0$ (the Kronecker delta) is 1 when $\gamma=0$ and is 0 otherwise.
In \cite{IF22}, and in this paper, we use this operation for two settings (note that in \cite{IF22} the operation was directly defined for these settings):

\begin{enumerate}

\item $L=m_w$, $\tau^0=g^0, ~\tau^1=g^1,\ldots,\tau^{m_w}=g^{m_w}$, which are the $m_w+1$ vector fields that appear in (\ref{EquationSystemDefinitionUIO})), and $\sigma=\nu$, which is the inverse of $\mu$ defined in Section \ref{SubSectionMuNu} (see Equation (\ref{EquationTensorMSynchro})). For simplicity, we denote this operation by $[\cdot]^\gamma$ (instead of $^\nu_g[\cdot]^\gamma$)

\item $L=m$, $\tau^0=g^0, ~\tau^1=g^1,\ldots,\tau^{m}=g^{m}$, which are the first $m+1$ vector fields that appear in (\ref{EquationSystemDefinitionUIO}) (once the unknown inputs are re-ordered, in accordance with the $\mathcal{R}$ operation, defined in Section \ref{SubSectionReorder}), and 
$\sigma=\nu$, which is the inverse of $\mu$ defined in Section \ref{SubSectionMumNum} (see Equation (\ref{EquationTensorMSynchrom})). For simplicity, also in this case, we denote this operation by $[\cdot]^\gamma$ (instead of $~^{\nu}_{g}[\cdot]^\gamma$).

\end{enumerate}

We have $L+1$ possible first order autobrackets ($\gamma=0,1,\ldots,L$).
By applying the autobracket $k$ consecutive times, we obtain $(L+1)^k$ vector fields that we denote by $^\sigma_\tau[f]^{(\gamma_1,\ldots,\gamma_k)}$ and we have $
^\sigma_\tau[f]^{(\gamma_1,\ldots,\gamma_{k-1},\gamma_k)}=
{\scriptsize\begin{array}{c}
 \sigma\\
\tau\\
\end{array}}\hskip-.25cm
\left[~^\sigma_\tau[f]^{(\gamma_1,\ldots,\gamma_{k-1})}\right]^{\gamma_k}
$.

Finally, given a distribution $\Delta$, we set $^\sigma_\tau\left[\Delta\right]^\gamma$ the span of all the vectors $^\sigma_\tau\left[f\right]^\gamma$, for any $f\in\Delta$.

\section{Minimal Invariant codistributions and distributions}\label{SectionInvariantDistAndCod}

Given a distribution $\Delta$ and a set of vector fields $\tau^1,\ldots,\tau^d$ defined on $\Mn$, we denote by
\[
\left.\left<\tau^1,\ldots,\tau^d~\right|~\Delta\right>
\]
the smallest distribution that contains $\Delta$ and such that, for any $f\in\Delta$, we have $\Li_{\tau^i}f=[\tau^i,~f]\in\left.\left<\tau^1,\ldots,\tau^d~\right|~\Delta\right>$, for any $i=1,\ldots,d$.
This minimal distribution 
can be easily computed by a simple recursive algorithm, which is \cite{Isi95}:

\begin{equation}\label{EquationAlgorithmsMinimalDist}
\left\{\begin{array}{ll}
 \Delta_0 &=  \Delta\\
 \Delta_{k+1} &=  \Delta_k+\sum_{i=1}^d\Li_{\tau^i}\Delta_k\\
\end{array}\right.,
\end{equation}
where we provide the initialization and the recursive step.
This algorithm converges at the smallest integer $j$ for which $\Delta_j=\Delta_{j-1}$ and $j\le n-\dim\{\Delta\}+1$. In addition, $\left.\left<\tau^1,\ldots,\tau^d~\right|~\Delta\right>=\Delta_{j-1}$\footnote{Throughout this paper, when we provide the convergence properties of iterative algorithms that build a family of distributions or codistributions (one distribution or codistribution constructed at each iterative step), we always refer to an open set where these distributions or codistributions are non singular.}.

Similarly, given a codistribution $\Omega$ and a set of vector fields $\tau^1,\ldots,\tau^d$ defined on $\Mn$, we denote by
\[
\left.\left<\tau^1,\ldots,\tau^d~\right|~\Omega\right>
\]
the smallest codistribution that contains $\Omega$ and such that, for any $\omega\in\Omega$, we have $\Li_{\tau^i}\omega\in\left.\left<\tau^1,\ldots,\tau^d~\right|~\Omega\right>$, for any $i=1,\ldots,d$.
This minimal codistribution 
can be easily computed by a simple recursive algorithm, which is \cite{Isi95}:

\begin{equation}\label{EquationAlgorithmsMinimalCod}
\left\{\begin{array}{ll}
 \Omega_0 &=  \Omega\\
 \Omega_{k+1} &=  \Omega_k+\sum_{i=1}^d\Li_{\tau^i}\Omega_k\\
\end{array}\right.,
\end{equation}
This algorithm converges at the smallest integer $j$ for which $\Omega_j=\Omega_{j-1}$ and $j\le n-\dim\{\Omega\}+1$. In addition, $\left.\left<\tau^1,\ldots,\tau^d~\right|~\Omega\right>=\Omega_{j-1}$.

Note that the execution of the above algorithms only requires to compute the Lie derivatives along $\tau^1,\ldots,\tau^d$ of the generators of the previous distribution / codistribution.



In the following, we also need two 
modified versions of the two above algorithms. In particular, these new algorithms compute the following two quantities:

\begin{enumerate}

\item
Given a distribution $\Delta$, a set of vector fields $\tau^0,\tau^1,\ldots,\tau^L$ defined on $\Mn$, and a non singular two-index tensor $\sigma$, we denote by
\begin{equation}\label{EquationDeltaInvAut}
\left.\left<~^\sigma_\tau\left[\cdot\right]^\cdot~\right|~\Delta\right>
\end{equation}
the smallest distribution that contains $\Delta$ and such that, for any $f\in\Delta$, we have $^\sigma_\tau\left[f\right]^\gamma\in\left.\left<~^\sigma_\tau\left[\cdot\right]^\cdot~\right|~\Delta\right>$, for any $\gamma=0,1,\ldots,L$
(where the operation $^\sigma_\tau\left[\cdot\right]^\cdot$ is defined at the end of Section
\ref{SectionAlgebraicOperations}, by Equation (\ref{EquationAutobracket})).
This minimal distribution 
can be easily computed by a simple recursive algorithm, which is:

\begin{equation}\label{EquationAlgorithmsMinimalDistAut}
\left\{\begin{array}{ll}
 \Delta_0 &=  \Delta\\
 \Delta_{k+1} &=  \Delta_k+\sum_{\gamma=0}^L ~^\sigma_\tau\left[\Delta_k\right]^\gamma\\
\end{array}\right.,
\end{equation}
where we provide the initialization and the recursive step.
This algorithm converges at the smallest integer $j$ for which $\Delta_j=\Delta_{j-1}$ and $j\le n-\dim\{\Delta\}+1$. In addition, $\left.\left<~^\sigma_\tau\left[\cdot\right]^\cdot~\right|~\Delta\right>=\Delta_{j-1}$.
Note that also the execution of the above algorithm only requires to compute the autobracket along $\gamma=0,1,\ldots,L$ of the generators of the previous distribution.

\item Given the codistribution $\Omega$ and the three sets of vector fields: $\tau^1,\ldots,\tau^{d_1}$, $\xi^1,\ldots,\xi^{d_2}$, and $\zeta^1,\ldots,\zeta^{d_3}$, we denote by
\begin{equation}\label{EquationOmgInvIf}
\left<\left.\tau^1,\ldots,\tau^{d_1}, \left\{ 
\begin{array}{lll}
 \xi^1, & \ldots, & \xi^{d_2}\\
 \zeta^1, & \ldots, & \zeta^{d_3}\\
\end{array}
\right\}~\right|\Omega\right>
\end{equation}
the codistribution computed by the algorithm in (\ref{EquationAlgorithmsMinimalCod}), where the sum $\sum_{i=1}^d$ at the recursive step 
includes all the $d_1$ vector fields $\tau^1,\ldots,\tau^{d_1}$. In addition, by denoting with $\omega$ a generator of $\Omega_k$, the codistribution $\Omega_{k+1}$ also includes, among its generators, all the $d_2$ Lie derivatives of $\omega$ along the vector fields $\xi^1,\ldots,\xi^{d_2}$ if and only if $\Li_{\zeta^l}\omega$ vanishes for all $l=1,\ldots,d_3$. The algorithm is then interrupted at the smallest integer $j$ such that $\Omega_j=\Omega_{j-1}$.

%
%
%
%
%
%

\end{enumerate}
 
\section{Unknown input degree of reconstructability}\label{SectionUIDeg}

The derivations in \cite{IF22} and in this paper need the definition of the {\it unknown input degree of reconstructability} (Definition \ref{DefinitionUIDegReconstrFromF} below), which is based on the definition of the {\it unknown input reconstructability matrix} (Definition \ref{DefinitionRM} below). These definitions were introduced in \cite{IF22}. For the sake of completeness, we report them below.

\begin{definition}[Unknown input reconstructability matrix 
]\label{DefinitionRM}
Given the system $\Sigma$ in (\ref{EquationSystemDefinitionUIO}) and $k$ scalar functions of the state, $\lambda_1(x),\ldots, \lambda_k(x)$, the unknown input reconstructability matrix of $\Sigma$ from $\lambda_1,\ldots, \lambda_k$ is defined as follows:

\begin{equation}\label{EquationRM}
\mathcal{RM}\left( \lambda_1,\ldots, \lambda_k\right)
:=
\left[\begin{array}{cccc}
\Li_{g^1} \lambda_1 & \Li_{g^2} \lambda_1 & \ldots & \Li_{g^{m_w}} \lambda_1 \\
\Li_{g^1} \lambda_2 & \Li_{g^2} \lambda_2 & \ldots & \Li_{g^{m_w}} \lambda_2 \\
\ldots &\ldots &\ldots &\ldots \\
\Li_{g^1} \lambda_k & \Li_{g^2} \lambda_k & \ldots & \Li_{g^{m_w}} \lambda_k \\
\end{array}
\right]
\end{equation}
\end{definition}

\begin{definition}[Unknown input degree of reconstructability
]\label{DefinitionUIDegReconstrFromF}
Given the system $\Sigma$ in (\ref{EquationSystemDefinitionUIO}), and the functions $\lambda_1,\ldots, \lambda_k$, the unknown input degree of reconstructability of $\Sigma$ from $\lambda_1,\ldots, \lambda_k$ is the rank of $\mathcal{RM}\left(\lambda_1,\ldots, \lambda_k\right)$. 
\end{definition}

By construction, given the system characterized by (\ref{EquationSystemDefinitionUIO}), its unknown input degree of reconstructability from any set of scalar functions cannot exceed $m_w$.
In \cite{IF22}, we introduced the definitions of {\it Canonic system with respect to its unknown inputs} and {\it system in Canonical Form with respect to its unknown inputs}. The former is a system such that its unknown input degree of reconstructability from all the observable functions is $m_w$. The latter is a system such that its unknown input degree of reconstructability from
the output functions ($h_1,\ldots,h_p$) is $m_w$. As the output functions are observable, a system in canonical form is certainly canonic. However, the viceversa does not hold, in general. Given a canonic system that is not in canonical form, we say that it has been set in canonical form, as soon as we are able  to somehow determine a set of observable functions such that the unknown input degree of reconstructability from them is $m_w$.

\section{Definition of the case study}\label{SectionCaseStudy}

\begin{figure}[htbp]
\begin{center}
\includegraphics[width=.5\columnwidth]{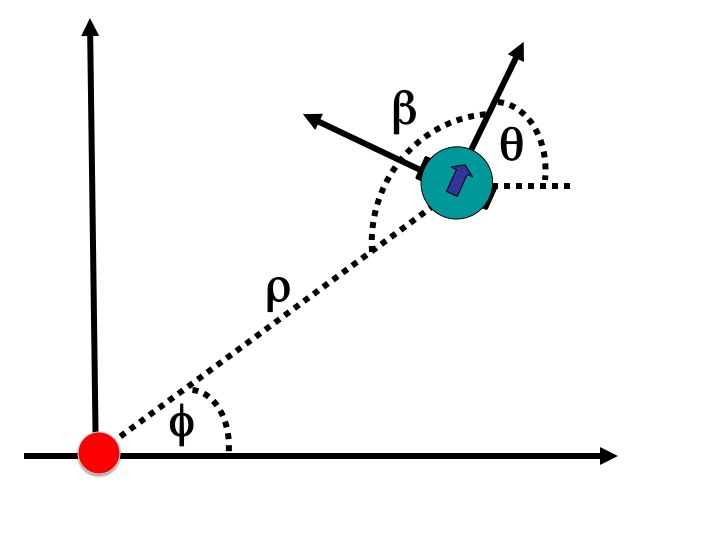}
\caption{The vehicle polar coordinates ($\rho,~\phi$), the orientation ($\theta$), and the angle measured by the camera ($\beta$).} \label{FigUnicycle}
\end{center}
\end{figure}

We introduce an elementary example of an ODE model that will be used in the rest of the paper to illustrate all the concepts and methods that will be introduced. 

We consider a vehicle that moves on a $2D$-environment. The configuration of the vehicle in a global reference frame can be characterized through the vector $[\rho, ~\phi, ~\theta]^T$, where $\rho$ and $\phi$ are the polar vehicle coordinates, and $\theta$ is the vehicle orientation (see Figure \ref{FigUnicycle} for an illustration). We assume that the dynamics of this vector satisfy the unicycle model. In polar coordinates we have:


\begin{equation}\label{EquationSimpleExampeDynamics}
\left[\begin{array}{ll}
  \dot{\rho} &= v \cos(\theta-\phi) \\
  \dot{\phi} &= \frac{v}{\rho} \sin(\theta-\phi) \\
  \dot{\theta} &= \omega \\
\end{array}
\right.
\end{equation}

\noindent where $v$ and $\omega$ are the linear and the rotational vehicle speed, respectively, and they are the system inputs. We assume that the vehicle is equipped with a monocular camera that provides the bearing angle of the origin in the vehicle local frame (at the origin there is a landmark perceived by the camera). In other words, the camera provides the angle $\beta$ in Figure \ref{FigUnicycle}.
We can express $\beta$ in terms of the vehicle configuration. We have: $\beta=\pi-(\theta-\phi)$.
Throughout this paper we will refer to the two scenarios defined by the following settings:

\begin{enumerate}

\item $v$ is known and $\omega$ is unknown.

\item $\omega$ is known and $v$ is unknown.

\end{enumerate}
In both scenarios, the system is directly a special case of (\ref{EquationSystemDefinitionUIO}). In particular, $x=[\rho, ~\phi, ~\theta]^T$, $g^0=[0, 0, 0]^T$, $m_u=m_w=1$, $p=1$, $h_1(x)=\phi-\theta$ (we ignore the constant term $\pi$ in $\beta$, which does not affect the observability of the state), and, in the first scenario $f^1=
\left[
\cos(\theta-\phi),~
\frac{\sin(\theta-\phi)}{\rho},~
0
\right]^T$, and
$g^1=
\left[
0,~
0,~
1
\right]^T$,
while, in the second scenario, 
$f^1=
\left[
0,~
0,~
1
\right]^T$, and
$g^1=
\left[
\cos(\theta-\phi),~
\frac{\sin(\theta-\phi)}{\rho},~
0
\right]^T$.

\chapter{Systematic procedure to perform the observability analysis}\label{ChapterSolutionUIO}

This section provides a very concise summary of the solution of the nonlinear unknown input observability problem. In particular, it is written to make a non specialist able to implement it. The interested reader can find all the theoretical foundation and proofs in \cite{IF22}.

The solution consists of a recursive procedure (Algorithm \ref{AlgoFull}, which uses Algorithm \ref{Algo***}, Algorithm \ref{AlgoOmegagNonAbelm} and Algorithm \ref{AlgoDeltaNonAbelm}\footnote{Often, when we refer to Algorithm \ref{AlgoFull}, we actually mean all the four algorithms (\ref{AlgoFull}-\ref{AlgoDeltaNonAbelm}).}). Given a system characterized by (\ref{EquationSystemDefinitionUIO}), Algorithm \ref{AlgoFull} automatically builds the entire observation space in a finite number of iterations. In particular, it builds the so-called {\it observability codistribution}, which is, by definition, the span of the gradients of all the observable functions.
Algorithm \ref{AlgoFull} builds the generators of the observability codistribution. To this regard, 
note that, in the algorithm, statements like "${\Omega}=\textnormal{span}\left\{\nabla h_1,\ldots,\nabla h_p\right\}$" (e.g., Line \ref{AlgoLineOmegaINIT} of Algorithm \ref{AlgoFull}) require no action. The algorithm builds the codistribution by building a set of its generators. When it recursively updates the codistribution it simply computes its new generators (and this is obtained by applying certain operations directly on the generators of the previous codistribution or by computing new covectors starting from the quantities that define the original system $\Sigma$ in (\ref{EquationSystemDefinitionUIO})).
In the following subsections, we provide all the ingredients necessary to implement Algorithm \ref{AlgoFull}.

\begin{algorithm}\caption{The algorithm to obtain the state observability in the presence of time varying parameters (or unknown inputs). Comments are in blue.}

\begin{algorithmic}[1]

\State Set $\Sigma$ the system in (\ref{EquationSystemDefinitionUIO}), 
$\Omega=\textnormal{span}\left\{\nabla h_1,\ldots,\nabla h_p\right\}$, and the boolean Continue=True.\label{AlgoLineOmegaINIT}

\Loop\Comment{{\blue This is the MAIN loop}}

\If{$\uideg \left({\Omega}\right)==m_w$ }
{\bf break} main loop.
\EndIf

\State $m=\uideg \left({\Omega}\right)$.

\State $[\tih_1,\ldots, \tih_m]=\mathcal{S}(\Sigma,~\Omega)$,~~
$\Sigma=\mathcal{R}(\Sigma,~\tih_1,\ldots, \tih_m)$.
\State \textnormal{Compute} $\mu$, $\nu$, $\widehat{g}^\alpha$ (Eqs (\ref{EquationTensorMSynchrom}-\ref{Equationgalpham})).

\State Run Algorithm \ref{Algo***}\label{AlgoLineRunAlgo***}

\If{Finish}
$\OBS=\Omega_*$, $\E=\Sigma$, Continue=false, and
{\bf break} main loop.
\EndIf

\If{$m_u>0$}

\State $\Sigma=\Sigma_*$, and Compute $\mu$, $\nu$, $\giat^\alpha$ (Eqs (\ref{EquationTensorMSynchrom}-\ref{Equationgalpham})).

\State Compute $\chi_*=\psi_{k_*-1}^{i_*}$ (Eq (\ref{Equationpsik+1}))\Comment{{\blue Carried out recursively in $\mathcal{W}$. Then, it is expressed in $\mathcal{V}$.}}
\State $\Omega=\Omega+
 \textnormal{span} \left \{\Li_{\chi_*} \nabla \tih_{q_*} \right \}$\Comment{{\blue When $k_*=1$ $\chi_*=f^{i_*}$}}

\State $\Omega=\left<\left.f^1,\ldots,f^{m_u}, \left\{ 
\begin{array}{lll}
 & \dtv{\Li}_{\giat}&\\
 \giat^{m+1}, & \ldots, & \giat^{m_w}\\
\end{array}
\right\}~\right|\Omega\right>$. \Comment{{\blue This is the nested loop of Algorithm 9 in \cite{IF22}}}\label{AlgoLineNested}

\Else

\State $\Omega=\left<\left. \left\{ 
\begin{array}{lll}
 & \dtv{\Li}_{\giat}&\\
 \giat^{m+1}, & \ldots, & \giat^{m_w}\\
\end{array}
\right\}~\right|\Omega\right>$. \Comment{{\blue This is the nested loop of Algorithm 9 in \cite{IF22}}}\label{AlgoLineNestedNonMisto}

\EndIf

\State \textnormal{Reset} $[\Sigma, ~\Omega]=\mathcal{A}^{-}(\Sigma,~\Omega)$.\Comment{{\blue This pass again in $\mathcal{W}$.}}


\EndLoop

\If{Continue}\label{AlgoLineFINIT}
\State $[\widetilde{h}_1,\ldots, \widetilde{h}_{m_w}]=\mathcal{S}(\Sigma,~\Omega)$.\label{AlgoLineFINITSelection}
\State \textnormal{Compute} $\mu$, $\nu$, $\widehat{g}^\alpha$ (Eqs (\ref{EquationTensorMSynchro}-\ref{Equationgalpha})), \textnormal{and} $\tobs$ (Eq. (\ref{EquationTOBSDef})).\label{AlgoLineFINITMuNuTOBS}
    \State $\OBS=\left<f^1,\ldots,f^{m_u}, \widehat{g}^0,\widehat{g}^1,\ldots,\widehat{g}^{m_w}~\left|~\Omega+\tobs\right.\right>$ \textnormal{and set} $\E=\Sigma$.\label{AlgoLineFINALSTEP}
\EndIf\label{AlgoLineFINITEND}

\end{algorithmic}
\label{AlgoFull}
\end{algorithm}

\begin{algorithm}
\caption{This algorithm returns the Boolean Finish. When Finish==True, it also returns $\Omega_*$. When Finish==False (and $m$ and $m_u$ are both non zero) it also returns $\Sigma_*$, $k_*$, $i_*$, and $q_*$.}

\begin{algorithmic}[1]

\If{$(m_u==0)|(m==0)$}
\State $\Omega_*=\left<\left. \giat^0, \ldots, \giat^m~\right|\Omega\right>$.
\If{($\giat^{m+1}\in\Omega_*^\bot) \And(\giat^{m+2}\in\Omega_*^\bot) \And\ldots\And(\giat^{m_w}\in\Omega_*^\bot$)}
\State Finish=True.
\Else
\State Finish=False.
\EndIf

\Else\label{AlgoLine***1}
\State Run Algorithms \ref{AlgoOmegagNonAbelm} and \ref{AlgoDeltaNonAbelm} to compute $\widehat{k}^m=s^m_x+r^m$.

\For{$k=1,\ldots,\widehat{k}^m$}

\For{$i=1,\ldots,m_u$~
$\And$~$q=1,\ldots,m$}

  \State $\Omega=\Omega+
 \sum_{\alpha_1=0}^m\ldots\sum_{\alpha_{k-1}=0}^m
 \textnormal{span} \left \{\mathcal{L}_{[f^i]^{(\alpha_1,\ldots,\alpha_{k-1})}} \nabla \tih_q \right \}$

\State $\Omega_*=\left<\left.f^1,\ldots,f^{m_u}, 
 \giat^0, \ldots, \giat^m~\right|\Omega\right>$.

\If{($\giat^{m+1}\in\Omega_*^\bot)\And(\giat^{m+2}\in\Omega_*^\bot)\And\ldots\And(\giat^{m_w}\in\Omega_*^\bot$)}
\State Compute $m$, $\mu$, $\nu$, $\widehat{g}^\alpha$ (Eqs (\ref{EquationTensorMSynchrom}-\ref{Equationgalpham})).
\Else
\State Finish=False. 
\State $\Sigma_*=\Sigma$, $k_*=k$, $i_*=i$, $q_*=q$.
\State {\bf return}
\EndIf

\EndFor
\EndFor

\State Finish=True.

\EndIf

\end{algorithmic}
\label{Algo***}
\end{algorithm}

\begin{algorithm}
\caption{The algorithm that returns the integer $s^m_x$ at Line \ref{AlgoLine***1} of Algorithm \ref{Algo***}.}

\begin{algorithmic}[1]
\State  \textnormal{Set $\Sigma'=\Sigma$, $m$, $\tih_1,\ldots,\tih_{m}$, as at Line \ref{AlgoLineRunAlgo***} of Algorithm \ref{AlgoFull}}.
 \State \textnormal{Set $k=0$, $\Omega^m_0=\textnormal{span}\left\{\nabla\tih_1,\ldots,\nabla\tih_{m}\right\}$, and $^x\Omega^m_0=\Omega^m_0$}.
\Loop
\State \textnormal{Set} $k=k+1$.
\State \textnormal{Reset} $[\Sigma', ~\Omega^m_{k-1}]=\mathcal{A}(\Sigma',~m, ~\Omega^m_{k-1})$\label{AlgoOmegaLineRESET}
\State $\Omega^m_k=\Omega^m_{k-1}+\dt{\Li}_{g^0} \Omega^m_{k-1}+\sum_{j=1}^{m}\Li_{g^j} \Omega^m_{k-1}$.\label{AlgoOmegaLineSum}

\State $^x\Omega^m_k=\mathcal{D}_x(\Omega^m_k)$.
 \If{$^x\Omega^m_k==~^x\Omega^m_{k-1}$}
 \State \textnormal{Set $s^m_x=k$ {\bf then exit}}.
 \EndIf
 \EndLoop
\end{algorithmic}\label{AlgoOmegagNonAbelm}
\end{algorithm}

\begin{algorithm}
\caption{The algorithm that returns the integer $r^m$ at Line \ref{AlgoLine***1} of Algorithm \ref{Algo***}.}
\begin{algorithmic}[1]
\State  \textnormal{Set $\Sigma'=\Sigma$ and $m$ as at Line \ref{AlgoLineRunAlgo***} of Algorithm \ref{AlgoFull}}.
 \State \textnormal{Set $k=0$, $\Delta_0=\textnormal{span}\left\{f^1,\ldots,f^{m_u} \right\}$}.

\Loop
\State \textnormal{Set} $k=k+1$.
\State \textnormal{Reset} $[\Sigma', ~\Delta_{k-1}]=\mathcal{A}(\Sigma',~m, ~\Delta_{k-1})$\label{AlgoDeltaLineRESET}

\State $\Delta_k=~\Delta_{k-1}+\sum_{\beta=0}^m ~\left[\Delta_{k-1}\right]^\beta$

 \If{$\Delta_k~==~\Delta_{k-1}$}
 \State \textnormal{Set $r^m=k-1$ {\bf then exit}}.
 \EndIf

\EndLoop

\end{algorithmic}\label{AlgoDeltaNonAbelm}
\end{algorithm}

\section{Basic operations in Algorithm \ref{AlgoFull}}\label{SectionUIOBasicOperations}

\subsection{$\boldsymbol{\deg_w(\Omega)}$ operation}\label{SubSectionDeg}

In Section \ref{SectionUIDeg}, we provided the definition of unknown input degree of reconstructability from a set of scalar functions.
Given an integrable codistribution
$\Omega=\textnormal{span}
\left\{\nabla\lambda_1,\ldots, \nabla\lambda_k\right\}$,
we define the unknown input degree of reconstructability of $\Sigma$ from $\Omega$ the unknown input degree of reconstructability from a set of generators of $\Omega$ (e.g., from $\lambda_1,\ldots, \lambda_k$). We denote it by $\uideg\left(\Omega\right)$.

 The main loop of Algorithm \ref{AlgoFull} starts by checking if $\uideg(\Omega)==m_w$.
 If this is the case, the main loop is interrupted.
Then, Lines \ref{AlgoLineFINIT}-\ref{AlgoLineFINALSTEP} of Algorithm \ref{AlgoFull} are executed and Algorithm \ref{AlgoFull} ends. For the clarity sake, we devote Section \ref{SectionSolutionUIOCanonic} to describe the behaviour of Algorithm  \ref{AlgoFull} in this case. Before, we provide the definition of further operations adopted by Algorithm \ref{AlgoFull}.

\subsection{$\boldsymbol{\mathcal{S}(\Sigma,~\Omega)}$ operation}\label{SubSectionSelection}

Let us consider a system $\Sigma$ that is characterized by (\ref{EquationSystemDefinitionUIO}) and an integrable codistribution $\Omega$. Let us set $m=\uideg\left(\Omega\right)$. The $\mathcal{S}(\Sigma,~\Omega)$ operation provides a set of $m$ scalar functions, denoted by $\widetilde{h}_1,\ldots, \widetilde{h}_m$, that make the unknown input reconstructability matrix full rank.
The operation is denoted by $\mathcal{S}$ because is the {\it Selection} of the aforementioned $\widetilde{h}_1,\ldots, \widetilde{h}_m$ from the generators of $\Omega$.
The execution of this operation is immediate because the generators of $\Omega$ are always available when executing Algorithm \ref{AlgoFull}.
In this paper, we use the notation $[\widetilde{h}_1,\ldots, \widetilde{h}_m]=\mathcal{S}(\Sigma,~\Omega)$ to denote the outputs of this operation.

\subsection{$\boldsymbol{\mathcal{R}(\Sigma,~\lambda_1,\ldots, \lambda_m)}$ operation}\label{SubSectionReorder}

Let us consider a system $\Sigma$ that is characterized by (\ref{EquationSystemDefinitionUIO}) and a set of $m<m_w$ scalar functions, $\lambda_1,\ldots, \lambda_m$, such that $\textnormal{rank}\left(\mathcal{RM}\left(\lambda_1,\ldots, \lambda_m\right)
\right)=m$. 
The $\mathcal{R}(\Sigma,~\lambda_1,\ldots, \lambda_m)$ operation provides a new system that is obtained from $\Sigma$ by reordering the unknown inputs. In particular, they are reordered in such a way that the square submatrix that consists of the first $m$ columns of the unknown input reconstructability matrix from 
$\lambda_1,\ldots, \lambda_m$, is non singular.
The operation is denoted by $\mathcal{R}$ because it is a {\it Reordering} of the unknown inputs, as explained above.
Its execution is immediate. It suffices to extract from
$\mathcal{RM}\left(\lambda_1,\ldots, \lambda_m \right)$ a set of $m$ independent columns.
In this paper, we use the notation $\Sigma'=\mathcal{R}(\Sigma,~\lambda_1,\ldots, \lambda_m)$ to denote the new ordered system. 

\subsection{$\boldsymbol{\mathcal{A}(\Sigma,~m)}$ operation}\label{SubSectionSigma++}
Let us consider a system $\Sigma$ that is characterized by (\ref{EquationSystemDefinitionUIO}) and an integer $m<m_w$. 
This operation provides a new extended system, 
defined as follows. It is obtained by introducing a 
new extended state that includes the last $d=m_w-m$ unknown inputs. We have:

\begin{equation}\label{EquationAugmentationState}
x\rightarrow
[x^T, w_{m+1}, \ldots, w_{m_w}]^T
\end{equation}

Starting from (\ref{EquationSystemDefinitionUIO}) we obtain the dynamics of the above extended state. We obtain a new system that still satisfies (\ref{EquationSystemDefinitionUIO}). It is still characterized by $m_u$ known inputs and $m_w$ unknown inputs. All the $m_u$ known inputs and the first $m$ unknown inputs
coincide with the original ones. The last $d$ unknown inputs are $\dt{w}_{m+j}$, $j=1,\ldots,d$.
Regarding the new vector fields that describe the dynamics we obtain:

\begin{equation}\label{EquationAugmentationSystem}
f^i \rightarrow
\left[\begin{array}{c}
   f^i \\
   0_d \\
\end{array}
\right]
,~~
g^0 \rightarrow
\left[\begin{array}{c}
   g^0 + \sum_{l=m+1}^{m_w}  g^l w_l\\
   0_d \\
\end{array}
\right]
,\\
\end{equation}
\[
g^k \rightarrow
\left[\begin{array}{c}
   g^k \\
   0_d \\
\end{array}
\right],~~
~~
g^{m+j} \rightarrow
\left[\begin{array}{c}
   0_n \\
   e^j \\
\end{array}
\right],
\]
with
$i=1,\ldots,m_u$,
$k=1,\ldots,m$, $j=1,\ldots,d$. In addition, $0_d$ and $0_n$ denote the zero $d$-column vector and the zero $n$-column vector, respectively, and $e^j$ denotes the $d$-column vector with the $j^{th}$ entry equal to 1 and the remaining $d-1$ entries equal to 0.

The operation is denoted by $\mathcal{A}$ because it consists of a state {\it Augmentation} and the consequent re-definition of all the key vector fields that characterize the new extended system, as specified above.
In this paper, we use the notation $\Sigma'=\mathcal{A}(\Sigma,~m)$ to denote the new extended system.


In Algorithms \ref{AlgoFull}, \ref{AlgoOmegagNonAbelm}, and \ref{AlgoDeltaNonAbelm}, we also use this operation by including further outputs and further inputs. Specifically, we consider the following two cases:

\begin{enumerate}

\item $[\Sigma', ~\Omega']=\mathcal{A}(\Sigma,~m,~\Omega)$, where the second output ($\Omega'$) is trivially the augmented codistribution obtained by extending all the covectors of $\Omega$ with $d$ zero entries.

\item $[\Sigma', ~\Delta']=\mathcal{A}(\Sigma,~m, ~\Delta)$, where the second output ($\Delta'$) is trivially the augmented distribution obtained by extending all the vectors of $\Delta$ with $d$ zero entries.

\end{enumerate}

%

We can apply this operation multiple consecutive times (and this is the case of
Algorithms \ref{AlgoFull}, \ref{AlgoOmegagNonAbelm}, and \ref{AlgoDeltaNonAbelm}, as the operation appears in a loop).
The resulting systems still satisfy (\ref{EquationSystemDefinitionUIO}) and they are all characterized by $m_u$ known inputs and $m_w$ unknown inputs. In \cite{IF22}, we called these extended systems {\it Finite Unknown Inputs Extensions}. In addition, in \cite{IF22}, we introduced the following definition:

\begin{definition}[Highest UI Degree of Reconstructability]\label{DefinitionHDegUIReconstr}
Given the system in (\ref{EquationSystemDefinitionUIO}), the highest unknown input degree of reconstructability is the largest unknown input degree of reconstructability of all its finite unknown input extensions.
\end{definition}

Finally, in \cite{IF22} we called a system {\it Canonizable with respect to its unknown inputs} if its highest unknown input degree of reconstructability is equal to $m_w$.

\subsection{$\boldsymbol{\mathcal{A}^-(\Sigma,~\Omega)}$ operation}\label{SubSectionSigma--}
This operation is only executed when, in the iteration of the main loop under execution, the Boolean Finish=False, and the remaining steps of this iteration determines a new observable function that increases the unknown input degree of reconstructability. In this case, the state can be augmented by including the unknown inputs with index larger than $m$ (through the operation $\mathcal{A}$). During this iteration of the main loop, the observable function, which increases the unknown input degree of reconstructability, is added to $\Omega$.
 Let us denote this function by $\theta$.
In general, $\theta$ also depends on the quantities $v_\alpha$, defined in (\ref{Equationvi}). 
The first operation executed by $\mathcal{A}^-$ is to express $\theta$ only in terms of the unknown inputs and the original state. In other words, all the $v_\alpha$ that appear in $\theta$ are expressed in terms of the unknown inputs (this is obtained by using (\ref{Equationvi}) and its time derivatives when $\theta$ also depends on the time derivatives of $v_\alpha$).
Then, all the unknown inputs that appear in $\theta$ are added to the original state. The resulting augmented state defines the new system $\Sigma$.

\section{Algorithm \ref{AlgoFull} for systems canonic with respect to the unknown inputs}\label{SectionSolutionUIOCanonic}

 When a system is canonic with respect to its unknown inputs, Algorithm \ref{AlgoFull} ends with the execution of Lines \ref{AlgoLineFINIT}-\ref{AlgoLineFINALSTEP} (the boolean variable {\it Continue} remains set to "true"). The system could be directly in canonic form with respect to its unknown inputs
or $\Omega$ was obtained in other parts of the algorithm~
and the system has been set in canonical form after the determination of one or more observable functions that are not among the outputs.
~The execution of Line \ref{AlgoLineFINITSelection} of Algorithm \ref{AlgoFull} provides the $m_w$ scalar functions $\widetilde{h}_1,\ldots, \widetilde{h}_{m_w}$
($[\widetilde{h}_1,\ldots, \widetilde{h}_{m_w}]=\mathcal{S}(\Sigma,~\Omega)$).
Then, the algorithm
computes 
the following quantities: $\mu$, $\nu$, $\widehat{g}^\alpha$, and the codistribution $\tobs$.

\subsection{$\boldsymbol{\mu}$ and $\boldsymbol{\nu}$}\label{SubSectionMuNu}

The two-index tensor $\mu$ is defined as follows:

\begin{equation}\label{EquationTensorMSynchro}
\mu^i_j = \mathcal{L}_{g^i}\widetilde{h}_j, ~~~i,~ j=1,\ldots,m_w
\end{equation}
\[
\mu^0_0 = 1,~
\mu^i_0=0,~
\mu^0_i = \frac{\partial\widetilde{h}_i}{\partial t}  +
\mathcal{L}_{g^0}\widetilde{h}_i,~  i=1,\ldots,m_w.
\]

Note that the entries with $i,~ j=1,\ldots,m_w$ are the same entries of the unknown input reconstructability matrix from $\widetilde{h}_1,\ldots, \widetilde{h}_{m_w}$, which is full rank. As a result, the tensor $\mu$ is also non singular and 
we denote by $\nu$ its inverse. 

\subsection{$\boldsymbol{\widehat{g}^{\alpha}}$}

Starting from $\nu$ the algorithm builds the new vector fields $\widehat{g}^0,\ldots,\widehat{g}^{m_w}$, defined as follows:
\begin{equation}\label{Equationgalpha}
   \widehat{g}^{\alpha}  = \sum_{\beta=0}^{m_w}\nu^{\alpha}_{\beta} g^{\beta},~~\alpha=0,1,\ldots,m_w.
\end{equation}

\subsection{The codistribution $\boldsymbol{\tobs}$}\label{SubSectionTOBS}
The codistribution $\tobs$ is defined as follows:

\begin{equation}\label{EquationTOBSDef}
\tobs:=
\sum_{q=1}^{m_w}
\sum_{j=0}^{\NO+\ND}
\sum_{\alpha_1=0}^{m_w}\ldots\sum_{\alpha_j=0}^{m_w}
\sum_{i=1}^{m_u}
\textnormal{span}\left\{
\nabla\Li_{[f^i]^{(\alpha_1,\ldots,\alpha_j)}}\tih_q
\right\},
\end{equation}

where the second sum is up to $s+r$, and the integers $s$ and $r$ are defined as follows:

\begin{itemize}

\item  $\NO$ is the smallest integer such that $\Omega_s=\Omega_{s-1}$, where $\Omega_k$ is the codistribution at the $k^{th}$ step of the algorithm in (\ref{EquationAlgorithmsMinimalCod}), when computing:

\[
\left<g^0,g^1,\ldots,g^{m_w}~\left|~\textnormal{span}\left\{\nabla\tih_1,\ldots,\nabla\tih_{m_w}\right\}\right.\right>
\]
Note that, for TV systems, the Lie derivative operator along $g^0$, i.e., $\Li_{g^0}$, must be replaced by  $\dt{\Li}_{g^0}:=\Li_{g^0}+\frac{\partial}{\partial t}$. Note that $\NO\le n-m_w+1$.

\item  $\ND$ is the smallest integer such that $\Delta_{r+1}=\Delta_r$, where $\Delta_k$ is the distribution at the $k^{th}$ step of the algorithm in (\ref{EquationAlgorithmsMinimalDistAut}), when computing:
%

\[
\left<\left[\cdot\right]^\cdot~\left|~\textnormal{span}\left\{f^1,\ldots,f^{m_u}\right.\right\}\right>
\]
Note that $\ND\le n-1$ (it is even
$\ND\le n-\textnormal{dim}\left\{\textnormal{span}\left\{f^1,\ldots,f^{m_u} \right\}\right\}$).

\end{itemize}

\subsection{Final step}

The last operation of Algorithm \ref{AlgoFull}, when the system is canonic with respect to its unknown inputs, is the computation of the observability codistribution:

\[
\OBS=\left<f^1,\ldots,f^{m_u}, \widehat{g}^0,\widehat{g}^1,\ldots,\widehat{g}^{m_w}~\left|~\Omega+\tobs\right.\right>
\]

This is obtained by running the algorithm in (\ref{EquationAlgorithmsMinimalCod}) for the specific case (note that, for TV systems, the Lie derivative operator along $\widehat{g}^0$, i.e., $\Li_{\widehat{g}^0}$, must be replaced by  $\dt{\Li}_{\widehat{g}^0}:=\Li_{\widehat{g}^0}+\frac{\partial}{\partial t}$). The convergence
of the algorithm in (\ref{EquationAlgorithmsMinimalCod}) is attained 
at the smallest integer $j$ such that 
$\Omega_{j}=\Omega_{j-1}$. Note  that $j$ cannot exceed $n-\dim\left(\Omega+\tobs\right)+1$.

\section{Algorithm \ref{AlgoFull} for systems that are not in canonical form}\label{SectionSolutionUIONonCanonic}

Let us back to the first part of Algorithm \ref{AlgoFull}, 
at the beginning of the main loop, 
and let us consider now the case when 
$\uideg \left({\Omega}\right)<m_w$.
The system is not in canonical form with respect to its unknown inputs. 
The algorithm continues by computing
the set of scalar functions 
$\widetilde{h}_1,\ldots, \widetilde{h}_m$. They are obtained by executing the operation $\mathcal{S}(\Sigma,~\Omega)$ defined in Section \ref{SubSectionSelection}, and the unknown inputs are reordered accordingly ($\mathcal{R}(\Sigma,~\widetilde{h}_1,\ldots, \widetilde{h}_m)$, Section \ref{SubSectionReorder}).
Then, the algorithm
computes the following quantities: $\mu$, $\nu$, and $\widehat{g}^\alpha$.

\subsection{$\boldsymbol{\mu}$ and $\boldsymbol{\nu}$}\label{SubSectionMumNum}

The two-index tensor $\mu$ is defined as follows:

\begin{equation}\label{EquationTensorMSynchrom}
\mu^i_j = \mathcal{L}_{g^i}\widetilde{h}_j, ~~~i,~ j=1,\ldots,m
\end{equation}
\[
\mu^0_0 = 1,~
\mu^i_0=0,~
\mu^0_i = \frac{\partial\widetilde{h}_i}{\partial t}  +
\mathcal{L}_{g^0}\widetilde{h}_i,~  i=1,\ldots,m.
\]

Note that the entries with $i,~ j=1,\ldots,m$ are the same entries of the matrix that consists of the first $m$ columns of the unknown input reconstructability matrix from $\widetilde{h}_1,\ldots, \widetilde{h}_{m}$. By construction, this submatrix is full rank. As a result, the tensor $\mu$ is also non singular and 
we denote by $\nu$ its inverse. 

\subsection{$\boldsymbol{\widehat{g}^{\alpha}}$}

Starting from $\nu$ we build the new vector fields $\widehat{g}^0,\ldots,\widehat{g}^{m}$ and $\giat^{m+1},\ldots,\giat^{m_w}$. They are defined as follows:
\begin{equation}\label{Equationgalpham}
   \widehat{g}^{\alpha}  = \sum_{\beta=0}^m~\nu^{\alpha}_{\beta} ~g^{\beta},~~\alpha=0,1,\ldots,m,~~~~\giat^k:=g^k-\sum_{\alpha=0}^m\giat^\alpha\Li_{g^k}\tih_\alpha, ~~k=m+1,\ldots,m_w
\end{equation}

%
%

\subsection{Algorithm \ref{Algo***}}\label{SubSectionAlgo***}

This algorithm returns the Boolean Finish. When Finish==True, it means that $\Omega_*$ is the observability codistribution. When Finish==False, the execution of the algorithm only tells us that there exists an observable function that increases the unknown inputs degree of reconstructability. Let us denote this function by $\theta$. Algorithm \ref{Algo***} does not provide $\theta$. $\theta$ is determined by the remaining steps of the iteration of the main loop of Algorithm \ref{AlgoFull}, under execution. Note that, when Finish==False, Algorithm \ref{Algo***} 
returns $\Sigma_*$, which  can differ from $\Sigma$ and is the suitable system to compute $\theta$. In addition, when Finish==False, Algorithm \ref{Algo***} provides
$k_*$, $i_*$, and $q_*$, which are used in the main loop of Algorithm \ref{AlgoFull} to compute $\theta$.

\subsection{The $\boldsymbol{\giat}$ vector field}\label{SubSectionDeg}

It appears at Line \ref{AlgoLineNested} of Algorithm \ref{AlgoFull}. It is defined as follows:

\begin{equation}\label{Equationgiatdinfty}
~\giat\triangleq\sum_{\beta=0}^m ~\giat^\beta~v_\beta
\end{equation}
with ($\beta=0,1,\ldots,m$)

\begin{equation}\label{Equationvi}
v_\beta\triangleq
\left[\begin{array}{ll}
\sum_{\gamma=0}^m
~\mu^\gamma_\beta~w_\gamma + \sum_{k=m+1}^{m_w}(\Li_{g^k}\tih_\beta)~ w_k& k_*=1\\
\sum_{\gamma=0}^m
~\mu^\gamma_\beta~w_\gamma & k_*>1\\
\end{array}
\right.
\end{equation}

\noindent The operator $\dtv{\Li}_\chi$ is:

\begin{equation}\label{EquationDTV}
\dtv{\Li}_{\chi}\triangleq\left(
\Li_\chi + \sum_{\alpha=0}^m\sum_{i_\alpha=0}^\infty
v_\alpha^{(i_\alpha+1)}
\frac{\partial}{\partial v_\alpha^{(i_\alpha)}} 
\right).
\end{equation}
For time varying systems $\dtv{\Li}_{\chi}\rightarrow \dtv{\Li}_{\chi}+\frac{\partial}{\partial t}$.

\vskip.1cm
\noindent The expression of $\psi_k^i$ is obtained automatically and recursively.
We have:

\begin{equation}\label{Equationpsik+1}
\psi_0^i=f^i,~~~~~\psi_{k+1}^i=\sum_{\gamma=0}^m[g^\gamma,~\psi^i_q]w_\gamma
+\sum_{\alpha=0}^m\sum_{i_\alpha=0}^k\frac{\partial \psi_k^i}{\partial w_\alpha^{(i_\alpha)}}w_\alpha^{(i_\alpha+1)}.
\end{equation}
For time-varying systems $\psi_{k+1}^i\rightarrow\psi_{k+1}^i+\frac{\partial \psi_k^i}{\partial t}$.

Once determined, we must eliminate $w_\alpha$, $w_\alpha^{(1)}$, $\ldots$, $w_\alpha^{(i_\alpha+1)}$ by expressing them in terms of 
$v_\alpha$, $v_\alpha^{(1)}$, $\ldots$, $v_\alpha^{(i_\alpha+1)}$. This is achieved by using the inverse of Equation (\ref{Equationvi}) above, i.e.,
$w_\alpha= \sum_{\beta=0}^m~\nu_\alpha^\beta v_\beta$.

\subsection*{Algorithm \ref{AlgoOmegagNonAbelm}}
It computes the codistribution $^x\Omega^m$. 
This algorithm uses a new operation denoted by $\mathcal{D}_x(\Omega)$. The input of this operation is a codistribution that is defined in a given augmented space (in the algorithm, before this operation, the $\mathcal{A}(\Sigma,~m)$ operation is applied). The operation $\mathcal{D}_x(\Omega)$, first detects a basis of $\Omega$.
As $\Omega$ is an integrable codistribution, it detects a set of scalar functions: $\theta_1,\ldots,\theta_D$, where $D$ is the dimension of $\Omega$.
In other words, 
$\Omega=\textnormal{span}\left\{\nabla\theta_1,\ldots,\nabla\theta_D\right\}$, where $\nabla$ is the gradient with respect to the new extended state (i.e., the state defined at the last $\mathcal{A}(\Sigma,~m)$ operation). The output of the operation $\mathcal{D}_x(\Omega)$, is $^x\Omega=\textnormal{span}\left\{\partial_x\theta_1,\ldots,\partial_x\theta_D\right\}$, where $\partial_x$ is the gradient with respect to the original state, i.e., the state of $\Sigma$
defined at the first line of Algorithm \ref{AlgoOmegagNonAbelm}.

The initialization step sets the codistribution equal to the span of the gradients of the selected functions 
$\widetilde{h}_1,\ldots, \widetilde{h}_{m}$. As a result, the dimension of $\Omega^m_0$ is $m$. Then, each iteration of the loop executes the following operations:

\begin{enumerate}

\item System augmentation, as explained in Section \ref{SubSectionSigma++}.

\item Computation of $\Omega^m_k$ by adding to $\Omega^m_{k-1}$ the term $\dt{\Li}_{g^0} \Omega^m_{k-1}+\sum_{j=1}^{m}\Li_{g^j} \Omega^m_{k-1}$.

\item Computation of $^x\Omega^m_k=\mathcal{D}_x(\Omega^m_k)$.

\end{enumerate}

The convergence of Algorithm \ref{AlgoOmegagNonAbelm} occurs at the smallest integer $j$ such that $^x\Omega^m_j=~^x\Omega^m_{j-1}$ and $j\le n - m+1$. We set $s^m_x=j$.
Note that, because of the presence of the state augmentation in the loop (Line \ref{AlgoOmegaLineRESET} of Algorithm \ref{AlgoOmegagNonAbelm}), proving the validity of the above convergence property is very demanding (see Appendix E of \cite{IF22} and in particular the proof of Proposition 11 in that appendix).

\subsection*{Algorithm \ref{AlgoDeltaNonAbelm}}
It computes the distribution $\Delta$. 
The initialization step sets the distribution equal to the span of the vector fields $f^1,\ldots,f^{m_u}$. Then, the recursive step adds to $\Delta_{k-1}$ the term
$\sum_{\beta=0}^{m_w}\left[\Delta_{k-1}\right]^\beta$.
The convergence is attained at the smallest integer $j$ such that $\Delta_j=\Delta_{j-1}$ and $j\le n-\textnormal{dim}\left\{\textnormal{span}\left\{f^1,\ldots,f^{m_u} \right\}\right\}+1$.
We set $r^m+1=j$.
Note that, because of the presence of the state augmentation in the loop (Line \ref{AlgoDeltaLineRESET} of Algorithm \ref{AlgoDeltaNonAbelm}), proving the validity of the above convergence property is non trivial (see Section 6.2.1 of \cite{IF22} and in particular the proof of Proposition 1 given in Appendix D of \cite{IF22}).

\vskip .2cm

Note that, the system augmentation performed by the $\mathcal{A}$ operation executed by Algorithms \ref{AlgoOmegagNonAbelm} and \ref{AlgoDeltaNonAbelm}, does not reset the system in Algorithm \ref{AlgoFull}. 
The execution of Algorithm \ref{AlgoOmegagNonAbelm} and Algorithm \ref{AlgoDeltaNonAbelm} only returns the integer $\widehat{k}^m=s^m_x+r^m$. All remaining quantities remain unchanged.


\section{Illustration by the case study}\label{SectionCaseStudyObservability}

We execute Algorithm 
\ref{AlgoFull} to obtain the observability codistribution in the two scenarios of our case study defined in Section \ref{SectionCaseStudy}. Note that, in both scenarios the system is in canonical form. The systems investigated in Chapters \ref{ChapterHIV} and \ref{ChapterCovid} are not in canonical form and, to see the implementation of  Algorithm \ref{AlgoFull} on a system that is not in canonical form, the reader is addressed to Section \ref{SectionHIVObs} or to Section \ref{SectionCovidObs}.

\subsection*{First scenario}
We have:

\[
g^0=
\left[
\begin{array}{c}
0\\
0\\
0\\
\end{array}
\right],~~
f^1=
\left[
\begin{array}{c}
\cos(\theta-\phi)\\
\frac{\sin(\theta-\phi)}{\rho}\\
0\\
\end{array}
\right],~~
g^1=
\left[
\begin{array}{c}
0\\
0\\
1\\
\end{array}
\right],
\]
and $h_1=\phi-\theta$.
We run Algorithm \ref{AlgoFull}. 
Line \ref{AlgoLineOmegaINIT} provides $\Omega=\textnormal{span}\{\nabla h_1\}=\textnormal{span}\{[0, 1, -1]\}$, and
$\uideg(\Omega)=1$, as the unknown input observability matrix from $h_1$ is the $1\times1$ matrix equal to $\Li_{g^1}h_1=-1\neq0$ (its rank is 1). The system is in canonical form with respect to its unknown input. Line \ref{AlgoLineFINITSelection} trivially selects $\widetilde{h}_1=h_1$. Then, Line \ref{AlgoLineFINITMuNuTOBS} provides:

\[
\mu=\nu=\left[
\begin{array}{cc}
1&0\\
0&-1\\
\end{array}
\right],
\]

obtained from (\ref{EquationTensorMSynchro}). In addition, from (\ref{Equationgalpha}) we obtain: $\widehat{g}^0=\nu^0_0g^0+\nu^0_1g^1=g^0=[0,0, 0]^T$, and
$\widehat{g}^1=\nu^1_0g^0+\nu^1_1g^1=-g^1=[0,0, -1]^T$.
Finally, we need to compute $\tobs$ from (\ref{EquationTOBSDef}). This requires, first of all, to compute the two integers $s$ and $r$:

\begin{itemize}

\item  $\NO$ is the smallest integer such that $\Omega_s=\Omega_{s-1}$, where $\Omega_k$ is the codistribution at the $k^{th}$ step of the algorithm in (\ref{EquationAlgorithmsMinimalCod}), when computing:

\[
\left<g^1~\left|~\textnormal{span}\left\{\nabla\tih_1\right\}\right.\right>=
\left.\left<\left[
\begin{array}{c}
0\\
0\\
1\\
\end{array}
\right]
~\right|~\textnormal{span}\left\{[0, 1, -1]\right\}\right>
\]
We immediately obtain $\NO=1$.

\item  $\ND$ is the smallest integer such that $\Delta_{r+1}=\Delta_r$, where $\Delta_k$ is the distribution at the $k^{th}$ step of the algorithm in (\ref{EquationAlgorithmsMinimalDistAut}), when computing:
%

\[
\left<\left[\cdot\right]^\cdot~\left|~\textnormal{span}\left\{f^1\right.\right\}\right>=
\left<\left[\cdot\right]^\cdot~\left|~\textnormal{span}\left\{\left[
\begin{array}{c}
\cos(\theta-\phi)\\
\frac{\sin(\theta-\phi)}{\rho}\\
0\\
\end{array}
\right]\right.\right\}\right>.
\]

We obtain $r=1$ (see the calculation details below).

\end{itemize}

\subsection*{Detail of computation of $\boldsymbol{r}$}
The algorithm in (\ref{EquationAlgorithmsMinimalDistAut}), for the specific case, provides: $\Delta_0=\textnormal{span}\left\{f^1\right\}$. We need to compute $\left[\Delta_0\right]^\alpha$, for $\alpha=0,1$. On the other hand, 
$\left[f^1\right]^0=\nu^0_0[g^0,~f^1]+\nu^0_1[g^1,~f^1]$ vanishes (the first term vanishes because $g^0$ is null, the second term vanishes because $\nu^0_1=0$).
$\left[f^1\right]^1=\nu^1_0[g^0,~f^1]+\nu^1_1[g^1,~f^1]=-[g^1,~f^1]=\chi^1$, with $\chi^1:=\left[\sin(\theta-\phi),~-\cos(\theta-\phi)/\rho,~0\right]^T$. Hence, $\Delta_1=\textnormal{span}\left\{f^1,~\chi^1\right\}$. Finally, $\left[\chi^1\right]^0$ vanishes and $\left[\chi^1\right]^1=-f^1$. Therefore, $\Delta_2=\Delta_1$ and $r+1=2$.

\vskip.2cm

Let us compute $\tobs$. From
(\ref{EquationTOBSDef}), and from the above computation, we obtain that the generators of $\tobs$ are $\Li_{f^1}h_1=\frac{\sin(\theta-\phi)}{\rho}$ and $\Li_{\chi^1}h_1=-\frac{\cos(\theta-\phi)}{\rho}$. Hence:

\[
\tobs=\textnormal{span}\left\{
\left[
-\frac{\sin(\theta-\phi)}{\rho^2},~-\frac{\cos(\theta-\phi)}{\rho}, ~\frac{\cos(\theta-\phi)}{\rho}
\right],
\right.
\]
\[
\left.
\left[
\frac{\cos(\theta-\phi)}{\rho^2},~-\frac{\sin(\theta-\phi)}{\rho}, ~\frac{\sin(\theta-\phi)}{\rho}.
\right]
\right\}
\]

The final step of Algorithm \ref{AlgoFull} is the execution of Line \ref{AlgoLineFINALSTEP}, namely:

\[
\OBS=
\left.\left<\left[
\begin{array}{c}
\cos(\theta-\phi)\\
\frac{\sin(\theta-\phi)}{\rho}\\
0\\
\end{array}
\right],\left[
\begin{array}{c}
0\\
0\\
-1\\
\end{array}
\right]~\right|~\textnormal{span}\left\{[0, 1, -1]\right\}+\tobs\right>
\]
as $\widehat{g}^0$ vanishes and $\widehat{g}^1=-g^1$. The algorithm 
in (\ref{EquationAlgorithmsMinimalCod}) converges at the first step ($\Omega_1=\Omega_0$), and $\OBS=$

\[
\textnormal{span}\left\{[0, 1, -1], ~\left[
-\frac{\sin(\theta-\phi)}{\rho^2},~-\frac{\cos(\theta-\phi)}{\rho}, ~\frac{\cos(\theta-\phi)}{\rho}
\right]\right\}.
\]

The above codistribution contains all the observability properties of our system. We postpone a discussion about them to Section \ref{SubSectionCaseStudySymmetries}.


\subsection*{Second scenario}
With respect to the previous scenario, we now have:

\[
g^1=
\left[
\begin{array}{c}
\cos(\theta-\phi)\\
\frac{\sin(\theta-\phi)}{\rho}\\
0\\
\end{array}
\right],~~
f^1=
\left[
\begin{array}{c}
0\\
0\\
1\\
\end{array}
\right]
\]

We run Algorithm \ref{AlgoFull}. Line \ref{AlgoLineOmegaINIT} provides $\Omega=\textnormal{span}\{\nabla h_1\}=\textnormal{span}\{[0, 1, -1]\}$, and 
 $\uideg(\Omega)=1$, as the unknown input observability matrix from $h_1$ is the $1\times1$ matrix equal to $\Li_{g^1}h_1=\frac{\sin(\theta-\phi)}{\rho}\neq0$ (its rank is 1). 
The system is in canonical form with respect to its unknown input. Line \ref{AlgoLineFINIT} trivially selects $\widetilde{h}_1=h_1$. Then, Line \ref{AlgoLineFINITMuNuTOBS} provides:

\begin{equation}\label{EquationCaseStudyMuNu}
\mu=\left[
\begin{array}{cc}
1&0\\
0&\sin(\theta-\phi)/\rho\\
\end{array}
\right], ~~
\nu=\left[
\begin{array}{cc}
1&0\\
0&\rho/\sin(\theta-\phi)\\
\end{array}
\right],
\end{equation}

obtained from (\ref{EquationTensorMSynchro}). In addition, from (\ref{Equationgalpha}) we obtain: $\widehat{g}^0=\nu^0_0g^0+\nu^0_1g^1=g^0=[0,0, 0]^T$, and
$\widehat{g}^1=\nu^1_0g^0+\nu^1_1g^1=[\rho\frac{\cos(\theta-\phi)}{\sin(\theta-\phi)},~1,~0]^T$.
Finally, we need to compute $\tobs$ from (\ref{EquationTOBSDef}). This requires, first of all, to compute the two integers $s$ and $r$:

\begin{itemize}

\item  $\NO$ is the smallest integer such that $\Omega_s=\Omega_{s-1}$, where $\Omega_k$ is the codistribution at the $k^{th}$ step of the algorithm in (\ref{EquationAlgorithmsMinimalCod}), when computing:

\[
\left<g^1~\left|~\textnormal{span}\left\{\nabla\tih_1\right\}\right.\right>=
\left.\left<
\left[
\begin{array}{c}
\cos(\theta-\phi)\\
\frac{\sin(\theta-\phi)}{\rho}\\
0\\
\end{array}
\right]
~\right|~\textnormal{span}\left\{[0, 1, -1]\right\}\right>
\]
By a direct computation, we obtain $\Omega_2=\Omega_1=\textnormal{span}\left\{\nabla h_1, \nabla\frac{\sin(\theta-\phi)}{\rho}\right\}$ and, consequently, $\NO=2$.

\item  $\ND$ is the smallest integer such that $\Delta_{r+1}=\Delta_r$, where $\Delta_k$ is the distribution at the $k^{th}$ step of the algorithm in (\ref{EquationAlgorithmsMinimalDistAut}), when computing:
%

\[
\left<\left[\cdot\right]^\cdot~\left|~\textnormal{span}\left\{f^1\right.\right\}\right>=
\left<\left[\cdot\right]^\cdot~\left|~\textnormal{span}\left\{\left[
\begin{array}{c}
0\\
0\\
1\\
\end{array}
\right]\right.\right\}\right>.
\]

We obtain $r=2$. We do not provide the detail of the computation. We only mention that, in this case, the generators of $\Delta$ are:
$f^1$, $[\rho,~-\frac{\cos(\theta-\phi)}{\sin(\theta-\phi)},~0]^T$, and $[\rho\frac{\cos(\theta-\phi)}{\sin(\theta-\phi)},~-\frac{\cos^2(\theta-\phi)}{\sin^2(\theta-\phi)},~0]^T$

\end{itemize}

From
(\ref{EquationTOBSDef}), by a direct computation we obtain:

\[
\tobs=\textnormal{span}\left\{\nabla\frac{\cos(\theta-\phi)}{\sin(\theta-\phi)}\right\}.
\]

The final step of Algorithm \ref{AlgoFull} is the execution of Line \ref{AlgoLineFINALSTEP}, namely:

\[
\OBS=
\left.\left<
\left[
\begin{array}{c}
0\\
0\\
1\\
\end{array}
\right],~
\left[
\begin{array}{c}
\rho\frac{\cos(\theta-\phi)}{\sin(\theta-\phi)}\\
1\\
0\\
\end{array}
\right]~\right|~\textnormal{span}\left\{[0, 1, -1]\right\}\right>,
\]

as $\tobs\subseteq\textnormal{span}\left\{[0, 1, -1]\right\}$, and
$\widehat{g}^0$ is null. The algorithm 
in (\ref{EquationAlgorithmsMinimalCod}) converges at the first step ($\Omega_1=\Omega_0$), and:

\[
\OBS=\textnormal{span}\left\{[0, 1, -1]\right\}.
\]

The above codistribution contains all the observability properties of our system. We postpone a discussion about them to Section \ref{SubSectionCaseStudySymmetries}.


\chapter{Systematic procedure to perform the identifiability analysis}\label{ChapterUIRec}
This section 
is devoted to the problem of the identifiability of the time-varying parameters, or, equivalently, to the reconstructability of the system unknown inputs. As we will see, this problem is strongly related with the problem of state observability in the presence of unknown inputs (whose solution was provided in Chapter \ref{ChapterSolutionUIO}).
We provide the general solution to obtain the reconstructability of the unknown inputs. 

This section consists of four subsections. In \ref{SectionUIRecObservableState}, we extend a preliminary result provided in \cite{IF22} and we obtain the answer to our problem when the state that characterizes our system is observable. In \ref{SectionSymmetryUnobservability}, we remind the reader of the concept of continuous symmetry introduced in \cite{SIAMbook}. In \ref{SectionUIRec}, we extend this concept to the unknown input vector and we provide the complete answer to our problem. In practice, we obtain the
general systematic procedure that solves the problem of the identifiability of the time-varying parameters. 
The procedure is Algorithm \ref{AlgoFullIDE}.
Finally, in \ref{SectionCaseStudyIdentifiability}, we illustrate all these concepts by using our case study.


\section{Identifiability when the state is observable}\label{SectionUIRecObservableState}
 
In \cite{IF22} we proved
that when the state that characterizes the system is observable, if the system is canonic with respect to its unknown inputs, then all the unknown inputs can be reconstructed (Theorem 4 in \cite{IF22}).
It also holds the viceversa and, consequently, when the state is observable, we obtain a complete answer to our problem which is given by the following theorem: 


\begin{theorem}\label{TheoremUIRecObservable}
If the state of the system characterized by (\ref{EquationSystemDefinitionUIO}) is observable, the unknown inputs can be reconstructed if and only if the system is canonic with respect to its unknown inputs.
\end{theorem}

Note that, checking if a system is canonic with respect to its unknown input is immediate. It suffices to compute the rank of the reconstructability matrix from any set of generators of the observability codistribution (and Algorithm \ref{AlgoFull} provides a set of generators of this codistribution). Therefore, Theorem \ref{TheoremUIRecObservable}
provides a full answer to the problem of unknown input reconstruction when the state is observable.

\proof{Based on the result stated by Theorem 4 in \cite{IF22}, it remains to prove that if the unknown inputs can be reconstructed then the system is canonic with respect to its unknown inputs. 

We proceed by contradiction. We assume that the system is not canonic with respect to its unknown inputs.

As the state is observable, all its components belong to the observation space and we have: $\OBS=\textnormal{span}\left\{
\nabla x_1, \ldots, \nabla x_n\right\}$.
As we assumed that the system is not canonic, $\uideg\left(\OBS\right)<m_w$. Therefore, the matrix $\RM\left( x_1,\ldots, x_n\right)$ has rank smaller than $m_w$. On the other hand, by an explicit computation we obtain

\[
\RM\left( x_1,\ldots, x_n\right)
=
\left[\begin{array}{cccc}
~[g^1]_1 & [g^2]_1 & \ldots & [g^{m_w}]_1 \\
~[g^1]_2 & [g^2]_2 & \ldots & [g^{m_w}]_2 \\
\ldots &\ldots &\ldots &\ldots \\
~[g^1]_n & [g^2]_n & \ldots & [g^{m_w}]_n \\
\end{array}
\right]=
\]
\[
\left[\begin{array}{cccc}
g^1 & g^2 & \ldots & g^{m_w} \\
\end{array}
\right]
\]
where $[g^i]_j$ is the $j^{th}$ component of the vector field $g^i$ ($i=1,\ldots,m_w$ and $j=1,\ldots,n$).
Therefore, as rank$\left(\RM\left( x_1,\ldots, x_n\right)\right)<m_w$, the vectors $g^1 ~g^2 ~\ldots ~ g^{m_w}$ are linearly dependent. 
We denote by $\widehat{n}$ a non trivial vector that belongs to the null space of the above matrix. 
Let us consider the dynamics:

\[
\dot{x} =   g^0+\sum_{k=1}^{m_u}f^k  u_k +  \sum_{j=1}^{m_w}g^j  w_j.
\]

For any $w=[w_1,~w_2,\ldots,w_{m_w}]$, we obtain the same dynamics by transforming the UI as follows:

\[
w\rightarrow w'=w+\widehat{n}
\]

As result, $w$ cannot be distinguished from $w'$ and, consequently, cannot be reconstructed. $\blacktriangleleft$}

\vskip.2cm

As we said, when the state is observable, the above theorem provides a full answer to our problem. It remains to find the solution when the state is not observable. Unfortunately, we need a complete different approach. In other words, in the presence of the state unobservability, we cannot exploit the result stated by Theorem \ref{TheoremUIRecObservable}. We provide an explanation of this fact.

Let us suppose that the state is not observable. In order to exploit the result stated by Theorem \ref{TheoremUIRecObservable}, we need to characterize our system by an observable state.
Algorithm \ref{AlgoFull} provides the entire observable space. 
We could introduce a new state whose components are the generators of this space, i.e., the generators of $\OBS$\footnote{Clearly, with generators, here we mean the scalar functions whose gradients generate $\OBS$.}. This new state is certainly observable and its components are expressed in terms of the old state.
~To use Theorem \ref{TheoremUIRecObservable}, we need to
describe our system by using this new state 
starting from the description that is characterized by the old state. This means to obtain a set of equations of the same format of (\ref{EquationSystemDefinitionUIO}) where the state $x$ is the new state. In other words, the dynamics of $x$, precisely as in (\ref{EquationSystemDefinitionUIO}), must be expressed only in terms of the components of the new $x$ (and not in terms of the old state).
~On the other hand, when we try to obtain this description, we often encounter two fundamental difficulties that make it critical, and even useless:

\begin{itemize}

\item {\bf Critical issues}: This description cannot be determined automatically. In some cases, its determination can be a very demanding and laborious task, based on enormous inventiveness (e.g., see Section \ref{SectionHIVLocDec} where, although the state is small, the determination of this description requires ability and inventiveness).

\item {\bf Uselessness}: In many cases, obtaining this description requires a re-definition of the unknown inputs. In other words, this task cannot be performed by maintaining the same original unknown inputs (see the discussion of the second scenario of our case study in Section \ref{SubSectionCaseStudyLocDec}, and our applications in Sections \ref{SectionHIVLocDec} and \ref{SectionCovidLocDec}).
When the final goal is precisely the reconstruction of the {\it original} unknown inputs, performing this task (i.e., describing the system by an observable state) can be useless. 

\end{itemize}


The above discussion showed that, in general, we cannot exploit the result stated by Theorem \ref{TheoremUIRecObservable} to deal with systems characterized by an unobservable state.
In this case, we need a different approach. This 
method will be introduced in Section 
\ref{SectionUIRec}. It is based
on the concept of continuous symmetry introduced in \cite{TRO11} and here summarized in Section \ref{SectionSymmetryUnobservability}.



\section{Unobservability of the State, State Symmetries and Indistinguishable States}\label{SectionSymmetryUnobservability}


In \cite{TRO11}, and more exhaustively in \cite{SIAMbook}, we introduced the concept of system symmetry in the presence of unobservability. 
When we are in the presence of unobservability, the dimension of the observability codistribution ($\OBS$) is smaller than the state dimension, $n$. Let us denote it by $n_o$.
The null space of $\OBS$ is not empty and it is a distribution (the orthogonal distribution, from now on denoted by $\OBS^\bot$), with dimension $n-n_o$. 
In \cite{SIAMbook,TRO11},
we defined {\it symmetry} any vector field that belongs to $\OBS^\bot$.
In Section 4.6 of \cite{SIAMbook} we showed that the generators of $\OBS^\bot$ are the generators of a Lie algebra associated to a Lie group.
Given a state $x$, all its indistinguishable states can be obtained by the action of this Lie group on $x$.
Specifically, in \cite{SIAMbook}, we showed that, starting from a system symmetry $\xi\in\OBS^\bot$ and a given state $\overline{x}$ we can generate a set of states which are indistinguishable from $\overline{x}$
by solving the following differential equation:

\begin{equation}\label{EquationDiffEqStates}
\left\{\begin{array}{ll}
  \frac{dx}{d\tau} &=   \xi(x(\tau))\\
  x(0) &= \overline{x}\\
\end{array}\right.
\end{equation}

In particular, if this equation admits solution on the interval $\tau\in[0, ~\mathcal{T}]$, then all the states $x(\tau)$, obtained by integrating the above equation up to $\tau$, are indistinguishable from each other.
The above procedure defines a one parameter Lie group. The parameter is $\tau$
and its action on the state $\overline{x}$ returns the state $x(\tau)$:

\begin{equation}\label{EquationSymmetryStateFIN}
\overline{x}\rightarrow x(\tau)
\end{equation}

As usual, when dealing with Lie groups, it is very useful to refer to the action of the group when the parameter takes infinitesimal values. The above transformation becomes:

\begin{equation}\label{EquationSymmetryState}
\overline{x}\rightarrow \overline{x}+\epsilon\xi
\end{equation}

where $\epsilon$ is an infinitesimal\footnote{The reader is addressed to \cite{Nonstandard Analysis1,Nonstandard Analysis2,Nonstandard Analysis3} for the definition of infinitesimal (or infinitesimal number) and its use in calculus.}. Given a symmetry $\xi$ of the system in (\ref{EquationSystemDefinitionUIO}), the two states $\overline{x}$ and $\overline{x}+\epsilon\xi$ are indistinguishable.
Note that we have a symmetry if and only if the distribution $\OBS^\bot$ is non trivial, i.e., if and only if the dimension of $\OBS$ is strictly smaller than $n$. Hence, we can say that the state is observable if and only we do not have symmetries. The symmetries can be computed automatically (it suffices to compute the nullspace of $\OBS$). Starting from them, we can also try to obtain the indistinguishable states, by solving the differential equation in (\ref{EquationDiffEqStates}). On the other hand, this last step cannot be performed automatically. When it can be performed analytically, it provides further useful insights on the ODE model.
Finally, note that when the dimension of $\OBS^\bot$ is strictly larger the 1, the states indistinguishable from a given state $\overline{x}$ can be obtained by the action of a multi parameters Lie group (instead of a simple one-parameter Lie group). We do not provide details on this Lie group. We only mention that a set of generators of its associated Lie algebra is precisely the set of generators of $\OBS^\bot$ (note that, as $\OBS$ is an integrable codistribution, because of the Frobenius Theorem, $\OBS^\bot$ is involutive, i.e., closed with respect to the Lie brackets). We also emphasize that, in this case of multiple independent symmetries, we can still consider each symmetry independently and obtaining 1-dimensional indistinguishable sets by the action of the corresponding one-parameter Lie group (i.e., by solving the differential equation in (\ref{EquationDiffEqStates})). We get a 1-dimensional indistinguishable set for each symmetry.

In Section \ref{SectionUIRec}, we extend the concept of symmetry to the unknown input vector and we obtain a full answer to the identifiability problem.


%
%
%
%
%
%
%
%
%
%
%

\section{Identifiability for any system}\label{SectionUIRec}

We start by running Algorithm \ref{AlgoFull}. By doing this, we automatically obtain the observability codistribution, $\OBS$. In addition, the algorithm provides a new system, which is an unknown input extension of the original system with the highest unknown input degree of recostructability.
Note that, the unknown input extension returned by Algorithm \ref{AlgoFull} does not guarantee that the following condition is met:

\begin{equation}\label{EquationConditionGk}
\Li_{g^k}\tih_i=0, ~~\forall i=1,\ldots,m,~~\forall k=m+1,\ldots,m_w.
\end{equation}
On the other hand, when for a given $k\in[m+1,\ldots,m_w]$ there exists $i\in[1,\ldots,m]$ such that $\Li_{g^k}\tih_i\neq0$, it suffices to consider the unknown input extension that also includes $w_k$.

Hence, from now on we consider the unknown input extension where the condition in (\ref{EquationConditionGk}) is honoured.
From now on, we refer to this system. Its structure still satisfies (\ref{EquationSystemDefinitionUIO}). We denote the dimension of its state by $n$ and we denote its unknown input degree of recostructability (which is the highest) by $m$. Note that, a set of $m$ observable functions, $\widetilde{h}_1,\ldots,\widetilde{h}_m$, is also available, after the execution of Algorithm \ref{AlgoFull}.
The new system could coincide with the original system, meaning that the UI degree of reconstructability of the original system is already the highest. In this case, Algorithm \ref{AlgoFull}
returns the original system where only the order of the unknown inputs could have changed. Indeed, when $m<m_w$, the system returned by Algorithm \ref{AlgoFull} is ordered in accordance with the $\mathcal{R}$ operation, defined in Section \ref{SubSectionReorder}. In other words, the rank of the first $m$ columns of $\RM\left(\widetilde{h}_1,\ldots,\widetilde{h}_m\right)$ is $m$.

In this section we provide the expression of all the symmetries that characterize the unknown input vector of the system returned by Algorithm \ref{AlgoFull}. In other words, we provide the expression of all the vectors $\chi=[\chi_1,\ldots,\chi_{m_w}]^T$ such that:

\begin{equation}\label{EquationSymmetryUI}
w'=w+\epsilon~\chi
\end{equation}

is indistinguishable from $w$.
In Section \ref{SubSectionUIRecMainCan}, we provide the expression of a first set of such symmetries that exist when the system is not canonic with respect to its unknown inputs (i.e., $m<m_w$). They will be denoted by $^c\chi$, where $c$ stands for {\it system non-Canonicity}. In Section \ref{SubSectionUIRecMain} we provide the expression of a second set of such symmetries that may exist when the state is unobservable. They will be denoted by $^u\chi$, where $u$ stands for {\it state Unobservability}.


Note that, the characterization of unknown input symmetry provided by ({\ref{EquationSymmetryUI}}) was also very recently introduced in \cite{Shi20,Shi21}. In these works, this definition was used by working in the augmented space, i.e., by characterizing the system with a state that includes the unknown inputs together with their time derivatives up to a given order. As we mentioned in Chapter \ref{ChapterIntroduction}, using this extended state has two limitations: (i) the computational burden can easily become prohibitive, and (ii) some of the symmetries could be actually spurious because we did not include enough time derivatives of the unknown inputs (and no criterion to establish which is the maximum order to be included was provided).
In addition, working with the augmented state does not allow us to distinguish between the two aforementioned categories of unknown input symmetries (i.e., to distinguish between $^c\chi$ and $^u\chi$) and can also make more difficult the physical interpretation of a given symmetry.

\subsection{Unknown input symmetries due to the system non canonicity}\label{SubSectionUIRecMainCan}

We provide the expression of a set of symmetries of the unknown input vector that certainly exist when the system is not canonic with respect to its unknown inputs. In particular, this set consists of $m_w-m$ independent symmetries.

\begin{theorem}\label{TheoremIdentifiabilityCan}
Let us consider the unknown input extension where the unknown input degree of reconstructability coincides with its highest unknown input degree of reconstructability and that honours the condition in (\ref{EquationConditionGk})\footnote{This is the system returned by Algorithm \ref{AlgoFull} and, if the condition in (\ref{EquationConditionGk}) is not honoured by a given $k$, it suffices to also include $w_k$ in the state, as explained above.}. Let us suppose that $m<m_w$ (system not canonic with respect to the unknown inputs). For this system, the following $m_w-m$ column vectors:

\begin{equation}\label{EquationIndwCanonic}
^c\chi^i=
\left[\begin{array}{l}
 0_m \\
 e_i \\
\end{array}\right],~~~
 i=1,\ldots,m_w-m,
\end{equation}
with $0_m$ the zero $m-$column vector and $e_i$ the unit vector of dimension $m_w-m$ with zero everywhere except that the $i^{th}$ component which is 1,
are $m_w-m$ independent symmetries of the unknown input vector.
\end{theorem}

\proof{The proof is given in Appendix \ref{AppendixTheoremCan}.}

\subsection{Possible unknown input symmetries due to the state unobservability}\label{SubSectionUIRecMain}

%
%


When the state is unobservable, we have one or more nonnull symmetries of the state. 
We have the following fundamental result.

\begin{theorem}\label{TheoremIdentifiability}
Let us consider the unknown input extension where the unknown input degree of reconstructability coincides with its highest unknown input degree of reconstructability and that honours the condition in (\ref{EquationConditionGk})\footnote{This is the system returned by Algorithm \ref{AlgoFull} and, if the condition in (\ref{EquationConditionGk}) is not honoured by a given $k$, it suffices to also include $w_k$ in the state, as explained above.}. Let us denote by $\xi\in\OBS^\bot$ a given symmetry of the state. The $m_w-$ column vector $^u\chi$, with components:

\begin{equation}\label{EquationSymmetryUIObs}
\left\{\begin{array}{ll}
 ^u\chi_k = -\sum_{i=1}^m\nu^i_k \left(\xi^0_i +  \sum_{j=1}^{m}\xi^j_i w_j 
 \right) &k=1,\ldots,m\\
^u\chi_k =0  ~~~~~~~~~~k=m+1,\ldots,m_w\\
\end{array}\right.
\end{equation}

where, for any $i=1,\ldots,m$:

\begin{equation}\label{EquationSymmetryUIObsCoeff}
\xi^\alpha_i=\left\{
\begin{array}{ll}
\Li_\xi
\left(
\Li_{g^0}\widetilde{h}_i+\frac{\partial\widetilde{h}_i}{\partial t}
\right)
& \alpha=0\\
\Li_\xi\left(\Li_{g^\alpha}\widetilde{h}_i
\right)& \alpha=1,\ldots,m\\
\end{array}
\right.
\end{equation}

is a symmetry of $w$.


\end{theorem}

\proof{The proof is given in Appendix \ref{AppendixTheoremObs}.}

\vskip .2cm

\noindent We provide the following remarks.


\begin{rem}
Given a symmetry $\xi$ of the state, the corresponding symmetry $^u\chi$ in (\ref{EquationSymmetryUIObs}) can be trivial, i.e., can be a null vector. This is certainly the case when all the quantities $\xi^\alpha_i$ in (\ref{EquationSymmetryUIObsCoeff}) vanish. If for all the symmetries of the state the corresponding $^u\chi$ vanish, the first $m$ component of $w$ can be reconstructed. This means that there are cases where the first $m$ unknown inputs can be reconstructed even in the presence of state unobservability. Hence, it is better to call the vector $^u\chi$ in (\ref{EquationSymmetryUIObs}) a potential symmetry of the unknown input vector.
\end{rem}

\begin{rem}
We have a potential symmetry of the unknown input vector for any symmetry of the state.
\end{rem}


\begin{rem}
If the state is observable, there are not state symmetries and the first $m$ unknown inputs can be reconstructed. This is consistent with the result stated by Theorem \ref{TheoremUIRecObservable}. Indeed, when the state is observable we only have the unknown input symmetries that characterize the last $m_w-m$ unknown inputs. These symmetries disappear if and only if $m=m_w$ (i.e., the system is canonic with respect to the UIs).
\end{rem}

\subsection{The systematic procedure}\label{SubSectionUIRecProcedure}

The systematic procedure to automatically obtain the identifiability of all the time-varying parameters (or the reconstructability of all the unknown inputs) is Algorithm \ref{AlgoFullIDE}. It is based on the results stated by Theorems \ref{TheoremIdentifiabilityCan} and \ref{TheoremIdentifiability}.
It starts by running Algorithm \ref{AlgoFull} on the system under investigation (Line \ref{AlgoIDELineAlgo1Run}).
This provides the observability codistribution $\OBS$, the highest UI degree of reconstructability with respect to the unknown inputs, $m$, and a new system $\E$ which is an unknown input extension of the original system with the highest UI degree of reconstructability. Note that, the identifiability analysis regards the new system $\E$. Hence, we investigate the reconstructability of the unknown inputs of this system. These unknown inputs may differ from the original ones for the following two reasons:
\begin{enumerate}
\item Their order may have been changed.
\item Some of them could be the time derivatives (of a given order) of the original ones.
\end{enumerate}

Note that, in any case, we also obtain the reconstructability properties of the original unknown inputs. Indeed, if a given UI of $\E$ is the time derivative of a given UI of the original system, it certainly means that the original UI was included in the state.
To check its identifiability it suffices to verify if its differential belongs to $\OBS$. Algorithm \ref{AlgoFullIDE} will tell us if the corresponding UI of $\E$ is identifiable (i.e., if the time derivative of a given order of the original UI is identifiable).

Line \ref{AlgoIDELineChiCSet} of Algorithm \ref{AlgoFullIDE} provides a set of $m_w-m$ symmetries of the unknown input, which are the $m_w-m$ unit vectors defined in (\ref{EquationIndwCanonic}). This set is empty if and only if the system is canonic with respect to its unknown inputs ($m=m_w$).


Lines \ref{AlgoIDELineObot} and \ref{AlgoIDELineObotBasis} compute the orthogonal distribution $\OBS^\bot$ and a basis of it. In particular, its generators are $\xi^1,\ldots,\xi^{S}$.
Then, 
the loop at Lines \ref{AlgoIDELineFORtoComputeChiU}-\ref{AlgoIDELineFORtoComputeChiUEND} computes the corresponding symmetries of the unknown input vector, i.e., $^u\chi^1,\ldots,~^u\chi^{S}$, on the basis of Theorem \ref{TheoremIdentifiability}.

At this point, we have obtained all the symmetries that characterize the unknown input vector. They are the following $S+m_w-m$ vectors:

\[
^c\chi^1,\ldots,~^c\chi^{m_w-m},~^u\chi^1,\ldots,~^u\chi^{S}.
\]

Finally, the loop at Lines \ref{AlgoIDELineFORtoCheckRec}-\ref{AlgoIDELineFORtoCheckRecEnd} checks the reconstructability of all the unknown inputs $w_1,\ldots,w_{m_w}$. Specifically,
the $j^{th}$ unknown input ($w_j$) can be reconstructed if and only if:

\[
^c\chi^1_j=\ldots=~^c\chi^{m_w-m}_j=~^u\chi^1_j=\ldots=~^u\chi^{S}_j=0,
\]
i.e., the $j^{th}$ components of all the unknown input symmetries, vanish, simultaneously.
%
%
%
%

Note that, when $S=0$ (and this is the case when the state that characterizes $\E$ is observable) the first $m$ unknown inputs, $w_1,\ldots,w_m$, can be reconstructed. 
If in addition the system is canonic with respect to its unknown inputs, all the unknown inputs can be reconstructed. 
Note that the observability of the state is a sufficient (but not a necessary) condition to have the reconstructability of the first $m$ unknown inputs. Indeed, there are cases where the state is unobservable (i.e., $S>0$) but all the symmetries $^u\chi^1,\ldots,~^u\chi^{S}$ vanish, simultaneously (e.g., this occurs in the first scenario of our case study, as discussed in Section \ref{SubSectionCaseStudyIdentifiability}).

%
%
%
%
%

\begin{algorithm}
\caption{The algorithm that builds the symmetries of the unknown input vector and check the reconstructability of all its components ($w_j,~j=1,\ldots,m_w$).}

\begin{algorithmic}[1]

\State  \textnormal{Run Algorithm \ref{AlgoFull}
and compute $\OBS$, $m$, $\E$. 
In the case, extends $\E$ by including $w_k$ to honour (\ref{EquationConditionGk})}.\label{AlgoIDELineAlgo1Run}



\If{$m<m_w$}\label{AlgoIDELineIFCanonic}
 \State \textnormal{Set $^c\chi^1,\ldots,~^c\chi^{m_w-m}$, as in (\ref{EquationIndwCanonic})}.\label{AlgoIDELineChiCSet}
 \State \textnormal{Set $^c\chi^i=
\left[\begin{array}{l}
 0_m \\
 e_i \\
\end{array}\right],~~~
 i=1,\ldots,m_w-m$, as in (\ref{EquationIndwCanonic})}.

\EndIf

\State  \textnormal{Compute $\OBS^\bot$}.\label{AlgoIDELineObot}
 
\State  \textnormal{Set $\OBS^\bot=$span$\left\{\xi^1,\ldots,\xi^{S}
\right\}$}.\label{AlgoIDELineObotBasis}
 
\State \textnormal{Set $\Delta^u=\left\{\right\}$}\label{AlgoIDELineDeltaInit}

\For{$i_s=1:S$}\label{AlgoIDELineFORtoComputeChiU}

 
\State  \textnormal{Set $\xi=\xi^{i_s}$, and compute the corresponding $^u\chi$ (from (\ref{EquationSymmetryUIObs}) and (\ref{EquationSymmetryUIObsCoeff})), and denote it by $^u\chi^{i_s}$}.\label{AlgoIDELineChiUCompute}

\State \textnormal{Set $\Delta^u=\Delta^u+$span$\left\{^u\chi\right\}$}\label{AlgoIDELineDeltaRecursive}

\EndFor\label{AlgoIDELineFORtoComputeChiUEND}

\State \textnormal{Set $\Delta^u=$span$\left\{
^u\chi^1,\ldots,~^u\chi^D
\right\}$}\label{AlgoIDELineDeltaBasis}


\For{$j=1:m_w$}\label{AlgoIDELineFORtoCheckRec}
 
 \If{$^c\chi^1_j=\ldots=~^c\chi^{m_w-m}_j=~^u\chi^1_j=\ldots=~^u\chi^{S}_j=0$}\label{AlgoIDELineIFCheckRec}
 \State \textnormal{$w_j$ can be reconstructed}\label{AlgoIDELineIFCheckRecSI}
 \Else
  \State \textnormal{$w_j$ cannot be reconstructed}\label{AlgoIDELineIFCheckRecNO}
  \EndIf

 \EndFor\label{AlgoIDELineFORtoCheckRecEnd}
 
\end{algorithmic}\label{AlgoFullIDE}
\end{algorithm}

\section{Illustration by the case study}\label{SectionCaseStudyIdentifiability}

We illustrate all the concepts introduced by this section by applying them to our case study. Note that, for educational purposes, we provide the indistinguishable states and we also obtain a description of our system by an observable state. On the other hand, the implementation of Algorithm \ref{AlgoFullIDE} requires neither the determination of the indistinguishable states nor a description of the system by an observable state. It only requires the computation of all the independent symmetries of the state, i.e., a basis of the orthogonal distribution $\OBS^\bot$.

\subsection{Symmetries and Indistinguishable states}\label{SubSectionCaseStudySymmetries}

\subsection*{First scenario}
From the expression of $\OBS$, provided in Section \ref{SectionCaseStudyObservability}, we obtain:

\[
\OBS^\bot=\textnormal{span}\left\{
\left[
\begin{array}{c}
0\\
1\\
1\\
\end{array}
\right]\right\},
\]

which is the generator of the rotations around the vertical axis. This agrees with our intuition as all the measurements (the linear velocity, $v$, and the bearing angle, $\beta$) are invariant under any rotation of the system around the vertical axis.

Let us consider a given state, $[\overline{\rho}, ~\overline{\phi}, ~\overline{\theta}]^T$ and let us compute its indistinguishable states. 
This computation cannot be performed automatically, as it requires to solve a differential equation. On the other hand, it is only provided for educational purposes.
From Equation (\ref{EquationDiffEqStates}), we need to solve the following differential equation:

\[
\left\{\begin{array}{ll}
  \frac{d\rho}{d\tau} &=  0\\
  \frac{d\phi}{d\tau} &=   1\\
  \frac{d\theta}{d\tau} &=   1\\
  \rho(0)&= \overline{\rho}, ~\phi(0)=\overline{\phi}, ~\theta(0)=\overline{\theta}\\
\end{array}\right.
\]

We obtain that all the states:

\[
\left[\begin{array}{c}
\overline{\rho}\\
\overline{\phi}+\tau\\
\overline{\theta}+\tau\\
\end{array}\right]
\]

are indinstinguishable. For a given $\tau$, the above state is obtained by rotating the state $[\overline{\rho}, ~\overline{\phi}, ~\overline{\theta}]^T$ by an angle of magnitude $\tau$ around the vertical axis.

\subsection*{Second scenario}
From the expression of $\OBS$, provided in Section \ref{SectionCaseStudyObservability}, we obtain:

\[
\OBS^\bot=\textnormal{span}\left\{
\left[
\begin{array}{c}
0\\
1\\
1\\
\end{array}
\right],~~
\left[
\begin{array}{c}
1\\
0\\
0\\
\end{array}
\right]\right\}.
\]

The first generator is again the
generator of the rotations around the vertical axis. 
It agrees with our intuition as also the measurements of the angular speed
$\omega$ are invariant under any rotation of the system around the vertical axis. The new symmetry is the generator of the absolute scale transformations. Its presence is not surprising. When the linear speed is unknown, all the available measurements are angular measurements ($\omega$ and $\beta$) and there is no metric information. In other words, all the measurements are also invariant to the scale.

Let us consider a given state, $[\overline{\rho}, ~\overline{\phi}, ~\overline{\theta}]^T$ and let us compute the indistinguishable states obtained by using the second generator. Equation (\ref{EquationDiffEqStates}) becomes:

\[
\left\{\begin{array}{ll}
  \frac{d\rho}{d\tau} &=  1\\
  \frac{d\phi}{d\tau} &=   0\\
  \frac{d\theta}{d\tau} &=   0\\
  \rho(0)&= \overline{\rho}, ~\phi(0)=\overline{\phi}, ~\theta(0)=\overline{\theta}\\
\end{array}\right.
\]

We obtain that all the states:

\[
\left[\begin{array}{c}
\overline{\rho}+\tau\\
\overline{\phi}\\
\overline{\theta}\\
\end{array}\right]
\]

are indinstinguishable. 
As $\rho$ is the only metric component of the state (the other two, $\theta$ and $\phi$, are angles), the above state can be regarded as a scale transformation of 
$[\overline{\rho}, ~\overline{\phi}, ~\overline{\theta}]^T$ (in particular, it is rescaled by the factor $\frac{\overline{\rho}+\tau}{\overline{\rho}}$).
Note that, a symmetry is defined up to a multiplying factor (for any vector in $\OBS^\bot$ also the vector obtained by multiplying it by a given scalar is in $\OBS^\bot$). Hence, 
the transformation on the state defined by $\xi=[1,~0,~0]^T$ can also be defined by the same symmetry multiplied by $\rho$. In other words, we could consider the symmetry:
$[\rho,~0,~0]^T$. By using this symmetry, Equation (\ref{EquationDiffEqStates}) provides, $\frac{d\rho}{d\tau}=\rho$, whose solution is $\rho(\tau)=\overline{\rho}e^\tau$.
This second option makes more readable that it is a scale transformation (in particular, the scale factor is $e^\tau$).

\subsection{Characterization by an observable state}\label{SubSectionCaseStudyLocDec}

This step cannot be performed automatically but, as the system is very simple, its execution is trivial. In addition, as we mentioned, this step is not required for the implementation of Algorithm \ref{AlgoFullIDE}.

\subsection*{First scenario}
The dimension of the observability codistribution is $2$ and two independent generators are: $\rho$, and  $\psi:=\theta-\phi$. Hence, a possible observable state is precisely $x=[\rho, ~\psi]^T$. From (\ref{EquationSimpleExampeDynamics}), we easily obtain:

\begin{equation}\label{EquationCaseStudyLocDec1}
\left[\begin{array}{ll}
  \dot{\rho} &= v \cos\psi \\
  \dot{\psi} &= \omega-\frac{v}{\rho} \sin\psi \\
\end{array}
\right.
\end{equation}

where $\omega(=w)$ is the unknown input and $v(=u)$ the known input. 

\subsection*{Second scenario}
The dimension of the observability codistribution is $1$ and one generator is $\psi$. Hence, a possible observable state is precisely $x=[\psi]$. From (\ref{EquationSimpleExampeDynamics}), we easily obtain: 
$\dot{\psi} = \omega-\frac{v}{\rho} \sin\psi$, 
where $v(=w)$ is the unknown input and $\omega(=u)$ the known input. Note that, by using $w$, we cannot obtain a description of the system in terms of the
solely observable state, because the expression of the dynamics needs the quantity $\rho$. We must redefine the unknown input. A possible choice is $\widetilde{w}:=\frac{w}{\rho}=\frac{v}{\rho}$.
The system dynamics become:

\begin{equation}\label{EquationCaseStudyLocDec2}
\left[\begin{array}{ll}
  \dot{\psi} &= \omega- \widetilde{w}\sin\psi \\
\end{array}
\right.
\end{equation}

\subsection{Identifiability}\label{SubSectionCaseStudyIdentifiability}

We use Algorithm \ref{AlgoFullIDE} to obtain the identifiability of the time-varying parameters for our case study. In particular, for the first scenario, the unknown time-varying parameter is the angular speed $\omega(t)$ while, in the second scenario, is the linear speed $v(t)$.
In both scenarios, the system is canonic  with respect to its unknown input. Hence, we can only have the unknown input symmetries due to the state unobservability ($^u\chi$).

\subsection*{First scenario}

We have a single symmetry of the state that is $\xi=[0,~1,~1]^T$. 
We compute the corresponding symmetry of the unknown input, $^u\chi$, by using (\ref{EquationSymmetryUIObs}). We need, first of all, to compute the terms $\xi^\alpha_i$ by using 
(\ref{EquationSymmetryUIObsCoeff}), for $\alpha=0,1$, and $i=1$. We have $\widetilde{h}_1=\widetilde{h}=\phi-\theta$.
As $g^0$ is null, $\Li_{g^0}\widetilde{h}=0$. Hence:

\[
\xi^0_1=0
\]

Let us compute $\xi^1_1$. We have $g^1=[0,~0,~1]^T$. We obtain: $\Li_{g^1}\widetilde{h}=-1$. As a result, 

\[
\xi^1_1=0
\]

From (\ref{EquationSymmetryUIObs}), we obtain that the symmetry of the unknown input vanishes.

Therefore, we do not have non trivial symmetries of the unknown input. This means that the unknown input can be reconstructed (or, in other words, the time-varying parameter $\omega(t)$ is identifiable).\\
Note that the same result is obtained by using Theorem \ref{TheoremUIRecObservable} on the system in (\ref{EquationCaseStudyLocDec1}). Indeed, the state that characterizes this system is observable and the system is canonic with respect to its unknown input. Hence, the unknown input can be reconstructed.

\subsection*{Second scenario}

We have two independent symmetries: $\xi^1=[0,~1,~1]^T$, and $\xi^2=[\rho,~0,~0]^T$. Note that, for the second symmetry, on the basis of the remark at the end of Section \ref{SubSectionCaseStudySymmetries}, we use $[\rho,~0,~0]^T$ instead of $[1,~0,~0]^T$. 
To compute the corresponding $^u\chi$, we need the terms $\xi^0_1$ and $\xi^1_1$. For both symmetries, we need to compute 
$\Li_{g^0}\widetilde{h}_1$ and $\Li_{g^1}\widetilde{h}_1$. On the other hand,
$g^0$ is null, $g^1=\left[\cos(\theta-\phi), ~\frac{\sin(\theta-\phi)}{\rho}, ~0\right]^T$, and
$\widetilde{h}_1=\widetilde{h}=\phi-\theta$.
Hence, we obtain:

\[
\Li_{g^0}\widetilde{h}=0, ~~\Li_{g^1}\widetilde{h}=\frac{\sin(\theta-\phi)}{\rho}.
\]

Let us consider the first symmetry: $\xi^1=[0,~1,~1]^T$. From (\ref{EquationSymmetryUIObsCoeff}), and the above expressions of $\Li_{g^0}\widetilde{h}$ and $\Li_{g^1}\widetilde{h}$,
we obtain:

\[
\xi^0_1=\xi^1_1=0.
\]

From (\ref{EquationSymmetryUIObs}), we obtain that the corresponding symmetry of the unknown input vanishes, i.e., $^u\chi^1=0$.

Let us consider the second symmetry $\xi^2=[\rho,~0,~0]^T$.
From (\ref{EquationSymmetryUIObsCoeff}), and the above expressions of $\Li_{g^0}\widetilde{h}$ and $\Li_{g^1}\widetilde{h}$,
we obtain:

\[
\xi^0_1=0, ~~~\xi^1_1=\frac{\partial \Li_{g^1}\widetilde{h}}{\partial x}\xi^2=-\frac{\sin(\theta-\phi)}{\rho}.
\]

We use (\ref{EquationSymmetryUIObs}) to compute $^u\chi$. Note that, in this case
$m=m_w=1$, and,
from (\ref{EquationCaseStudyMuNu}),
 $\nu^1_1=\nu^1_1=\frac{\rho}{\sin(\theta-\phi)}$. We obtain:

\[
^u\chi^2=v.
\]

This means that, although we have two symmetries of the state, we have a single non trivial symmetry of the unknown input.
The unknown input cannot be reconstructed (or, in other words, the time-varying parameter $v(t)$ is not identifiable). 

In this case, by applying Theorem \ref{TheoremUIRecObservable} on (\ref{EquationCaseStudyLocDec2}), we only conclude that the unknown input $\widetilde{w}=\frac{v(t)}{\rho(t)}$ can be reconstructed. However, the Theorem does not allow us to take a conclusion about the identifiability of the original time-varying parameter $w=v(t)$.

\chapter{Indistinguishable states and parameters and minimal missing information to achieve observability/identifiability}\label{ChapterIndistinguishableStatesAndUIsAndMI}


%
%

In this chapter we focus our attention to the systems that are unobservable and/or unidentifiable. We start by providing a general procedure that generate indistinguishable states and parameters (Section \ref{SectionIndistinguishableStatesAndUIs}). Then, in Section \ref{SectionProcedureExternalInformation} we exploit the determination of these indistinguishable states and parameters to characterize the minimal missing information to achieve observability and/or identifiability. In particular, we provide a preliminary result that allows us to obtain the above characterization in the presence of a single symmetry. 

\section{Indistinguishable states and parameters}\label{SectionIndistinguishableStatesAndUIs}

Algorithm \ref{AlgoFullIDE} can be executed automatically and provides a full answer to the problem of the identifiability of the time-varying parameters (or unknown input reconstruction). 
It uses the results stated by Theorems \ref{TheoremIdentifiabilityCan} and \ref{TheoremIdentifiability}, which are based on the powerful concept of infinitesimal transformation. 
In this section, we want to discuss the case of finite transformations. In other words, we want to extend the same analysis provided in Section \ref{SectionSymmetryUnobservability} to the unknown inputs. In that case, the infinitesimal transformation was given by Equation (\ref{EquationSymmetryState}) and the finite transformation was
given by Equation (\ref{EquationSymmetryStateFIN}), where $x(\tau)$ is the solution of the differential equation in (\ref{EquationDiffEqStates}).
Note that, the solution of the differential equation in (\ref{EquationDiffEqStates}) cannot be obtained automatically, and, in many cases, it can only be determined numerically. On the other hand, its determination is unnecessary to perform both the observability and the identifiability analysis. 
The same remarks hold for the differential equations given in this section (i.e., Equations (\ref{EquationDiffEqUICanonic}) and (\ref{EquationDiffEqStatesUI})), which provide the finite transformations to build indistinguishable unknown inputs and initial states, simultaneously.
On the other hand, their solution can provide useful insights on a given ODE model (in Sections \ref{SectionHIVIdentifiability}
we provide the analytical solution for a model in the framework of HIV dynamics and, in Section \ref{SectionCovidIdentifiability}, we provide the analytical solution for a model in the framework of Covid-19 dynamics).

Let us consider the system $\E$ returned by Algorithm \ref{AlgoFull}. It still satisfies Equation (\ref{EquationSystemDefinitionUIO}). 
Let us assume that the initial time is $t=0$ and the initial state is $x(0)=x_0$.  The state at time $t$ is obtained by solving the following differential equation (where all the quantities are the ones that characterize the system $\E$ returned by Algorithm \ref{AlgoFull}):

\begin{equation}\label{EquationSystemTrue}
\left\{\begin{array}{ll}
  \dot{x}(t) &=   g^0(x, t)+\sum_{k=1}^{m_u}f^k (x, t) u_k(t) +  \\
  &~~~~~~~~~~~~~~~~~~~~~~~~~~~~~~~~~+\sum_{j=1}^{m_w}g^j (x, t) w_j(t)  \\
  x(0)&=x_0 \\
\end{array}\right.
\end{equation}

We consider any choice of the known inputs, $u_1(t),\ldots,u_{m_u}(t)$, such that the above differential equation admits a solution in the time interval $[0,~T]$.

Now, we introduce the state $x'(t)$ obtained by solving the following differential equation:

\begin{equation}\label{EquationIndistSystem}
\left\{\begin{array}{ll}
  \dot{x}'(t) &=   g^0(x', t)+\sum_{k=1}^{m_u}f^k (x', t) u_k(t) \\
  &~~~~~~~~~~~~~~~~~~~~~~~~~~~~~~~~~+  \sum_{j=1}^{m_w}g^j (x', t) w_j'(t)  \\
  x'(0)&=x_0' \\
\end{array}\right.
\end{equation}

We emphasize that the state $x'(t)$ is obtained by solving the same differential equation that provides $x(t)$ but with the following two differences:

\begin{itemize}

\item The initial state is $x_0'$ instead of  $x_0$.

\item The unknown input is $w'(t)=[w_1'(t),\ldots,w_{m_w}'(t)]^T$ instead of $w(t)=[w_1(t),$ $\ldots,w_{m_w}(t)]^T$ (and this second difference involves the unknown input at any time in the considered time interval and not simply the initial time).

\end{itemize}

We are interested in detecting all the initial states $x_0'$ and all the unknown input functions $w'(t)$ such that, independently of the choice of the known input functions, the states $x'(t)$ produce exactly the same outputs of the states $x(t)$, i.e.:
\[
   h_i\left(x(t), ~t\right)=h_i\left(x'(t), ~t\right), ~~~i=1,\ldots,p,
\]

for any $t\in[0,~T]$.

On the basis of Theorems \ref{TheoremIdentifiabilityCan} and \ref{TheoremIdentifiability}, we have two sets of solutions, which are provided by the two types of unknown input symmetries, $^c\chi$ and $^u\chi$. 
Specifically, on the basis of the result stated by Theorem \ref{TheoremIdentifiabilityCan}, we obtain the following first set of solutions:

\begin{equation}\label{EquationTransformationINFUICanonic}
x_0'=x_0,~~~w'(t)=w(t)+\epsilon~^c\chi^i, ~~i=1,\ldots,m_w-m
\end{equation}

On the basis of the result stated by Theorem \ref{TheoremIdentifiability}, we obtain the following second set of solutions:

\begin{equation}\label{EquationTransformationINFUOBS}
x_0'=x_0+\epsilon~\xi^i,~~~w'(t)=w(t)+\epsilon~^u\chi^i, ~~i=1,\ldots,S
\end{equation}
where $\xi^i\in\OBS^\bot$ is the symmetry of the state that provides $^u\chi^i$, in accordance with (\ref{EquationSymmetryUIObs}) (i.e., $^u\chi^i$ is obtained by setting $\xi=\xi^i$ in (\ref{EquationSymmetryUIObsCoeff})).

Starting from the infinitesimal transformations given in (\ref{EquationTransformationINFUICanonic}) and in (\ref{EquationTransformationINFUOBS}), we can compute the corresponding finite transformations.
For the $m_w-m$ infinitesimal transformations given in (\ref{EquationTransformationINFUICanonic}), the corresponding finite transformations are the following. Regarding $x_0'$, it trivially remains the same for all of them, i.e.:

\[
x_0'=x_0
\]

Regarding the unknown inputs, we have $m_w-m$ independent $w'=w'(t,\tau)$, which are the solutions of the following differential equations:

\begin{equation}\label{EquationDiffEqUICanonic}
\left\{\begin{array}{ll}
  \frac{dw'}{d\tau} &=   ~^c\chi^i\\
  w'(t,~0)&=w(t),\\
\end{array}\right.
\end{equation}
$i=1,\ldots,m_w-m$.
In the above equation, the time $t$ is a fixed parameter. In other words, to obtain $w'(t,~\tau)$, we first set the time to the desired value $t$.
Due to the trivial expression of $^c\chi$ in (\ref{EquationIndwCanonic}) (and in particular its independence of the specific case), this differential equation can be solved once forever and provides:

\[
\left[\begin{array}{ll}
  w'_j(t,~\tau)=w_j(t),&j\neq m+i\\
  w'_j(t,~\tau)=w_j(t)+\tau,&j=m+i\\
\end{array}\right.
\]
%

For the $S$ infinitesimal transformations given in (\ref{EquationTransformationINFUOBS}), the corresponding finite transformations 
now involve simultaneously the initial state and the unknown inputs.
To analytically compute the corresponding finite transformations, we need the following condition to be satisfied:

\begin{equation}\label{EquationCONDITIONCommutativitaTempoSimmetria}
\left[
\xi,~g^0+\sum_{i=1}^{m_u}f^iu_i+\sum_{j=1}^{m_w}g^jw_j
\right]
+\sum_{j=1}^mg^j~^u\chi_j=0,
\end{equation}
where the square brackets are the Lie brackets.
This condition expresses the commutativity between the temporal shift and the shift generated by the considered symmetry.

If the condition in (\ref{EquationCONDITIONCommutativitaTempoSimmetria}) is honoured, the result of many consecutive applications of the $S$ infinitesimal transformations in (\ref{EquationTransformationINFUOBS}) can be analytically computed. Specifically, for the $i^{th}$ transformation in 
(\ref{EquationTransformationINFUOBS}) we obtain the new $x_0'=x_0'(\tau)$, and $w'=w'(t,\tau)$ by solving the following differential equations:

%
%

\begin{equation}\label{EquationDiffEqStatesUI}
\left\{\begin{array}{ll}
  \frac{dx'}{d\tau} &=   \xi^i(x'(t,~\tau))\\
  \frac{dw'}{d\tau} &=   ~^u\chi^i(x'(t,~\tau),~w'(t,~\tau))\\
  x'(t,~0) &=x(t),~~w'(t,~0)=w(t),\\
\end{array}\right.
\end{equation}
$i=1,\ldots,S$.
Again, in the above equation, the time $t$ is a fixed parameter. In other words, to obtain $x'(t,~\tau)$, and $w'(t,~\tau)$, we first set the time to desired value $t$.
If, for any $t\in[0,~T]$, the above equation admits a solution in the interval $[0,~\mathcal{T}]$, then, for any $\tau\in[0,~\mathcal{T}]$, the unknown input vector $w'(t,\tau)$ is indistinguishable from 
the unknown input vector $w(t)$. Note that, in the above equation, the value of $x(t)$, which appears at the initial condition  $x'(t,~0) =x(t)$, is the solution of (\ref{EquationSystemTrue}) at the time $t$.
From $x'(t,~\tau)$ we obtain the finite transformation of the initial state, which is the value of this function at the initial time, i.e., $x_0'=x_0'(\tau)=x'(0,~\tau)$.

The solution of (\ref{EquationDiffEqStatesUI}) is obtained by first determining $x'(t,~\tau)$ (i.e., by first solving $ \frac{dx'}{d\tau} =   \xi^i(x'(t,~\tau))$, with initial condition $x'(t,~0) =x(t)$), and then by using it  in $ \frac{dw'}{d\tau} =   ~^u\chi^i(x'(t,~\tau),~w'(t,~\tau))$, which becomes a differential equation of only $w'(t,~\tau)$. 
Note that the solution of
$ \frac{dx'}{d\tau} =   \xi^i(x'(t,~\tau))$, with initial condition $x'(t,~0) =x(t)$, is precisely the solution 
of (\ref{EquationDiffEqStates}) with $\overline{x}=x(t)$ and for the same state symmetry (i.e., $\xi=\xi^i$).

\section{Minimal missing information to achieve observability/identifiability}\label{SectionProcedureExternalInformation}
We discuss a fundamental issue that arises every time we are in the presence of unobservability and/or unidentifiability. We wish to answer to the following practical question:
{\it What is the minimal 
external information (external to the available measurements of the system outputs) required to make observable the state and identifiable all the model parameters?}

In particular, we would like an automatic answer, which holds for any system,
as in the case of the observability and identifiability analysis that are obtained by simply running Algorithm \ref{AlgoFull} and \ref{AlgoFullIDE}.

Unfortunately, we do not have a general answer yet (this will be the matter of a future investigation). 
%
%
For the moment, we have a preliminary result that provides an answer to our question when the state is characterized by a single symmetry (i.e., the orthogonal distribution $\OBS^\bot$ has a single generator).
The key is that, in the presence of a single symmetry, we are able to determine {\it all} the indistinguishable states by solving the differential equations in (\ref{EquationDiffEqStates}) (or even to determine {\it all} the indistinguishable states and unknown inputs by solving  (\ref{EquationDiffEqStatesUI})).

Let us denote the components of the true state at time $t$ by $x_1(t), ~x_2(t),~\ldots,~x_n(t)$. As the system is characterized by a single symmetry, all the indistinguishable states are:

\begin{equation}\label{EquationExtInformationSingleSymmetry}
\left[\begin{array}{c}
x'_1(t,~\tau)= s_1(x_1(t), ~x_2(t),~\ldots,~x_n(t),~\tau)\\
x'_2(t,~\tau)= s_2(x_1(t), ~x_2(t),~\ldots,~x_n(t),~\tau)\\
\ldots\\
x'_n(t,~\tau)= s_n(x_1(t), ~x_2(t),~\ldots,~x_n(t),~\tau)\\
\end{array}\right.
\end{equation}
where $s_1,\ldots,s_n$ are specific functions determined by solving the differential equations in (\ref{EquationDiffEqStates}) (or (\ref{EquationDiffEqStatesUI}), in which case the system in (\ref{EquationExtInformationSingleSymmetry}) includes the components of the state and the unknown inputs).

We know that, by using all the available measurements, we cannot determine $x_1(t), ~x_2(t),~\ldots,~x_n(t)$, but we can determine $x_1'(t,~\tau), ~x_2'(t,~\tau),~\ldots,~x_n'(t,~\tau)$. The latter 
are related to the former by the equation system in (\ref{EquationExtInformationSingleSymmetry}). In particular, this system can be regarded as a system of $n$ equations in the $n+1$ unknowns, which are the $n$ components of the true state and $\tau$. Note that we have such a system of equations for any $t$ in our time interval.
On the other hand, while the components of the state are in general time dependent, $\tau$ is independent of $t$. Let us suppose that we add a measurement, at a single time $t^*$, on one of the unobservable components of the true state. Without loss of generality, let us assume that this is the last component of the state, (i.e., we are measuring $x_n(t^*)$). The fact that $x_n$ is unobservable, means that the function $s_n$ depends on $\tau$. Now we consider our equation system in (\ref{EquationExtInformationSingleSymmetry}) at $t=t^*$. This system consists of $n$ equations in $n$ unknowns, which are the first $n-1$ components of the true state at time $t^*$, and $\tau$. In general, this allows us to obtain the $n$ unknowns, and in particular, to obtain $\tau$. But once $\tau$ has been determined, our equation system in (\ref{EquationExtInformationSingleSymmetry}) for any $t\neq t^*$ consists of $n$ equations in $n$ unknowns, which are the $n$ components of the true state at time $t$. In general, this allows us to obtain the $n$ unknowns, i.e., to obtain the components of the state at any time $t$.

In Section \ref{SectionCaseStudyIndAndMI} we provide a trivial application of this procedure for the case study.
In Section \ref{SectionHIVMinimalExtInfo} we will see a more complex application of this procedure for a model that characterizes the HIV dynamics.


\section{Illustration by the case study}\label{SectionCaseStudyIndAndMI}


We determine the set of indistinguishable unknown inputs for our case study. Note that, when the system parameters are identifiable, this set is trivial and consists of a single (the true) unknown input. Hence, we only consider the second scenario for which we obtained a single non trivial symmetry of the unknown input vector. Specifically, in Section \ref{SubSectionCaseStudyIdentifiability} we obtained $\xi=[\rho,~0,~0]^T$ and $^u\chi=v$.
By an explicit computation, it is possible to check that the condition in (\ref{EquationCONDITIONCommutativitaTempoSimmetria}) is honoured.
The differential equation in (\ref{EquationDiffEqStatesUI}) becomes:

\begin{equation}\label{EquationCaseStudyDiffEqStateUI}
\left\{\begin{array}{ll}
  \frac{d\rho'}{d\tau} &=  \rho'(t,~\tau)\\
  \frac{d\phi'}{d\tau} &=   0\\
  \frac{d\theta'}{d\tau} &=   0\\
  \frac{dv'}{d\tau} &= v'(t,~\tau)\\
  \rho'(t,~0)&= \rho(t), ~\phi'(t,~0)=\phi(t), ~\theta'(t,~0)=\theta(t)\\
  v'(t,~0)&=v(t)\\
\end{array}\right.
\end{equation}

and its solution is:

\begin{equation}\label{EquationCaseStudyIndStateUI}
\left[
\begin{array}{ll}
\rho'(t,~\tau)&=\rho(t)~e^\tau,\\
\phi'(t,~\tau)&=\phi(t),\\
\theta'(t,~\tau)&=\theta(t),\\
v'(t,~\tau)&=v(t)~e^\tau,\\
\end{array}
\right.
\end{equation}

%
%
%
%
%
%
which is a scale transformation with factor $e^\tau$.
As expected, the unknown input $v(t)$ can only be reconstructed up to a scale.
By computing the above $\rho'(t,~\tau)$ at $t=0$ we obtain that the initial state is only observable up to the same scale, $\rho'_0=\rho'(0,~\tau)=\rho_0~e^\tau$. 

Note that, by using $\xi=[1,~0,~0]^T$ instead of $\xi=[\rho,~0,~0]^T$, we would obtain the same result. The corresponding $^u\chi$ is $\frac{v}{\rho}$ and the differential equation in (\ref{EquationDiffEqStatesUI}), becomes:

\[
\left\{\begin{array}{ll}
  \frac{d\rho'}{d\sigma} &=  1\\
  \frac{d\phi'}{d\sigma} &=   0\\
  \frac{d\theta'}{d\sigma} &=   0\\
  \frac{dv'}{d\sigma} &= \frac{v'(t,~\sigma)}{\rho'(t,~\sigma)}\\
  \rho'(t,~0)&= \rho(t), ~\phi'(t,~0)=\phi(t), ~\theta'(t,~0)=\theta(t)\\
  v'(t,~0)&=v(t)\\
\end{array}\right.
\]
where we used $\sigma$ instead of $\tau$. The solution is:

\[
\left[
\begin{array}{ll}
\rho'(t,~\sigma)&=\rho(t)+\sigma,\\
\phi'(t,~\sigma)&=\phi(t),\\
\theta'(t,~\sigma)&=\theta(t),\\
v'(t,~\sigma)&=\frac{v(t)}{\rho(t)}(\rho(t)+\sigma),\\
\end{array}
\right.
\]

which coincides with the previous solution by setting $\sigma=\rho(t)(e^\tau-1)$.

We conclude by remarking that,
in accordance with the above results, the ratio $\frac{v}{\rho}$ is unaffected by the above scale transformation. Indeed, we have:

\[
\frac{v}{\rho}\rightarrow\frac{ve^\tau}{\rho e^\tau}=\frac{v}{\rho}.
\]

This means that this ratio is observable. This result is consistent with the result that we obtain by applying Theorem \ref{TheoremUIRecObservable} on the system in (\ref{EquationCaseStudyLocDec2}), which is characterized by an observable state. As this system is also canonic with respect to its unknown input, it means that the unknown input $\widetilde{w}=\frac{v}{\rho}$ can be reconstructed.


\vskip.2cm

We conclude this section by characterizing the minimal missing information to achieve observability and identifiability. Clearly, due to the simplicity of the system, we obtain an answer by proceeding intuitively.
Specifically, we know that to have observability and identifiability we need to know the absolute scale. This is provided by a single measurement of the distance $\rho$ or by a single measurement of the speed $v$. We show that we obtain the same result by applying the procedure provided in Section \ref{SectionProcedureExternalInformation}. Note that we are allowed to use this procedure because the system is characterized by a single symmetry.

Equation (\ref{EquationExtInformationSingleSymmetry}) becomes, in this case, (\ref{EquationCaseStudyIndStateUI}). In order to determine the true state and the true parameter, we must determine the value of $\tau$. This can be determined 
by knowing $\rho$ or $v$ at a single time ($t^*$). In the first case, $\tau$ is obtained from the first equation  in (\ref{EquationCaseStudyIndStateUI}), which provides $\rho(t^*)e^\tau$. The knowledge of $\rho$ at $t^*$, i.e. $\rho(t^*)$, makes immediate the determination of $e^\tau$, and then $\tau$.
Once $\tau$ has been determined, we determine the remaining quantities at any $t\in\mathcal{I}$.

\chapter{HIV infection}\label{ChapterHIV}

We investigate the observability and the identifiability properties of a simple ODE model widely used to describe
HIV dynamics in HIV-infected patients with antiretroviral treatment
\cite{Miao11,Villa19b,Per99,Per02,Chen08}.
The model is characterized by the following three equations and the following two outputs:

\begin{equation}\label{EquationHIVSystem}
\left\{\begin{array}{ll}
\dot{T}_U &= \lambda -\rho T_U -\eta(t) T_U V\\
\dot{T}_I &= \eta(t) T_U V -\delta T_I\\
\dot{V} &= N\delta T_I - cV\\
y &= [V,~T_U+T_I], \\
\end{array}\right.
\end{equation}
where, $T_U$ is the concentration of uninfected cells, $T_I$ the concentration of infected cells, and $V$ the viral load. The model is also characterized by the time-varying parameter $\eta(t)$ and the five constant parameters $\lambda,~\rho,~\delta,~N,~c$. They are defined as follows:

\begin{itemize}

\item $\eta(t)$ is the infection rate, which is a function of the antiviral treatment efficacy. 
\item $\lambda$ is the source rate of uninfected cells. 
\item $\rho$ is the death rate of uninfected cells. 
\item $\delta$ is the death rate of infected cells. 
\item $N$ is the average number of virions produced by a single infected cell during its lifetime. 
\item $c$ is the clearance rate of free virions.

\end{itemize}

All these parameters are assumed to be unknown.

\vskip.2cm

To proceed, we need, first of all, to introduce a state that includes both the time-varying quantities (i.e., $T_U,~T_I,~V$) and the constant parameters (i.e., $\lambda,~\rho,~\delta,~N,~c$). We set:

\begin{equation}\label{EquationHIVState}
x=[T_U,~T_I,~V,~\lambda,~\rho,~\delta,~N,~c]^T
\end{equation}

The system is directly a special case of (\ref{EquationSystemDefinitionUIO}). In particular, 
$m_u=0$, $m_w=1$, $p=2$, $h_1(x)=V$, $h_2(x)=T_U+T_I$,

\begin{equation}\label{EquationHIVg0g1}
g^0=
\left[
\begin{array}{c}
 \lambda -\rho T_U\\
 -\delta T_I\\
N\delta T_I - cV\\
0\\
0\\
0\\
0\\
0\\
\end{array}
\right],~~
g^1=
\left[
\begin{array}{c}
 -T_UV\\
 T_UV\\
0\\
0\\
0\\
0\\
0\\
0\\
\end{array}
\right]
\end{equation}

\section{Observability analysis}\label{SectionHIVObs}

We run Algorithm \ref{AlgoFull} to obtain the observability codistribution.
Line \ref{AlgoLineOmegaINIT} provides:
 
\[
\Omega=\textnormal{span}\{\nabla h_1,~\nabla h_2\}
=\textnormal{span}\{[0, 0, 1, 0, 0, 0, 0, 0],~[1, 1, 0, 0, 0, 0, 0, 0]\}.
\]

  $\uideg \left({\Omega}\right)=0~(<m_w)$, as $\Li_{g^1}h_1=\Li_{g^1}h_2=0$ (and the rank of the unknown input reconstructability matrix from $h_1,~h_2$ vanishes).
The system is not in canonical form with respect to its unknown input.


We have $m=0$ and
$\mu^0_0=~\nu^0_0=1$ obtained from (\ref{EquationTensorMSynchrom}) (these tensors have a single entry as the indices only take the value 0).
 In addition, from (\ref{Equationgalpham}) we obtain: $\widehat{g}^0=~\nu^0_0g^0=g^0$, and $\giat^1=g^1$.
 
 Algorithm \ref{Algo***} provides Finish=False.
We have:
\[
\Omega=
\left<\left.\left\{ 
\begin{array}{l}
 \dtv{\Li}_{\giat} \\
 \giat^1\\
\end{array}
\right\}~\right|\textnormal{span}\left\{\nabla h_1,~\nabla h_2\right\}\right>=
\left<\left.\left\{ 
\begin{array}{l}
 g^0 \\
 g^1\\
\end{array}
\right\}~\right|\textnormal{span}\left\{\nabla h_1,~\nabla h_2\right\}\right>,
\]
as $\giat=\widehat{g}^0=g^0$.
By a direct computation, we obtain:

\[
\Omega=\textnormal{span}\left\{\nabla h_1,~\nabla h_2,~\nabla\Li_{g^0}h_1,~\nabla\Li_{g^0}h_2\right\}=
\]
\[
\textnormal{span}\left\{\nabla h_1,~\nabla h_2,~\nabla h_3,~\nabla h_4\right\}
\]
where we adopted the following notation:
\[
h_3:= \Li_{g^0}h_1,~~
h_4:= \Li_{g^0}h_2
\]

Now, $\uideg\left(\Omega\right)=1=m_w$. Hence, the execution of Algorithm \ref{AlgoFull} continues with Line \ref{AlgoLineFINITSelection}. The operation at this line ($[\widetilde{h}_1,\ldots, \widetilde{h}_{m_w}]=\mathcal{S}(\Sigma,~\Omega)$) sets 

\[
\widetilde{h}_1=h_4=\Li_{g^0}h_2=\lambda-\rho T_U-\delta T_I
\]

(we would obtain the same final result by setting  $\widetilde{h}_1=h_3=\Li_{g^0}h_1$).
Then, Line \ref{AlgoLineFINITMuNuTOBS} provides:

\[
\mu=\left[
\begin{array}{cc}
1&0\\
T_I\delta^2 - \rho(\lambda - T_U\rho)& -T_UV(\delta - \rho)\\
\end{array}
\right],
\]
\begin{equation}\label{EquationHIVnu}
\nu=\left[
\begin{array}{cc}
1&0\\
\frac{T_I\delta^2 + T_U\rho^2 - \lambda\rho}{T_UV(\delta - \rho)}& -\frac{1}{T_UV(\delta - \rho)}\\
\end{array}
\right],
\end{equation}

obtained from (\ref{EquationTensorMSynchro}). In addition, from (\ref{Equationgalpha}) we obtain:

\[
\widehat{g}^0=
\left[
\begin{array}{c}
-\frac{\delta(T_I\delta - \lambda + T_U\rho)}{\delta - \rho}\\
 \frac{\rho(T_I\delta - \lambda + T_U\rho)}{\delta - \rho}\\
 NT_I\delta - Vc\\
0\\
0\\
0\\
0\\
0\\
\end{array}
\right],~~
\widehat{g}^1=
\left[
\begin{array}{c}
 \frac{1}{\delta - \rho}\\
 -\frac{1}{\delta - \rho}\\
0\\
0\\
0\\
0\\
0\\
0\\
\end{array}
\right]
\]

Finally, from (\ref{EquationTOBSDef}) we obtain that, when $m_u=0$, the codistribution $\tobs$ vanishes. 
The final step of Algorithm \ref{AlgoFull} is the execution of Line \ref{AlgoLineFINALSTEP}, namely:

\[
\OBS=
\left<
\left.
\widehat{g}^0,~
\widehat{g}^1
\right|
~\textnormal{span}\left\{\nabla h_1,~\nabla h_2,~\nabla h_3,~\nabla h_4\right\}\right>.
\]

The algorithm 
in (\ref{EquationAlgorithmsMinimalCod}) converges at the third step ($\Omega_3=\Omega_2$), and $\OBS=$

\[
\textnormal{span}\left\{
\nabla h_1,~\nabla h_2,~\nabla h_3,~\nabla h_4,~\nabla\Li_{\widehat{g}^0}h_3,~\nabla\Li_{\widehat{g}^1}h_3,~\nabla\Li_{\widehat{g}^0}^2h_3
\right\}.
\]

Its dimension is $7$, which is smaller than the dimension of the state in (\ref{EquationHIVState}). Hence, the state is not observable. 
In particular, regarding the constant parameters, it is immediate to check that two of them ($\delta$ and $N$) are unidentifiable (it suffices to show that their gradient, i.e., the two covectors $[0,0,0,0,0,1,0,0]$ and $[0,0,0,0,0,0,1,0]$, does not belong to $\OBS$).
This result contradicts the result available in the state of the art (e.g., see Section 6.2 of \cite{Miao11} or Section 3.3 of \cite{Villa19b}).
 In Section \ref{SectionHIVComparisonSOTA}, we discuss the above result by referring to the same data set adopted in \cite{Villa19b}.

\section{Symmetries and indistinguishable states}\label{SectionHIVSymmetries}

Once we have $\OBS$, the symmetries are immediately obtained by computing the orthogonal distribution. We have:

\begin{equation}\label{EquationHIVSymmetry}
\OBS^\bot=\textnormal{span}\left\{
\left[\begin{array}{l}
T_I\delta\\
-T_I\delta\\
0\\
0\\
0\\
\delta(\delta-\rho)\\
N\rho\\
0\\
\end{array}\right]\right\}.
\end{equation}

As for the case study, starting from the generators of $\OBS^\bot$ we compute the indistinguishable states (although the determination of the indistinguishable states is not requested by Algorithm \ref{AlgoFullIDE}).
We need to solve the differential equation in (\ref{EquationDiffEqStates}). 
Given a state $[\overline{T}_U, ~\overline{T}_I, ~\overline{V}, ~\overline{\lambda}, ~\overline{\rho}, ~\overline{\delta}, ~\overline{N}, ~\overline{c}]^T$, the solution of this differential equation provides states which are indistinguishable. 
For the specific case, by using the generator in (\ref{EquationHIVSymmetry}) for $\xi$, 
Equation (\ref{EquationDiffEqStates}) becomes:

\begin{equation}\label{EquationHIVDiffEqStates}
\left\{\begin{array}{ll}
  \frac{dT_U}{d\tau} &=  T_I\delta\\
  \frac{dT_I}{d\tau} &=  -T_I\delta\\
  \frac{dV}{d\tau} &=  0\\
  \frac{d\lambda}{d\tau} &=  0\\
  \frac{d\rho}{d\tau} &=  0\\
  \frac{d\delta}{d\tau} &=  \delta(\delta-\rho)\\  
  \frac{dN}{d\tau} &=   N\rho\\
  \frac{dc}{d\tau} &=  0\\
 T_U(0)&=\overline{T}_U, ~T_I(0)= \overline{T}_I, ~V(0)= \overline{V}, ~\lambda(0)= \overline{\lambda},\\
 \rho(0)&= \overline{\rho}, ~\delta(0)= \overline{\delta}, ~N(0)= \overline{N}, ~c(0)= \overline{c} \\
\end{array}\right.
\end{equation}

This equation can be solved analytically. The solution is:

\begin{equation}\label{EquationHIVIndistinguishableStates}
x(\tau)=
\left[\begin{array}{l}
\overline{T}_U+\overline{T}_I-\frac{\overline{T}_I}{\overline{\rho}}\left(
\overline{\delta}e^{-\overline{\rho}\tau}+\overline{\rho}-\overline{\delta}
\right)\\
\frac{\overline{T}_I}{\overline{\rho}}\left(
\overline{\delta}e^{-\overline{\rho}\tau}+\overline{\rho}-\overline{\delta}
\right)\\
\overline{V}\\
\overline{\lambda}\\
\overline{\rho}\\
\overline{\delta}\overline{\rho}\left/
\left((\overline{\rho}-\overline{\delta})e^{\overline{\rho}\tau}+\overline{\delta}\right)\right.\\
\overline{N}e^{\overline{\rho}\tau}\\
\overline{c} \\
\end{array}\right]
\end{equation}

All the above states ($x(\tau)$) are indistinguishable for any $\tau$ (in particular, they are indistinguishable from $[\overline{T}_U, ~\overline{T}_I, ~\overline{V}, ~\overline{\lambda}, ~\overline{\rho}, ~\overline{\delta}, ~\overline{N}, ~\overline{c}]^T$). In Section \ref{SectionHIVComparisonSOTA} we explicitly show that they produce the same outputs.
%
%
%
%
%

\section{Characterization by an observable state}\label{SectionHIVLocDec}

Note that this task is not requested by Algorithm \ref{AlgoFullIDE}, to obtain the identifiability of the parameters. In addition, this task 
cannot be performed automatically.
On the other hand, we were able to obtain this description that can be useful for further investigations.

The original state has dimension 8 and the observability codistribution 7. This is the reason why we have a single symmetry ($8=7+1$). A possible observable state consists precisely of a set of generators of the observability codistribution. Algorithm \ref{AlgoFull} automatically returns a set of generators. We obtained: 
\[
h_1,~ h_2,~\Li_{g^0}h_1,~\Li_{g^0}h_2,~\Li_{\widehat{g}^0} \Li_{g^0}h_1,~\Li_{\widehat{g}^1}\Li_{g^0}h_1,~\Li_{\widehat{g}^0}^2\Li_{g^0}h_1.
\]
On the other hand, by setting the state entries equal to these functions, obtaining the expression of its dynamics in terms of the same state components is definitely prohibitive (although possible). 
To simplify this task, first of all, we try to determine 7 independent and simple functions, which are observable and that generate the observability codistribution. By using the expression of the above generators and the expression of the above symmetry, by using our inventiveness, we were able to select the following observable functions:

\begin{equation}\label{EquationHIVObservables}
\begin{array}{ll}
-~~ \Psi_1 &:= V\\
-~~ \Psi_2 &:= T_U+T_I\\
-~~ \Psi_3 &:= N\delta T_I\\
-~~ \mu &:= \frac{\delta-\rho}{N\delta}\\
-~~ \lambda&\\
-~~ c&\\
-~~ \rho&\\
\end{array}
\end{equation}

(by a direct calculation it is possible to verify that their gradients generate $\OBS$). 
As their expression is much easier than the expression of the generators automatically selected by Algorithm \ref{AlgoFull}, we have the possibility to describe our system by using them. After some trials, by using (\ref{EquationHIVSystem}),
we were able to obtain the following dynamics:

\begin{equation}\label{EquationHIVLocalDec}
\left\{\begin{array}{ll}
\dot{\Psi}_1 &= \Psi_3-c\Psi_1\\
\dot{\Psi}_2 &= \lambda-\rho\Psi_2-\Psi_3~\mu\\
\dot{\Psi}_3 &= \widetilde{\eta}\\
\dot{\mu} &=0\\
\dot{\lambda} &=0\\
\dot{c} &=0\\
\dot{\rho} &=0\\
y&=[\Psi_1,~\Psi_2]\\
\end{array}\right.
\end{equation}
The correctness of the above expression can be easily verified by using in 
(\ref{EquationHIVLocalDec})
the expressions of $\Psi_1, ~\Psi_2,~\Psi_3$, and $\mu$ given in (\ref{EquationHIVObservables}) and by using  Equation (\ref{EquationHIVSystem}).
Note that, in order to achieve this description, we had to redefine the unknown input ($\widetilde{\eta}:=
N\delta T_U\Psi_1\eta-\delta\Psi_3$). By using the original unknown input $\eta$, we would obtain $\dot{\Psi}_3 =N\delta T_U\Psi_1\eta-\delta\Psi_3$. This equation cannot be expressed only in terms of the components of the new observable state. It needs the product $N\delta T_U$ and $\delta$, which cannot be expressed in terms of the selected observable functions because both $N\delta T_U$ and $\delta$ are unobservable.

\section{Identifiability analysis}\label{SectionHIVIdentifiability}

We execute Algorithm \ref{AlgoFullIDE}.
First, as the system is canonic with respect to its unknown inputs, we do not have symmetries of the unknown input due to the system non canonicity ($^c\chi$). On the other hand, 
we have a single symmetry of the state ($\xi$), which is the single generator of $\OBS^\bot$ in (\ref{EquationHIVSymmetry}), and,
as a result, we may have a non vanishing symmetry $^u\chi$ of the unknown input. Let us compute it.
In accordance with Equation (\ref{EquationSymmetryUIObs}), we need to compute 
$\xi^0_1$ and $\xi^1_1$.
We have:
$\widetilde{h}=\widetilde{h}_1=\lambda-\rho T_U-\delta T_I$. From (\ref{EquationHIVg0g1}) we obtain:

\[
\Li_{g^0}\widetilde{h}=T_I\delta^2 - \rho(\lambda - T_U\rho),~~
\Li_{g^1}\widetilde{h}= -T_UV(\delta - \rho).
\]

From Equation (\ref{EquationSymmetryUIObsCoeff}), by using the generator of $\OBS^\bot$ in (\ref{EquationHIVSymmetry}) for $\xi$, we obtain:

\[
\xi^0_1=T_I\delta(\delta - \rho)^2,~~
\xi^1_1=-V\delta(T_I + T_U)(\delta - \rho).
\]

In addition, from Equation (\ref{EquationHIVnu}) and by knowing that $m=m_w=1$ (canonic system), we have $\nu^1_1=\nu^1_1=-\frac{1}{T_UV(\delta - \rho)}$.

By using (\ref{EquationSymmetryUIObs}) and the above expressions, we obtain:

\[
^u\chi = -\nu^1_1 ( \xi^0_1 + \xi^1_1 w_1 )=
\frac{1}{T_UV(\delta - \rho)}T_I\delta(\delta - \rho)^2
\]
\[
-\frac{1}{T_UV(\delta - \rho)}V\delta(T_I + T_U)(\delta - \rho)\eta.
\]

Hence:
\begin{equation}\label{EquationHIVSymmetryUI}
^u\chi=\frac{T_I\delta(\delta - \rho)}{T_UV}
-\frac{\delta(T_I + T_U)}{T_U}\eta.
\end{equation}

We have obtained a non trivial symmetry ($^u\chi$) of the unknown input. This means that the unknown input cannot be reconstructed (or, in other words, the time-varying parameter $\eta(t)$ is not identifiable).
This result contradicts the result available in the state of the art
(e.g., see Section 6.2 of \cite{Miao11} or Section 3.3 of \cite{Villa19b}). In Section \ref{SectionHIVComparisonSOTA}, we prove the validity of the above result by quantitatively providing more unknown inputs that are consistent with the same outputs.

\section{Indistinguishable states and unknown inputs}\label{SectionHIVIndStatesAndUI}

In order to obtain the indistinguishable states and unknown inputs
we need to solve the differential equation in (\ref{EquationDiffEqStatesUI}), for the specific case. We first need to check if the condition in (\ref{EquationCONDITIONCommutativitaTempoSimmetria}) is honoured.
By using the expression of $\xi$ (which is the generator of $\OBS^\bot$ in (\ref{EquationHIVSymmetry})) and the expression of $^u\chi$ in (\ref{EquationHIVSymmetryUI}), it is possible to check that 
this condition is satisfied and
the differential equation in (\ref{EquationDiffEqStatesUI}) becomes:

\begin{equation}\label{EquationHIVDiffEqStatesUI}
\left\{\begin{array}{ll}
  \frac{dT_U'}{d\tau} &=  T_I'\delta'\\
  \frac{dT_I'}{d\tau} &=  -T_I'\delta'\\
  \frac{dV'}{d\tau} &=  0\\
  \frac{d\lambda'}{d\tau} &=  0\\
  \frac{d\rho'}{d\tau} &=  0\\
  \frac{d\delta'}{d\tau} &=  \delta'(\delta'-\rho')\\  
  \frac{dN'}{d\tau} &=   N'\rho'\\
  \frac{dc'}{d\tau} &=  0\\
  \frac{d\eta'}{d\tau} &=  \frac{T_I'\delta'(\delta' - \rho')}{T_U'V'}
-\frac{\delta'(T_I' + T_U')}{T_U'}\eta'\\
 T_U'(t,~0)&=T_U(t), ~T_I'(t,~0)= T_I(t),\\
 V'(t,~0)&= V(t),\\
  \lambda'(0)= \lambda, ~\rho'(0)&= \rho, ~\delta'(0)= \delta, ~N'(0)= N, ~c'(0)= c\\
 \eta'(t,~0)&= \eta(t) \\
\end{array}\right.
\end{equation}

where we only provide the explicit dependence on $t$ and $\tau$ at the initial conditions and we omit it for the differential equations, for the simplicity sake.
These equations can be solved analytically. The solution for the first eight components (which are the components of the state) is the same solution of
(\ref{EquationHIVDiffEqStates}) given 
in (\ref{EquationHIVIndistinguishableStates}), 
with the appropriate values for the initial conditions. In other words, $x'(t,~\tau)$ is:

\begin{equation}\label{EquationHIVIndistinguishableStatesforUI}
\left\{\begin{array}{ll}
T_U'(t,~\tau)&=T_U(t)+ T_I(t)-\frac{ T_I(t)}{ \rho}\left(
 \delta e^{- \rho\tau}+ \rho- \delta
\right)\\
 T_I'(t,~\tau)&=\frac{ T_I(t)}{ \rho}\left(
 \delta e^{- \rho\tau}+ \rho- \delta
\right)\\
V'(t,~\tau)&=V(t)\\
\lambda'(\tau)&= \lambda\\
\rho'(\tau)&= \rho\\
\delta'(\tau)&= \delta \rho\left/
\left(( \rho- \delta)e^{ \rho\tau}+ \delta\right)\right.\\
N'(\tau)&= Ne^{ \rho\tau}\\
c'(\tau)&= c \\
\end{array}\right.
\end{equation}

Regarding the unknown input, we obtain:

\begin{equation}\label{EquationHIVIndistinguishableUI}
\eta'(t,~\tau)=\frac{ \eta(t) T_U(t)  V(t)  \rho e^{ \rho \tau} +\left[T_I(t)  \delta^2-  T_I(t)  \delta  \rho-  \eta(t) T_U(t)  V(t)  \delta  \right] \left[e^{ \rho \tau}-1 \right]}{ V(t) \left[ T_I(t)\delta +  T_U(t)  \rho \right] e^{ \rho \tau} -  V(t)T_I(t)  \delta}
\end{equation}


%

%


We have obtained the following fundamental result.
Let us consider the ODE model in (\ref{EquationHIVSystem}) and let us consider a given time interval $\mathcal{I}:=[t_0,~T]$.
Let us suppose that the initial conditions and the values of the model parameters are:

\begin{itemize}
\item $T_U(t_0)=T_{U_0}$, $T_I(t_0)=T_{I_0}$, and $V(t_0)=V_0$.

\item The values of the constant parameters are $\lambda,~\rho,~\delta,~N$, and $c$.

\item The time-varying parameter on $\mathcal{I}$ is $\eta(t)$.
\end{itemize}

Then, the following initial conditions and the values of the model parameters:

\begin{itemize}
%
%
%
%
%
%
%
%

\item 
\begin{equation}\label{EquationHIVIndistinguishableInitStates}
\left[\begin{array}{ll}
T_U'(t_0,~\tau)&=T_{U_0}+ T_{I_0}-\frac{ T_{I_0}}{ \rho}\left(
 \delta e^{- \rho\tau}+ \rho- \delta
\right)\\
 T_I'(t_0,~\tau)&=\frac{ T_{I_0}}{ \rho}\left(
 \delta e^{- \rho\tau}+ \rho- \delta
\right)\\
V'(t_0,~\tau)&=V_0\\
\lambda'(\tau)&= \lambda\\
\rho'(\tau)&= \rho\\
\delta'(\tau)&= \delta \rho\left/
\left(( \rho- \delta)e^{ \rho\tau}+ \delta\right)\right.\\
N'(\tau)&= Ne^{ \rho\tau}\\
c'(\tau)&= c \\
\end{array}\right.
\end{equation}

\item the time-varying parameter given in (\ref{EquationHIVIndistinguishableUI}), with $T_U(t), ~T_I(t),$ and $V(t)$ the values of $T_U$, $T_I$, and $V$ obtained by integrating (\ref{EquationHIVSystem}) with initial conditions $T_{U_0}$, $T_{I_0}$, and $V_0$, and unknown input $\eta(t)$,

\end{itemize}

produce exactly the same outputs $y_1(t), ~y_2(t)$, on the entire time interval $\mathcal{I}$.
This holds for any value of the parameter $\tau$ that belongs to an interval $\mathcal{U}\subseteq\mathbb{R}$, which includes $\tau=0$. In addition, the interval $\mathcal{U}$ is such that, for any $\tau\in\mathcal{U}$, the solution in (\ref{EquationHIVIndistinguishableInitStates}) must exist. 
On the other hand, even if the above requirements on $\mathcal{U}$ ensure the validity of our result (i.e., that the solution in (\ref{EquationHIVIndistinguishableUI}) and (\ref{EquationHIVIndistinguishableInitStates}) produces the same outputs at any time) we have a further requirement, due to the physical meaning of our states and parameters. For any $\tau\in\mathcal{U}$ the quantities $\delta'(\tau)$, $N'(\tau)$, $\eta'(t,~\tau)$, $T_U'(t,~\tau)$, and $T_I'(t,~\tau)$ must be positive.

In Section \ref{SectionHIVComparisonSOTA}, we test the validity of the above fundamental result by using the same data set adopted in \cite{Villa19b}.

\section{Comparison with the state of the art results}\label{SectionHIVComparisonSOTA}

As we aforementioned, our results contradict the result available in the state of the art for exactly the same ODE model characterized by 
(\ref{EquationHIVSystem}) (e.g., see Section 6.2 of \cite{Miao11}\footnote{The reader can find a detailed explanation of the serious error made by the authors of \cite{Miao11} in \cite{arXivErratum}.} or Section 3.3 of \cite{Villa19b})).
Specifically, in contrast with the state of the art, we obtained that both the state and the unknown input (or the time-varying parameter) $\eta(t)$, cannot be recovered from the outputs. In addition, we also obtained a one dimensional set of indistinguishable states, $x_0'(\tau)$, and a one parameter set of functions $\eta'(t,~\tau)$ such that, by only using the outputs, it is not possible to distinguish among the elements of these sets.
In other words, the outputs obtained by using the initial state $x_0'(\tau)$ and by using the unknown input function $\eta'(t,~\tau)$, are independent of the parameter $\tau$, and this holds for any $\tau\in\mathcal{U}$.

In this section, we explicitly provide these sets by referring to the same data available in the literature (e.g., see the electronic supplementary material of \cite{Villa19b}) and
we show that any choice in these sets produces the same outputs.

The data set is characterized as follows:

\begin{itemize}

\item The time interval is $\mathcal{I}=[0,~201]$ (in days).

\item $T_U(0)=600$, $T_I(0)=0$, $V(0)=10^5$, $\lambda=36$, $\rho=0.108$, $\delta=0.5$, $N=10^3$, and $c=3$.

\item The time-varying parameter is:
\begin{equation}\label{EquationHIVEtaDataSet}
\eta(t)=k\left(
1-0.9\cos\left(\frac{\pi}{1000}t\right)
\right),
\end{equation}
with $k=0.00009$.

\end{itemize}

\begin{figure}[htbp]
\begin{center}
\includegraphics[width=.9\columnwidth]{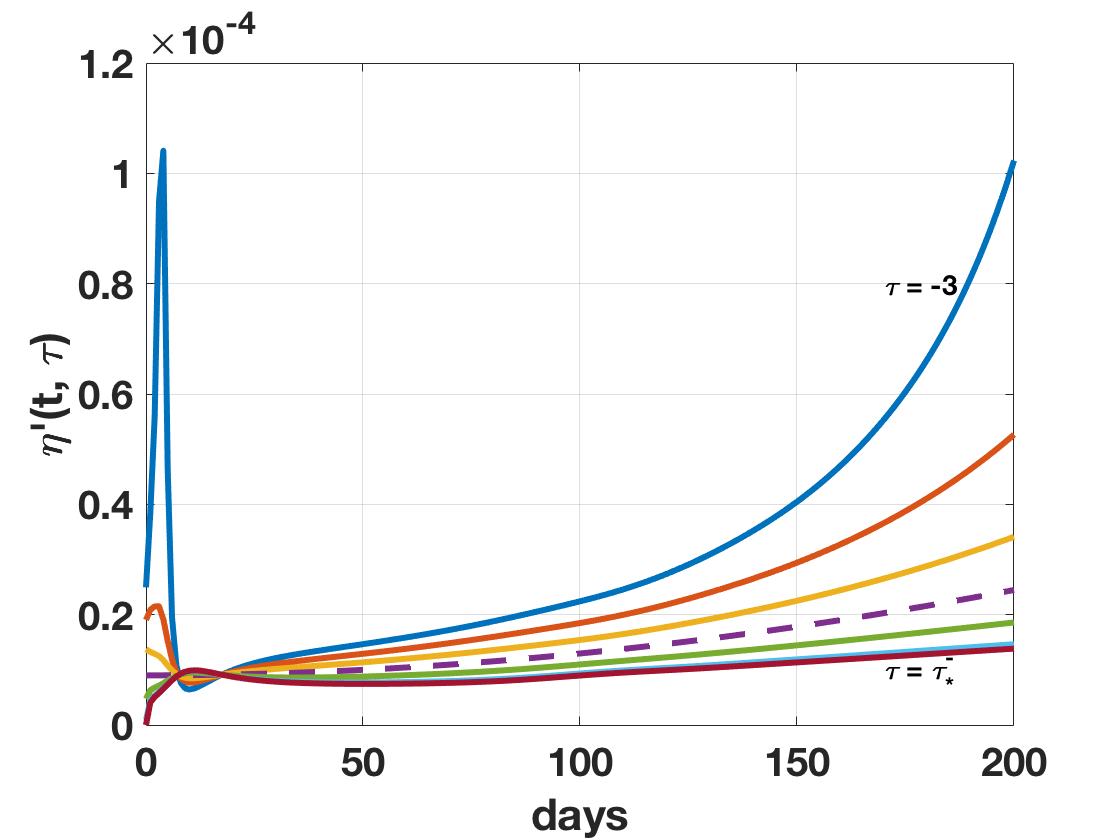}
\caption{Several indistinguishable profiles for the time-varying parameter $\eta'(t,~\tau)$ for several values of the parameter $\tau$ ranging from $-3$ up to $\tau_*\cong2.2532$. The dashed purple line is the profile given by (\ref{EquationHIVEtaDataSet}), i.e., $\eta'(t,~\tau)$ for $\tau=0$.} \label{FigEta}
\end{center}
\end{figure}

\begin{figure}[htbp]
\begin{center}
\includegraphics[width=.9\columnwidth]{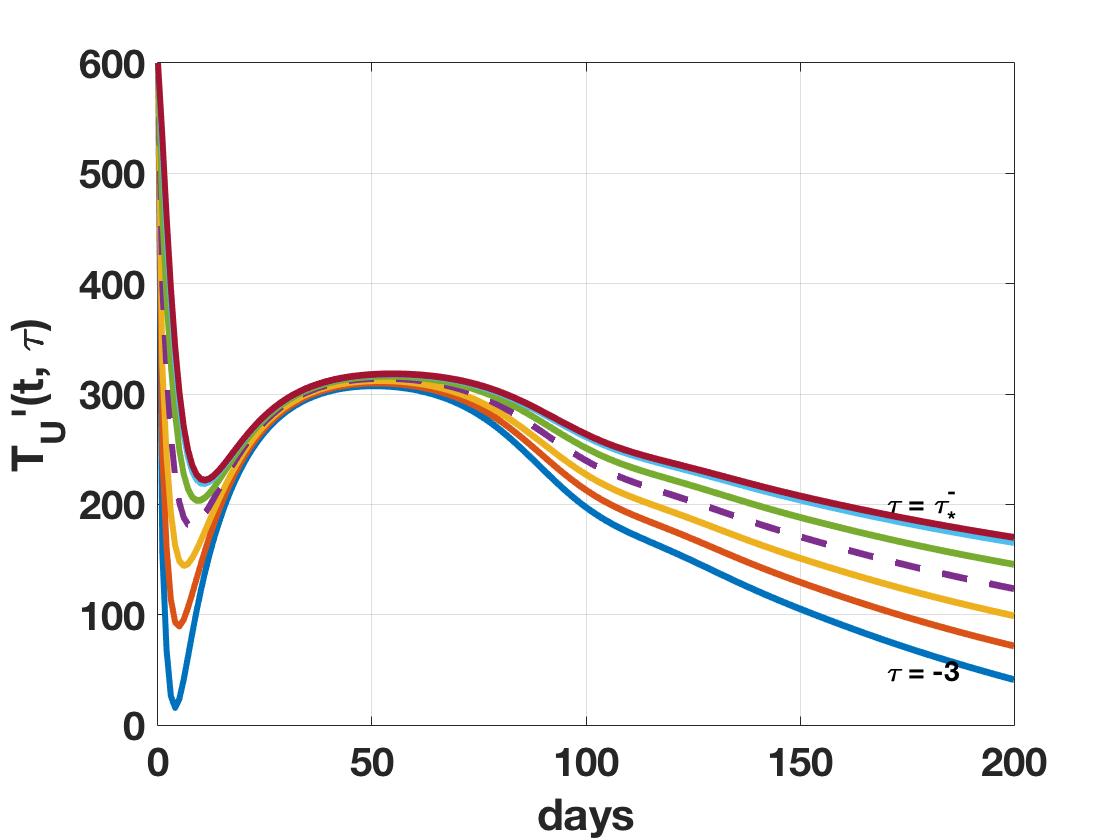}
\caption{The profiles $T_U'(t,~\tau)$ for several values of the parameter $\tau$ ranging from $-3$ up to $\tau_*\cong2.2532$. The dashed purple line is the profile for $\tau=0$.} \label{FigTu}
\end{center}
\end{figure}

\begin{figure}[htbp]
\begin{center}
\includegraphics[width=.9\columnwidth]{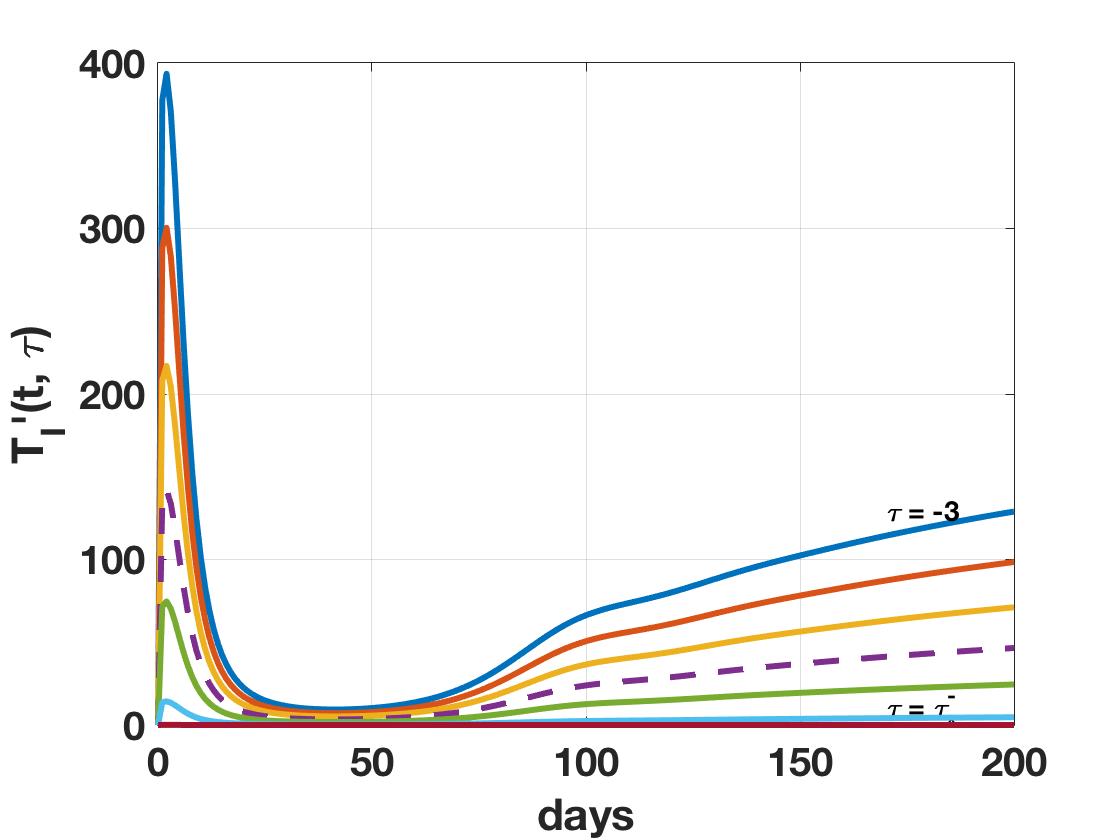}
\caption{The profiles $T_I'(t,~\tau)$ for several values of the parameter $\tau$ ranging from $-3$ up to $\tau_*\cong2.2532$. The dashed purple line is the profile for $\tau=0$.} \label{FigTi}
\end{center}
\end{figure}

%
%
%
%

We provide indistinguishable initial states and indistinguishable unknown inputs for several values of the parameter $\tau\in\mathcal{U}$.
First of all, let us characterize $\mathcal{U}$. 

From the expression of $\delta'(\tau)$ in (\ref{EquationHIVIndistinguishableInitStates}) and the values of our data set, we obtain that

\[
\lim_{\tau\rightarrow\tau_*^-}\delta'(\tau)=+\infty,~~
\lim_{\tau\rightarrow\tau_*^+}\delta'(\tau)=-\infty
\]

with $\tau_*:=\frac{\log\left(\frac{\delta}{\delta-\rho}\right)}{\rho}\cong 2.2532$.

In addition, we found that, for $\tau=-4$ the function $\eta'(t,~\tau)$ in (\ref{EquationHIVIndistinguishableUI}) takes negative values for $t\in[2,~6]s$.
Finally, we verified that in the interval
\[
\mathcal{U}_0:=\left[-3,~\frac{}{}\tau_*\right)
\]
all the quantities $\delta'(\tau)$, $N'(\tau)$, $\eta'(t,~\tau)$, $T_U'(t,~\tau)$, and $T_I'(t,~\tau)$ take positive values (the last three on the entire time interval $\mathcal{I}$).
The condition $\mathcal{U}_0\subset\mathcal{U}$ is ensured and we are allowed to use any $\tau\in\mathcal{U}_0$.
Hence, we consider the indistinguishable initial states and unknown inputs obtained by varying $\tau$ in this interval $\mathcal{U}_0$.


Figure \ref{FigEta} displays the profiles of $\eta'(t,~\tau)$. The profile for $\tau=0$ (purple dashed line) is precisely the one set in (\ref{EquationHIVEtaDataSet}). By varying $\tau$ we obtain a significant change of the profile meaning that this parameter is "strongly" unidentifiable.

Figures \ref{FigTu} and \ref{FigTi} display the profiles of $T_U'(t,~\tau)$ and $T_I'(t,~\tau)$, respectively. The initial conditions of $T_U$ and $T_I$ are obtained by these profiles at $t=0$. By chance, in this case, these initial conditions are independent of $\tau$. This because the part that depends on $\tau$ is linear in $T_{I_0}$ (see the first two equations in (\ref{EquationHIVIndistinguishableInitStates})) and we set $T_{I_0}=0$.
Note that, by changing the values of $T_{I_0}$, the initial conditions significantly depend on $\tau$ meaning that also the initial state is "strongly" unobservable.
In addition, by varying $\tau$, the two constant parameters that are not identifiable (i.e., $\delta$ and $N$) significantly change. For instance, in accordance with (\ref{EquationHIVIndistinguishableInitStates}),
for $\tau=-3$ and $\tau=\tau_*$ we obtain the following changes:

\[
\begin{array}{lll}
\delta=\delta'(0)=0.5 & \hskip-.2cm\rightarrow \delta'(-3)\cong0.25 &\hskip-.2cm\rightarrow \delta'(\tau\rightarrow\tau_*^-)=+\infty\\
N=N'(0)=10^3& \hskip-.2cm\rightarrow N'(-3)\cong 723 &\hskip-.2cm\rightarrow N'(\tau_*)\cong 1276.\\
\end{array}
\]

The variation of $\delta'$ is even divergent.

We verified that the two outputs, $y_1(t)=V$ and $y_2(t)=T_U+T_I$, are independent of $\tau$, at any $t$.
This is obtained by proceeding as follows.

\begin{figure}[htbp]
\begin{center}
\includegraphics[width=.9\columnwidth]{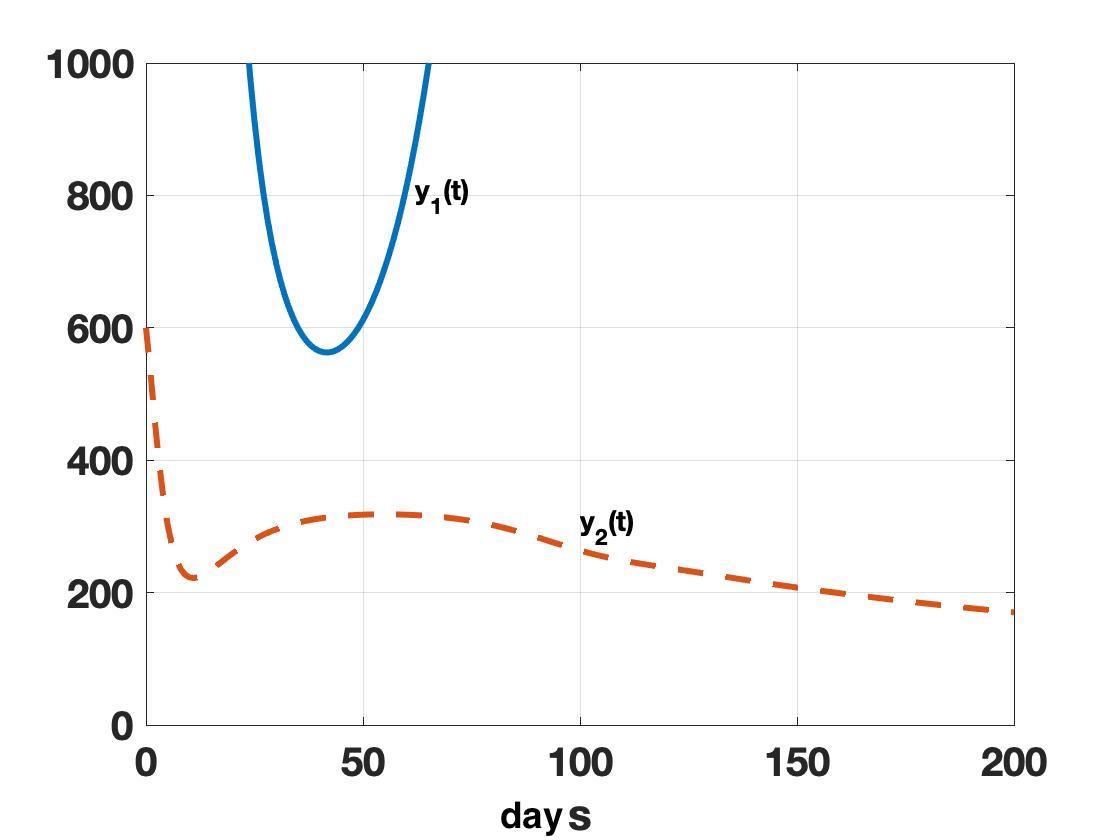}
\caption{The two outputs $y_1(t)=V$ (solid blue line) and $y_2(t)=T_U+T_I$ (dashed red line). The profiles are independent of the parameter $\tau$, as expected.} \label{FigY}
\end{center}
\end{figure}

We considered several values of $\tau$ in $\mathcal{U}_0$. For each $\tau$, we integrated the differential equation in (\ref{EquationHIVSystem}) with the initial conditions given in (\ref{EquationHIVIndistinguishableInitStates}) for that $\tau$ and by setting the  unknown time-varying parameter as in (\ref{EquationHIVIndistinguishableUI}) for the same value of $\tau$ (that is one of the profiles plotted in Figure \ref{FigEta}). Both the initial conditions in (\ref{EquationHIVIndistinguishableInitStates}) and the unknown input significantly depend on $\tau$. However, the two outputs (which are displayed in Figure \ref{FigY}) are independent, in accordance with our results.  

We also check the validity of the same results by considering other settings (e.g., by setting $T_{I_0}\neq0$).

%

\section{Minimal external information}\label{SectionHIVMinimalExtInfo}

Our analysis showed that only three parameters are identifiable. They are $\lambda,~\rho$, and $c$. The remaining constant parameters, $\delta$, and $N$, and the time varying parameter, $\eta(t)$, cannot be identified.
It is possible to prove that, by also including the output $y_3=T_U$ (or $y_3=T_I$), the state becomes observable and all the parameters identifiable. This can be proved by repeating the analysis for the new system characterized by this further output.
On the other hand, from our analysis, it is immediate to obtain a more interesting result. 

The system is characterized by a single symmetry. As a result, we can apply the general procedure described in Section \ref{SectionProcedureExternalInformation} and we can conclude that it suffices to have $T_U$ (or $T_I$) at a single time $t^*\in\mathcal{I}$ (and not necessarily on the entire time interval) to make possible the identification of all the parameters (and the observability of the state). 
Equation (\ref{EquationExtInformationSingleSymmetry}) becomes, in this case, (\ref{EquationHIVIndistinguishableStatesforUI}) together with (\ref{EquationHIVIndistinguishableUI}). In order to determine the true state and the true parameters, we must determine the value of $\tau$. This can be determined 
by knowing $T_U$ (or equivalently $T_I=y_2-T_U$) at a single time.
First of all, Equation (\ref{EquationHIVIndistinguishableStatesforUI}) tells us that we have the true $V,~\lambda,~\rho$, and $c$. Now let us suppose that we know $T_I$ at a single time. From the second equation in (\ref{EquationHIVIndistinguishableStatesforUI}) we obtain the quantity $\delta e^{-\rho\tau}-\delta=\frac{T_I'(t^*,~\tau)~\rho}{T_I(t^*)}-\rho$, which is an equation in the two unknowns $\delta$ and $\tau$. On the other hand, the sixth equation in (\ref{EquationHIVIndistinguishableStatesforUI}) is a further equation in the same two unknowns. From them, we obtain both $\delta$ and $\tau$. Finally, once $\tau$ has been determined, we determine the remaining quantities at any $t\in\mathcal{I}$.

\chapter{Compartmental SEIAR model of the epidemic dynamics of Covid-19}\label{ChapterCovid}

We consider a simple model that generalizes the SEIR model commonly used for virus disease. 
This more general ODE model includes a further compartment, $A(t)$, that is the asymptomatic infected, \cite{SEIARPribylova}. 
The other compartments are the same of a SEIR model, i.e.: S, Susceptible; E, Exposed; I, Infected (symptomatic infected); R, Removed (i.e., healed or dead).

The differential equations of this model 
describe the movement of individuals of the
population between the five aforementioned classes, i.e.: $S(t),~E(t),~I(t),~A(t)$ and $R(t)$. 

%
%

The dynamics and the outputs are given by the following set of equations (e.g., see \cite{SEIARPribylova}):

\begin{equation}\label{EquationCovidSystem}
\left\{\begin{array}{ll}
\dot{S} &= -\beta S(I+A) \\
\dot{E} &= \beta S(I+A) -\gamma E\\
\dot{I} &= \gamma pE-\mu_1I\\
\dot{A} &= \gamma(1-p)E-\mu_2A\\
\dot{R} &= \mu_1I+\mu_2A\\
y &= [I, ~A, ~S+E+R], \\
\end{array}\right.
\end{equation}

This model includes the following constant parameters, which are assumed to be unknown:

\begin{itemize}

\item $\mu_1$ and $\mu_2$ that are the remove rates from the infected symptomatic individuals ($I(t)$) and the infected asymptomatic individuals ($A(t)$), respectively.



\item $\gamma$ that is the rate
at which the exposed individuals ($E(t)$) are infected. 

\item $p$ that is the probability that infected individuals are symptomatic (and $1-p$  is the probability that infected individuals are asymptomatic).

\end{itemize}

The model also includes the following unknown time-varying parameter:

\begin{itemize}

\item $\beta=\beta(t)$ that is the probability of disease transmission in a single contact
times the average number of contacts per person, due to contacts between an individual of the class $S$ and an individual that belongs to one of the two classes $I$ and $A$. This parameter is assumed to be time-varying as it can be modified by government measures (e.g., , using masks, closing schools, remote working).

\end{itemize}

We assume that we can measure the infected symptomatic individuals ($I(t)$), and the infected asymptomatic individuals ($A(t)$).
Hence, our model is characterized by the first two outputs in (\ref{EquationCovidSystem}). 
Finally, as we know the total population, $S(t)+E(t)+I(t)+A(t)+R(t)$, by removing from this the above two measurements, we obtain the third output in (\ref{EquationCovidSystem}), i.e., $S(t)+E(t)+R(t)$.

\vskip.2cm

To proceed, we need, first of all, to introduce a state that includes both the time-varying quantities (i.e., $S,~E,~I,~A,~R$) and the constant parameters (i.e., $\mu_1, ~\mu_2, ~\gamma,~p$). We set:

\begin{equation}\label{EquationCovidState}
x=[S,~E,~I, ~A, ~R, ~\mu_1, ~\mu_2, ~\gamma,~p]^T
\end{equation}

The system is directly a special case of (\ref{EquationSystemDefinitionUIO}). In particular, 
$m_u=0$, $m_w=1$, $p=3$, $h_1(x)=I$, $h_2(x)=A$, $h_3(x)=S+E+R$,

\begin{equation}\label{EquationCovidg0g1}
g^0=
\left[
\begin{array}{c}
0\\
-\gamma E\\
\gamma pE-\mu_1I\\
\gamma(1-p)E-\mu_2A\\
\mu_1I+\mu_2A\\
0\\
0\\
0\\
0\\
\end{array}
\right],~~
g^1=
\left[
\begin{array}{c}
 -S(I+A)\\
 S(I+A)\\
0\\
0\\
0\\
0\\
0\\
0\\
0\\
\end{array}
\right]
\end{equation}

\section{Observability analysis}\label{SectionCovidObs}

We run Algorithm \ref{AlgoFull} to obtain the observability codistribution.
Line \ref{AlgoLineOmegaINIT} provides 
$\Omega=\textnormal{span}\{\nabla h_1,~\nabla h_2,~\nabla h_3\}$, and 
  $\uideg \left({\Omega}\right)=0~(<m_w)$, as $\Li_{g^1}h_1=\Li_{g^1}h_2=\Li_{g^1}h_3=0$ (and the rank of the unknown input reconstructability matrix from $h_1,~h_2,~h_3$ vanishes).
The system is not in canonical form with respect to its unknown input. 
We have $m=0$ and
$\mu^0_0=~\nu^0_0=1$ obtained from (\ref{EquationTensorMSynchrom}) (these tensors have a single entry as the indices only take the value 0).
 In addition, from (\ref{Equationgalpham}) we obtain: $\widehat{g}^0=~\nu^0_0g^0=g^0$, and $\giat^1=g^1$.
 
 Algorithm \ref{Algo***} provides Finish=False.

We have:
\[
\Omega=
\left<\left.\left\{ 
\begin{array}{l}
 \dtv{\Li}_{\giat} \\
 \giat^1\\
\end{array}
\right\}~\right|\textnormal{span}\left\{\nabla h_1,~\nabla h_2,~\nabla h_3\right\}\right>=
\left<\left.\left\{ 
\begin{array}{l}
 g^0 \\
 g^1\\
\end{array}
\right\}~\right|\textnormal{span}\left\{\nabla h_1,~\nabla h_2,~\nabla h_3\right\}\right>,
\]
as $\giat=\widehat{g}^0=g^0$.
By a direct computation, we obtain:

\[
\Omega=\textnormal{span}\left\{\nabla h_1,~\nabla h_2,~\nabla h_3,~\nabla\Li_{g^0}h_1,~\nabla\Li_{g^0}h_2\right\}=
\textnormal{span}\left\{\nabla h_1,~\nabla h_2,~\nabla h_3,~\nabla h_4,~\nabla h_5\right\}
\]
where we adopted the following notation:
\[
h_4:= \Li_{g^0}h_1,~~
h_5:= \Li_{g^0}h_2
\]

Now, $\uideg\left(\Omega\right)=1=m_w$. Hence, 
the execution of Algorithm \ref{AlgoFull} continues with Line 
\ref{AlgoLineFINITSelection}. The operation at this line ($[\widetilde{h}_1,\ldots, \widetilde{h}_{m_w}]=\mathcal{S}(\Sigma,~\Omega)$) sets

\[
\widetilde{h}_1=h_4=\Li_{g^0}h_1=\gamma pE-\mu_1I
\]

(we would obtain the same final result by setting  $\widetilde{h}_1=h_5=\Li_{g^0}h_2$).
Then, Line \ref{AlgoLineFINITMuNuTOBS} provides:

\[
\mu=\left[
\begin{array}{cc}
1&0\\
\mu_1 ( I \mu_1 - E \gamma p) - E \gamma^2 p& S \gamma p (A +  I)\\
\end{array}
\right],
\]
\[
\nu=\left[
\begin{array}{cc}
1&0\\
\frac{E p \gamma^2 + E p \gamma \mu_1 -  I \mu_1^2}{S \gamma p (A +  I)} & \frac{1}{S \gamma p (A +  I)}\\
\end{array}
\right],
\]

obtained from (\ref{EquationTensorMSynchro}). In addition, from (\ref{Equationgalpha}) we obtain:

\[
\widehat{g}^0=
\left[
\begin{array}{c}
-\frac{E p \gamma^2 + E p \gamma \mu_1 -  I \mu_1^2}{\gamma p}\\
-\frac{\mu_1 ( I \mu_1 - E \gamma p)}{\gamma p}\\
E \gamma p -  I \mu_1\\
- A \mu_2 - E \gamma (p - 1)\\
A \mu_2 +  I \mu_1\\
0\\
0\\
0\\
0\\
\end{array}
\right],~~
\widehat{g}^1=
\left[
\begin{array}{c}
-\frac{1}{\gamma p}\\
\frac{1}{\gamma p}\\
0\\
0\\
0\\
0\\
0\\
0\\
0\\
\end{array}
\right]
\]

Finally, from (\ref{EquationTOBSDef}) we obtain that, when $m_u=0$, the codistribution $\tobs$ vanishes. 
The final step of Algorithm \ref{AlgoFull} is the execution of Line \ref{AlgoLineFINALSTEP}, namely:

\[
\OBS=
\left<
\left.
\widehat{g}^0,~\widehat{g}^1~
\right|
~\textnormal{span}\left\{\nabla h_1,~\nabla h_2,~\nabla h_3,~\nabla h_4,~\nabla h_5\right\}\right>.
\]

The algorithm 
in (\ref{EquationAlgorithmsMinimalCod}) converges at the second step ($\Omega_2=\Omega_1$), and

\[
\OBS=\textnormal{span}\left\{
\nabla h_1,~\nabla h_2,~\nabla h_3,~\nabla h_4,~\nabla h_5,~\nabla\Li_{\widehat{g}^0}h_5,~\nabla\Li_{\widehat{g}^1}h_5
\right\}.
\]

Its dimension is $7$, which is smaller than the dimension of the state in (\ref{EquationCovidState}).
The state is unobservable.

\section{Symmetries and indistinguishable states}\label{SectionCovidSym}

Once we have $\OBS$, the symmetries are immediately obtained by computing the orthogonal distribution. We have:

\begin{equation}\label{EquationCovidSymmetry}
\OBS^\bot=\textnormal{span}\left\{
\left[\begin{array}{c}
1\\
0\\
0\\
0\\
-1\\
0\\
0\\
0\\
0\\
\end{array}\right],~
\left[\begin{array}{c}
E\\
-E\\
0\\
0\\
0\\
0\\
0\\
\gamma\\
0\\
\end{array}\right]
\right\}.
\end{equation}

As for the case study, starting from the generators of $\OBS^\bot$ we compute the indistinguishable states (although the determination of the indistinguishable states is not requested by Algorithm \ref{AlgoFullIDE}).
We need to solve the differential equation in (\ref{EquationDiffEqStates}). 
Given a state $[\overline{S}, ~\overline{E}, ~\overline{I}, ~\overline{A}, ~\overline{R}, ~\overline{\mu}_1, ~\overline{\mu}_2, ~\overline{\gamma}, ~\overline{p}]^T$, the solution of this differential equation provides states which are indistinguishable. 
For the specific case, 
we have two distinct independent symmetries that are the two generators of $\OBS^\bot$
in (\ref{EquationCovidSymmetry}). Let us discuss the two cases separately.

 \subsection{First symmetry}

Equation (\ref{EquationDiffEqStates}) becomes:

\begin{equation}\label{EquationCovidDiffEqStates1}
\left\{\begin{array}{ll}
  \frac{dS}{d\tau} &= 1 \\
  \frac{dE}{d\tau} &= 0 \\
  \frac{dI}{d\tau} &=  0\\
  \frac{dA}{d\tau} &=  0\\
  \frac{dR}{d\tau} &=  -1\\
  \frac{d\mu_1}{d\tau} &= 0 \\  
  \frac{d\mu_2}{d\tau} &=  0\\  
  \frac{d\gamma}{d\tau} &= 0 \\
  \frac{dp}{d\tau} &=  0\\
 S(0)&=\overline{S}, E(0)= \overline{E}, I(0)= \overline{I}, A(0)= \overline{A}, R(0)= \overline{R},\\
 \mu_1(0)&= \overline{\mu}_1, ~\mu_2(0)= \overline{\mu}_2, ~\gamma(0)= \overline{\gamma}, ~p(0)= \overline{p} \\
\end{array}\right.
\end{equation}

This equation can be trivially solved analytically. The solution is:

\begin{equation}\label{EquationCovidIndistinguishableStates1}
x(\tau)=
\left[\begin{array}{l}
\overline{S}+\tau\\
 \overline{E}\\
\overline{I}\\
\overline{A}\\
\overline{R}-\tau\\
\overline{\mu}_1\\
\overline{\mu}_2\\
\overline{\gamma}\\
\overline{p} \\
\end{array}\right]
\end{equation}

All the above states ($x(\tau)$) are indistinguishable for any $\tau$ (in particular, they are indistinguishable from $[\overline{S}, ~\overline{E}, ~\overline{I}, ~\overline{A}, ~\overline{R}, ~\overline{\mu}_1, ~\overline{\mu}_2, ~\overline{\gamma}, ~\overline{p}]^T$). 

\subsection{Second symmetry}

Equation (\ref{EquationDiffEqStates}) becomes:

\begin{equation}\label{EquationCovidDiffEqStates2}
\left\{\begin{array}{ll}
  \frac{dS}{d\tau} &= E \\
  \frac{dE}{d\tau} &= -E \\
  \frac{dI}{d\tau} &=  0\\
  \frac{dA}{d\tau} &=  0\\
  \frac{dR}{d\tau} &=  0\\
  \frac{d\mu_1}{d\tau} &= 0 \\  
  \frac{d\mu_2}{d\tau} &=  0\\  
  \frac{d\gamma}{d\tau} &= \gamma \\
  \frac{dp}{d\tau} &=  0\\
 S(0)&=\overline{S}, E(0)= \overline{E}, I(0)= \overline{I}, A(0)= \overline{A}, R(0)= \overline{R},\\
 \mu_1(0)&= \overline{\mu}_1, ~\mu_2(0)= \overline{\mu}_2, ~\gamma(0)= \overline{\gamma}, ~p(0)= \overline{p} \\
\end{array}\right.
\end{equation}

This equation can be solved analytically. The solution is:

\begin{equation}\label{EquationCovidIndistinguishableStates2}
x(\tau)=
\left[\begin{array}{l}
-\overline{E}~e^{-\tau}+\overline{S}+\overline{E}\\
\overline{E}~e^{-\tau}\\
\overline{I}\\
\overline{A}\\
\overline{R}\\
\overline{\mu}_1\\
\overline{\mu}_2\\
\overline{\gamma}~e^\tau\\
\overline{p} \\
\end{array}\right]
\end{equation}

All the above states ($x(\tau)$) are indistinguishable for any $\tau$ (in particular, they are indistinguishable from $[\overline{S}, ~\overline{E}, ~\overline{I}, ~\overline{A}, ~\overline{R}, ~\overline{\mu}_1, ~\overline{\mu}_2, ~\overline{\gamma}, ~\overline{p}]^T$). 

\section{Characterization by an observable state}\label{SectionCovidLocDec}

Note that this task is not requested by Algorithm \ref{AlgoFullIDE}, to obtain the identifiability of the parameters. In addition, this task 
cannot be performed automatically.
On the other hand, we were able to obtain this description that can be useful for further investigations.

The original state has dimension 9 and the observability codistribution 7. This is the reason why we have two independent symmetries ($9=7+2$). A possible observable state consists precisely of a set of generators of the observability codistribution. Algorithm \ref{AlgoFull} automatically returns a set of generators. We obtained: $h_1,~h_2,~h_3,~h_4,~h_5,~\Li_{\widehat{g}^0}h_5,~\Li_{\widehat{g}^1}h_5$. On the other hand, by setting the state entries equal to these functions, obtaining the expression of its dynamics in terms of the same state components is definitely prohibitive (although possible). 
To simplify this task, first of all, we try to determine 7 independent and simple functions, which are observable and that generate the observability codistribution. By using the expression of the generators of $\OBS$ and the expression of the generators of $\OBS^\bot$, by using our inventiveness we were able to select the following observable functions:

\begin{equation}\label{EquationCovidObservables}
\begin{array}{ll}
-~~ \Psi_1 &:= S+E+R\\
-~~ \Psi_2 &:= \gamma E\\
-~~ I&\\
-~~ A&\\
-~~ \mu_1&\\
-~~ \mu_2&\\
-~~ p&\\
\end{array}
\end{equation}

(by a direct calculation it is possible to verify that their gradients generate $\OBS$). 
As their expression is much easier than the expression of the generators automatically selected by Algorithm \ref{AlgoFull}, we have the possibility to describe our system by using them. After some trials, by using (\ref{EquationCovidSystem}),
we were able to obtain the following dynamics:

\begin{equation}\label{EquationCovidLocalDec}
\left\{\begin{array}{ll}
\dot{\Psi}_1 &=-\Psi_2 + \mu_1I+\mu_2A\\
\dot{\Psi}_2 &= \widetilde{\beta} \\
\dot{I} &= p\Psi_2-\mu_1I\\
\dot{A} &= (1-p)\Psi_2-\mu_2A\\
y &= [I, ~A, ~\Psi_1], \\
\end{array}\right.
\end{equation}

Note that, in order to achieve this description, we had to redefine the unknown input. By using the original unknown input $\beta$, we would obtain $\dot{\Psi}_2=\gamma\dot{E} = \beta S\gamma(I+A) -\gamma \Psi_2$. This equation cannot be expressed only in terms of the components of the new observable state. It needs the product $S\gamma$ and $\gamma$, which cannot be expressed in terms of the selected observable functions because they are unobservable.

\section{Identifiability analysis}\label{SectionCovidIdentifiability}

We execute Algorithm \ref{AlgoFullIDE}.
First, as the system is canonic with respect to its unknown inputs, we do not have symmetries of the unknown input due to the system non canonicity ($^c\chi$). On the other hand, 
we have two independent symmetries of the state, which are the generators of $\OBS^\bot$ in (\ref{EquationCovidSymmetry}), and,
as a result, we may have non vanishing symmetries $^u\chi$ of the unknown input. Let us compute them.
We use Equation (\ref{EquationSymmetryUIObs}) twice, by setting $\xi$ equal to the two generators of $\OBS^\bot$ in (\ref{EquationCovidSymmetry}), respectively.
In both cases, we have:

\[
\widetilde{h}:=\widetilde{h}_1=\gamma pE-\mu_1I.
\]

and from (\ref{EquationCovidg0g1}), we obtain:

\begin{equation}\label{EquationCovidLig0g1h}
\Li_{g^0}\widetilde{h}=\mu_1 ( I \mu_1 - E \gamma p) - E \gamma^2 p,~~
\Li_{g^1}\widetilde{h}=S \gamma p (A +  I).
\end{equation}

\subsection{First symmetry}
We set

\[
\xi=\xi^1=
\left[\begin{array}{c}
1\\
0\\
0\\
0\\
-1\\
0\\
0\\
0\\
0\\
\end{array}\right].
\]

In accordance with Equation (\ref{EquationSymmetryUIObs}), we need to compute 
$\xi^0_1$ and $\xi^1_1$.
From Equation (\ref{EquationSymmetryUIObsCoeff}), by using the above $\xi$ and (\ref{EquationCovidLig0g1h}), we obtain:

\[
\xi^0_1=0,~~~
\xi^1_1=\gamma p (A +  I).
\]

and, by substituting in  (\ref{EquationSymmetryUIObs}), we obtain:

\[
^u\chi = -~\nu^1_1 ( \xi^0_1 + \xi^1_1 w_1 )=
-\frac{\xi^1_1}{S \gamma p (A +  I)}  w_1=
-\frac{1}{S} \beta
\]
because $\nu^1_1=\nu^1_1= \frac{1}{S \gamma p (A +  I)}$.
Hence, we have the following first symmetry of the unknown input:

\begin{equation}\label{EquationCovidSymmetryUI1}
^u\chi^1 =
-\frac{1}{S} \beta
\end{equation}

\subsection{Second symmetry}
We set

\[
\xi=\xi^2=
\left[\begin{array}{c}
E\\
-E\\
0\\
0\\
0\\
0\\
0\\
\gamma\\
0\\
\end{array}\right].
\]

In accordance with Equation (\ref{EquationSymmetryUIObs}), we need to compute 
$\xi^0_1$ and $\xi^1_1$.
From Equation (\ref{EquationSymmetryUIObsCoeff}), by using the above $\xi$ and (\ref{EquationCovidLig0g1h}), we obtain:

\[
\xi^0_1=-\gamma^2~p~E,~~~
\xi^1_1=\gamma p (A +  I)(E+S).
\]

and, by substituting in  (\ref{EquationSymmetryUIObs}), we obtain:

\[
^u\chi = -\nu^1_1 ( \xi^0_1 + \xi^1_1 w_1 )=
-\frac{\xi^0_1}{S \gamma p (A +  I)}-\frac{\xi^1_1}{S \gamma p (A +  I)}  w_1=
\]
\[
\frac{\gamma~E}{S (A +  I)}-\frac{E+S}{S}  \beta
\]
because $\nu^1_1=\nu^1_1= \frac{1}{S \gamma p (A +  I)}$.
Hence, we have the following second symmetry of the unknown input:

\begin{equation}\label{EquationCovidSymmetryUI2}
^u\chi^2 =
\frac{\gamma~E}{S (A +  I)}-\frac{E+S}{S}  \beta
\end{equation}

\vskip.2cm
We conclude by remarking that the unknown input is characterized by two symmetries. Hence, $\beta(t)$ is unidentifiable. 

\section{Indistinguishable states and unknown inputs}\label{SectionCovidIndStatesAndUI}

We have two generators of $\OBS^\bot$ and we discuss the two cases separately.

\subsection{First set of solutions}\label{SubSectionCovidSetSolution1}

By using the first generator of $\OBS^\bot$ in (\ref{EquationCovidSymmetry})) and the expression of $^u\chi$ in (\ref{EquationCovidSymmetryUI1}), 
it is possible to check that 
the condition in (\ref{EquationCONDITIONCommutativitaTempoSimmetria}) is satisfied and
the differential equation in (\ref{EquationDiffEqStatesUI}) becomes:

\begin{equation}\label{EquationCovidDiffEqStatesUI1}
\left\{\begin{array}{ll}
  \frac{dS'}{d\tau} &= 1 \\
  \frac{dE'}{d\tau} &= 0 \\
  \frac{dI'}{d\tau} &=  0\\
  \frac{dA'}{d\tau} &=  0\\
  \frac{dR'}{d\tau} &=  -1\\
  \frac{d\mu_1'}{d\tau} &= 0 \\  
  \frac{d\mu_2'}{d\tau} &=  0\\  
  \frac{d\gamma'}{d\tau} &= 0 \\
  \frac{dp'}{d\tau} &=  0\\
   \frac{d\beta'}{d\tau} &=  -\frac{\beta'}{S'}\\
S'(t,~0)&=S(t), ~E'(t,~0)= E(t), ~I'(t,~0)= I(t), \\
A'(t,~0)&= A(t), ~R'(t,~0)= R(t), ~\mu_1'(0)= \mu_1,\\
 \mu_2'(0)&= \mu_2, ~\gamma'(0)= \gamma, ~p'(0)= p \\
 \beta'(t,~0)&= \beta(t) \\
\end{array}\right.
\end{equation}

where we only provide the explicit dependence on $t$ and $\tau$ at the initial conditions and we omit it for the differential equations, for the simplicity sake.
These equations can be solved analytically. The solution for the first nine components (which are the components of the state) is the same solution of
(\ref{EquationCovidDiffEqStates1}) given 
in (\ref{EquationCovidIndistinguishableStates1}), 
with the appropriate values for the initial conditions. In other words, $x'(t,~\tau)$ is:

\begin{equation}\label{EquationCovidIndistinguishableStatesforUI1}
\left\{\begin{array}{ll}
S'(t,~\tau)&=S(t)+\tau\\
E'(t,~\tau)&=E(t)\\
I'(t,~\tau)&=I(t)\\
A'(t,~\tau)&=A(t)\\
R'(t,~\tau)&=R(t)-\tau\\
~\mu_1'(\tau)&= \mu_1\\
\mu_2'(\tau)&= \mu_2\\
\gamma'(\tau)&= \gamma\\
p'(\tau)&= p \\
\end{array}\right.
\end{equation}

Regarding the unknown input, we obtain:

\begin{equation}\label{EquationCovidIndistinguishableUI1}
\beta'(t,~\tau)=\frac{\beta(t)S(t)}{S(t)+\tau}.
\end{equation}

We have obtained the following fundamental result.
Let us consider the ODE model in (\ref{EquationCovidSystem}) and let us consider a given time interval $[t_0,~T]$.
Let us suppose that the initial conditions and the values of the model parameters are:

\begin{itemize}
\item $S(t_0)=S_{0}$, $E(t_0)=E_{0}$, $I(t_0)=I_{0}$, $A(t_0)=A_{0}$, and $R(t_0)=R_0$.

\item The values of the constant parameters are $\mu_1,~\mu_2,~\gamma$, and $p$.

\item The time-varying parameter on the considered time interval $[t_0,~T]$ is $\beta(t)$.
\end{itemize}

Then, the following initial conditions and the values of the model parameters:

\begin{itemize}

\item 
\begin{equation}\label{EquationCovidIndistinguishableInitStates1}
\left[\begin{array}{ll}
S'(t_0,~\tau)&=S_0+\tau\\
E'(t_0,~\tau)&=E_0\\
I'(t_0,~\tau)&=I_0\\
A'(t_0,~\tau)&=A_0\\
R'(t_0,~\tau)&=R_0-\tau\\
~\mu_1'(\tau)&= \mu_1\\
\mu_2'(\tau)&= \mu_2\\
\gamma'(\tau)&= \gamma\\
p'(\tau)&= p \\
\end{array}\right.
\end{equation}

\item the time-varying parameter given in (\ref{EquationCovidIndistinguishableUI1}), with $S(t), ~E(t), ~I(t), ~A(t)$, and $R(t)$ the values of $S$, $E$, $I$, $A$, and $R$ obtained by integrating (\ref{EquationCovidSystem}) with initial conditions $S_{0}$, $E_{0}$, $I_{0}$, $A_{0}$, and unknown input $\beta(t)$,

\end{itemize}

produce exactly the same outputs $y_1(t), ~y_2(t), ~y_3(t)$, on the entire time interval $[t_0,~T]$.
This holds for any value of the parameter $\tau$ that belongs to an interval $\mathcal{U}\subseteq\mathbb{R}$, which includes $\tau=0$, and  all the values of $\tau$ for which $\beta'(t,~\tau)$, $S'(t,~\tau)$, and $R'(t,~\tau)$ are positive.

\subsection{Second set of solutions}\label{SubSectionCovidSetSolution2}

By using the second generator of $\OBS^\bot$ in (\ref{EquationCovidSymmetry})) and the expression of $^u\chi$ in (\ref{EquationCovidSymmetryUI2}), 
it is possible to check that 
the condition in (\ref{EquationCONDITIONCommutativitaTempoSimmetria}) is satisfied and the differential equation in (\ref{EquationDiffEqStatesUI}) becomes:

\begin{equation}\label{EquationCovidDiffEqStatesUI2}
\left\{\begin{array}{ll}
  \frac{dS'}{d\tau} &= E' \\
  \frac{dE'}{d\tau} &= -E' \\
  \frac{dI'}{d\tau} &=  0\\
  \frac{dA'}{d\tau} &=  0\\
  \frac{dR'}{d\tau} &=  0\\
  \frac{d\mu_1'}{d\tau} &= 0 \\  
  \frac{d\mu_2'}{d\tau} &=  0\\  
  \frac{d\gamma'}{d\tau} &= \gamma' \\
  \frac{dp'}{d\tau} &=  0\\
   \frac{d\beta'}{d\tau} &=  \frac{\gamma'~E'}{S' (A' +  I')}-\frac{(E'+S')}{S'}  \beta'\\
S'(t,~0)&=S(t), ~E'(t,~0)= E(t), ~I'(t,~0)= I(t), \\
A'(t,~0)&= A(t), ~R'(t,~0)= R(t), ~\mu_1'(0)= \mu_1,\\
 \mu_2'(0)&= \mu_2, ~\gamma'(0)= \gamma, ~p'(0)= p \\
 \beta'(t,~0)&= \beta(t) \\
\end{array}\right.
\end{equation}

where we only provide the explicit dependence on $t$ and $\tau$ at the initial conditions and we omit it for the differential equations, for the simplicity sake.
These equations can be solved analytically. The solution for the first nine components (which are the components of the state) is the same solution of
(\ref{EquationCovidDiffEqStates2}) given 
in (\ref{EquationCovidIndistinguishableStates2}), 
with the appropriate values for the initial conditions. In other words, $x'(t,~\tau)$ is:

\begin{equation}\label{EquationCovidIndistinguishableStatesforUI2}
\left\{\begin{array}{ll}
S'(t,~\tau)&=S(t)+E(t)~(1-e^{-\tau})\\
E'(t,~\tau)&=E(t)~e^{-\tau}\\
I'(t,~\tau)&=I(t)\\
A'(t,~\tau)&=A(t)\\
R'(t,~\tau)&=R(t)\\
~\mu_1'(\tau)&= \mu_1\\
\mu_2'(\tau)&= \mu_2\\
\gamma'(\tau)&= \gamma~e^{\tau}\\
p'(\tau)&= p \\
\end{array}\right.
\end{equation}

Regarding the unknown input, we obtain:

\begin{equation}\label{EquationCovidIndistinguishableUI2}
\beta'(t,~\tau)=\frac{\mu(t)(1-e^\tau)-S(t)\beta(t)}{E(t)-(E(t)+S(t))e^\tau},~~\mu:=\frac{\gamma E(t)}{A(t)+I(t)}.
\end{equation}

We have obtained the following fundamental result.
Let us consider the ODE model in (\ref{EquationCovidSystem}) and let us consider a given time interval $[t_0,~T]$.
Let us suppose that the initial conditions and the values of the model parameters are:

\begin{itemize}
\item $S(t_0)=S_{0}$, $E(t_0)=E_{0}$, $I(t_0)=I_{0}$, $A(t_0)=A_{0}$, and $R(t_0)=R_0$.

\item The values of the constant parameters are $\mu_1,~\mu_2,~\gamma$, and $p$.

\item The time-varying parameter on the considered time interval $[t_0,~T]$ is $\beta(t)$.
\end{itemize}

Then, the following initial conditions and the values of the model parameters:

\begin{itemize}

\item 
\begin{equation}\label{EquationCovidIndistinguishableInitStates2}
\left[\begin{array}{ll}
S'(t_0,~\tau)&=S_0+E_0~(1-e^{-\tau})\\
E'(t_0,~\tau)&=E_0~e^{-\tau}\\
I'(t_0,~\tau)&=I_0\\
A'(t_0,~\tau)&=A_0\\
R'(t_0,~\tau)&=R_0\\
~\mu_1'(\tau)&= \mu_1\\
\mu_2'(\tau)&= \mu_2\\
\gamma'(\tau)&= \gamma~e^{\tau}\\
p'(\tau)&= p \\
\end{array}\right.
\end{equation}

\item the time-varying parameter given in (\ref{EquationCovidIndistinguishableUI2}), with $S(t), ~E(t), ~I(t), ~A(t)$, and $R(t)$ the values of $S$, $E$, $I$, $A$, and $R$ obtained by integrating (\ref{EquationCovidSystem}) with initial conditions $S_{0}$, $E_{0}$, $I_{0}$, $A_{0}$, and unknown input $\beta(t)$,

\end{itemize}

produce exactly the same outputs $y_1(t), ~y_2(t), ~y_3(t)$, on the entire time interval $[t_0,~T]$.
This holds for any value of the parameter $\tau$ that belongs to an interval $\mathcal{U}\subseteq\mathbb{R}$, which includes $\tau=0$, and  all the values of $\tau$ for which $\gamma'(\tau),$ $\beta'(t,~\tau)$, $S'(t,~\tau)$, and $E'(t,~\tau)$ are positive.

%
%
%
%
%
%
%
%
%
%
%
%

\section{Numerical results}\label{SectionCovidNumerical}


In this section, we provide some numerical results. We consider a given data set for our ODE model in (\ref{EquationCovidSystem}) and we compute the outputs. Then, by using our results in Section 
\ref{SectionCovidIndStatesAndUI}, we produce other initial states and model parameters that produce exactly the same outputs. 
This unequivocally prove the state unobservability and the parameters unidentifiability.

The considered data set is characterized as follows. 
For the ground truth of the model parameters, we use the values available in \cite{SEIARPribylova}. We set:

\begin{itemize}

\item The time interval is $\mathcal{I}=[0,~200]$ (in days).

\item $S(0)=1-10^{-10}$, $E(0)=0$, $I(0)=10^{-10}$, , $A(0)=0$, $R(0)=0$,
$\mu_1=\frac{1}{3}$, $\mu_2=\frac{1}{10}$, $\gamma=\frac{1}{4}$, and $p=0.14$.

\item We consider separately two distinct profiles of the time-varying parameter, which are:

\begin{enumerate}

\item $\beta(t)=1$, i.e., independent of time.

\item 
\begin{equation}\label{EquationCovidEtaDataSet}
\beta(t)=\cos\left(\frac{\pi}{400}t\right).
\end{equation}

\end{enumerate}

\end{itemize}

We provide indistinguishable initial states and indistinguishable unknown inputs. We have two distinct sets of solutions, which are the ones provided in Sections \ref{SubSectionCovidSetSolution1}, and \ref{SubSectionCovidSetSolution2}, respectively. However, we only provide the results obtained by using the second set. This because, by using the first set, independently of the choice of $\tau\in\mathbb{R}$, the solution in (\ref{EquationCovidIndistinguishableStatesforUI1}) provides a negative first component ($S(t)+\tau$) and/or a negative fifth component ($R(t)-\tau$).
Indeed, with our data set, by integrating (\ref{EquationCovidSystem}) in the time interval $\mathcal{I}$, the solutions $S(t)$ and $R(t)$ take all the values in the interval $[0, ~1]$. As a result, there exists $t\in\mathcal{I}$ such that, for any $\tau\in\mathbb{R}$, at least one of the two components $S(t)+\tau$ and $R(t)-\tau$ is negative.
In few words, for the first set of solutions given in Section \ref{SubSectionCovidSetSolution1}, the interval $\mathcal{U}$ that characterizes the parameter $\tau$ is empty. Note that this holds for the specific data set.

Let us consider the second set of solutions given in Section  
\ref{SubSectionCovidSetSolution2}. We will discuss the two cases of constant $\beta$ ($\beta=1$) and $\beta(t)$ given in (\ref{EquationCovidEtaDataSet}), separately.

\subsection{$\beta=1$}
First of all, let us characterize $\mathcal{U}$. By performing several runs, we obtain that all the states and parameters in (\ref{EquationCovidIndistinguishableStatesforUI2}) and $\beta'(t,~\tau)$ in (\ref{EquationCovidIndistinguishableUI2}) take positive values, for any $t\in\mathcal{I}$, when $\tau\in[-0.05, ~\infty)$. Hence, we set:

\[
\mathcal{U}_0:=[-0.05, ~10].
\]

The condition $\mathcal{U}_0\subset\mathcal{U}$ is ensured and we are allowed to use any $\tau\in\mathcal{U}_0$.
We also find that the plots obtained with $\tau>10$ cannot be distinguished one each other, showing that we achieved a convergence.
We consider the indistinguishable initial states and unknown inputs obtained by varying $\tau$ in $\mathcal{U}_0$.

\begin{figure}[htbp]
\begin{center}
\includegraphics[width=.9\columnwidth]{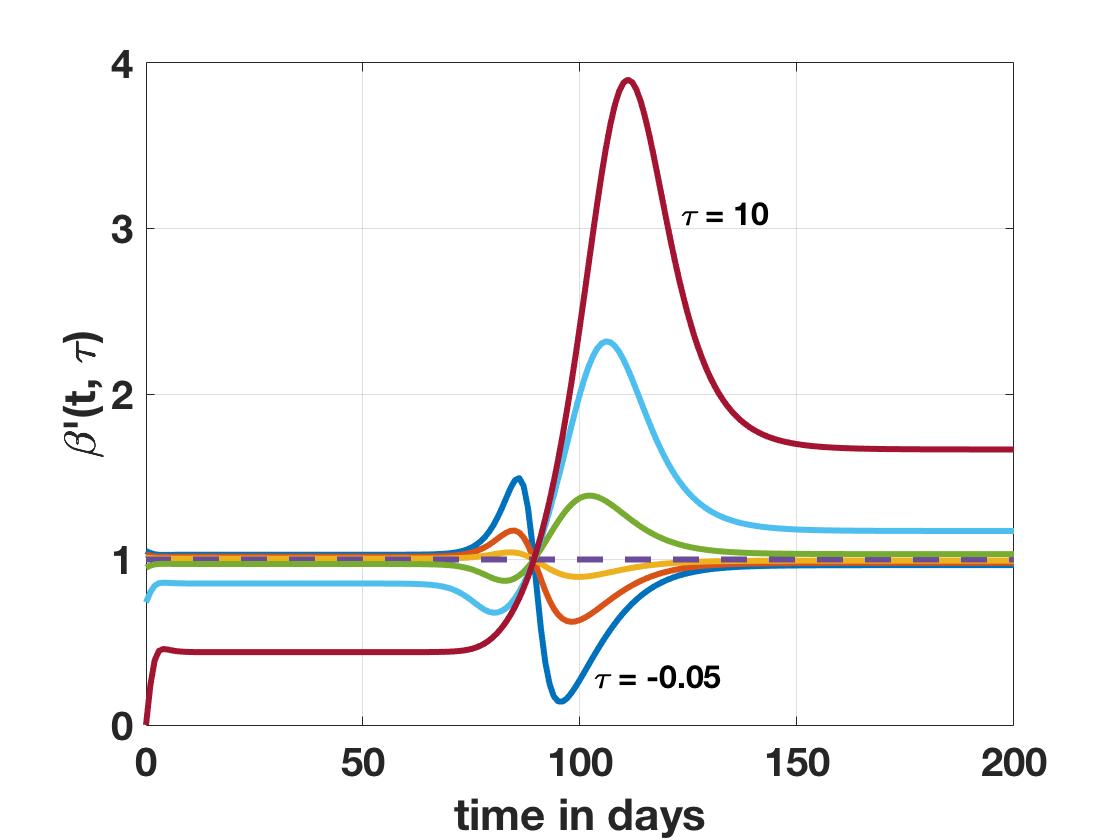}
\caption{Several indistinguishable profiles for the time-varying parameter $\beta'(t,~\tau)$ for several values of the parameter $\tau$ ranging from $-0.05$ up to $10$. The dashed purple line is the profile given by $\beta=1$, i.e., $\beta'(t,~\tau)$ for $\tau=0$.} \label{FigBeta1}
\end{center}
\end{figure}

\begin{figure}[htbp]
\begin{center}
\includegraphics[width=.9\columnwidth]{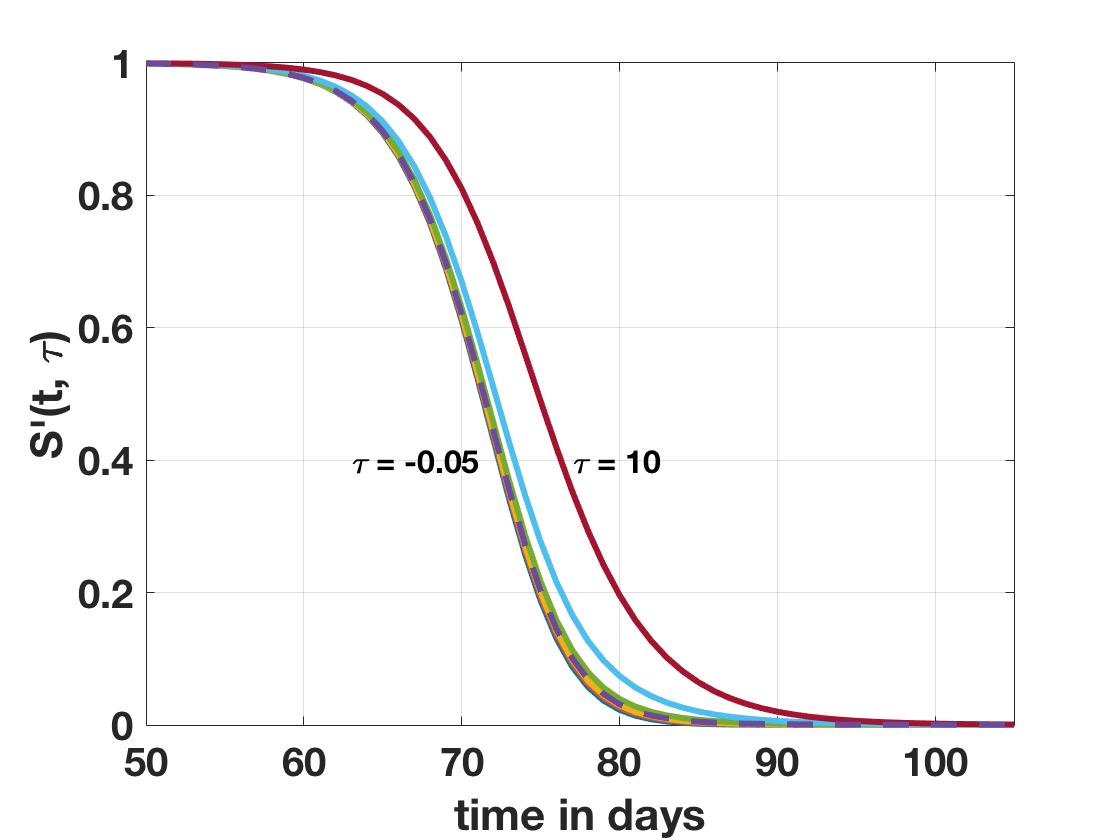}
\caption{The profiles $S'(t,~\tau)$ for several values of the parameter $\tau$ ranging from $-0.05$ up to $10$. The dashed purple line is the profile for $\tau=0$.} \label{FigSbeta1}
\end{center}
\end{figure}

\begin{figure}[htbp]
\begin{center}
\includegraphics[width=.9\columnwidth]{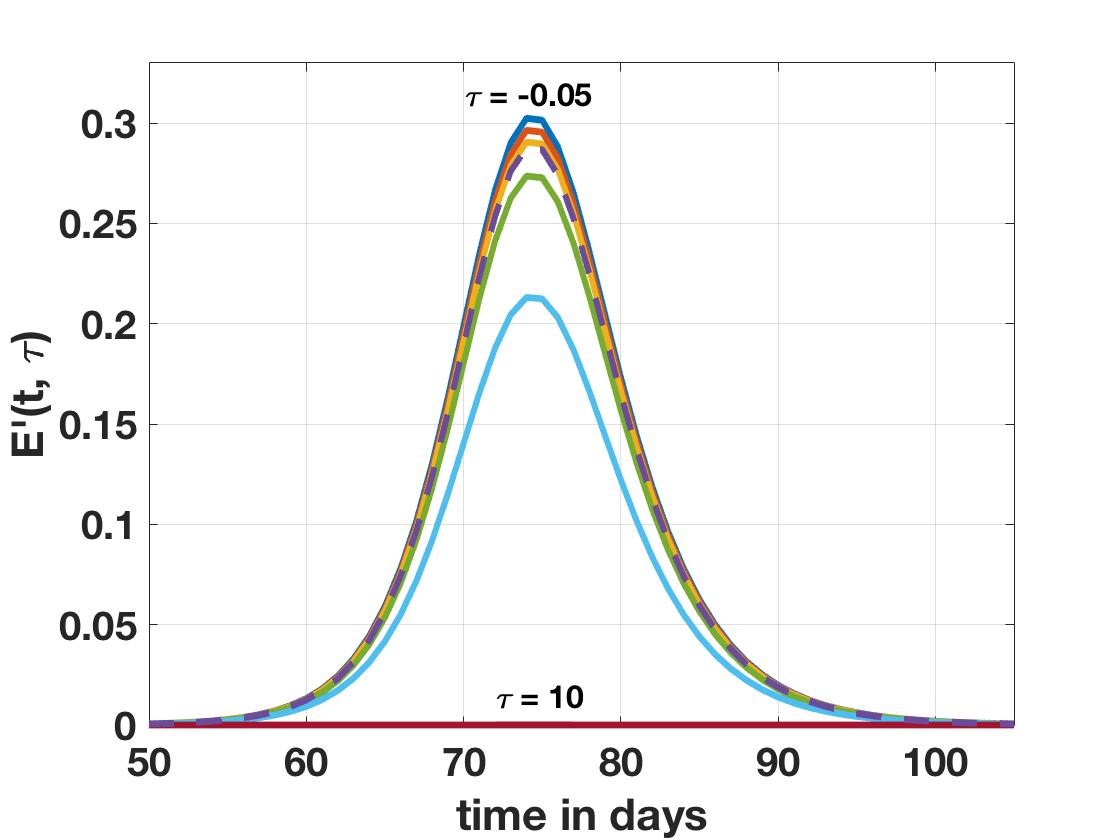}
\caption{The profiles $E'(t,~\tau)$ for several values of the parameter $\tau$ ranging from $-0.05$ up to $10$. The dashed purple line is the profile for $\tau=0$.} \label{FigEbeta1}
\end{center}
\end{figure}

Figure \ref{FigBeta1} displays the profiles of $\beta'(t,~\tau)$. The profile for $\tau=0$ (purple dashed line) is precisely the constant value $\beta(t)=1$. By varying $\tau$ we obtain a significant change of the profile meaning that this parameter is "strongly" unidentifiable.

Figures \ref{FigSbeta1} and \ref{FigEbeta1} display the profiles of $S'(t,~\tau)$ and $E'(t,~\tau)$, respectively. The initial conditions of $S$ and $E$ are obtained by these profiles at $t=0$.
In addition, by varying $\tau$, the constant parameter that is not identifiable (i.e., $\gamma$) significantly changes. For instance, in accordance with (\ref{EquationCovidIndistinguishableInitStates2}),
for $\tau=-0.05$ and $\tau=10$ we obtain the following changes:

\[
\begin{array}{lll}
\gamma=\gamma'(0)=0.25& \hskip-.25cm\rightarrow \gamma'(-0.05)\cong 0.2378 &\hskip-.25cm\rightarrow \gamma'(10)\cong 5506\\
\end{array}
\]

\begin{figure}[htbp]
\begin{center}
\includegraphics[width=1.\columnwidth]{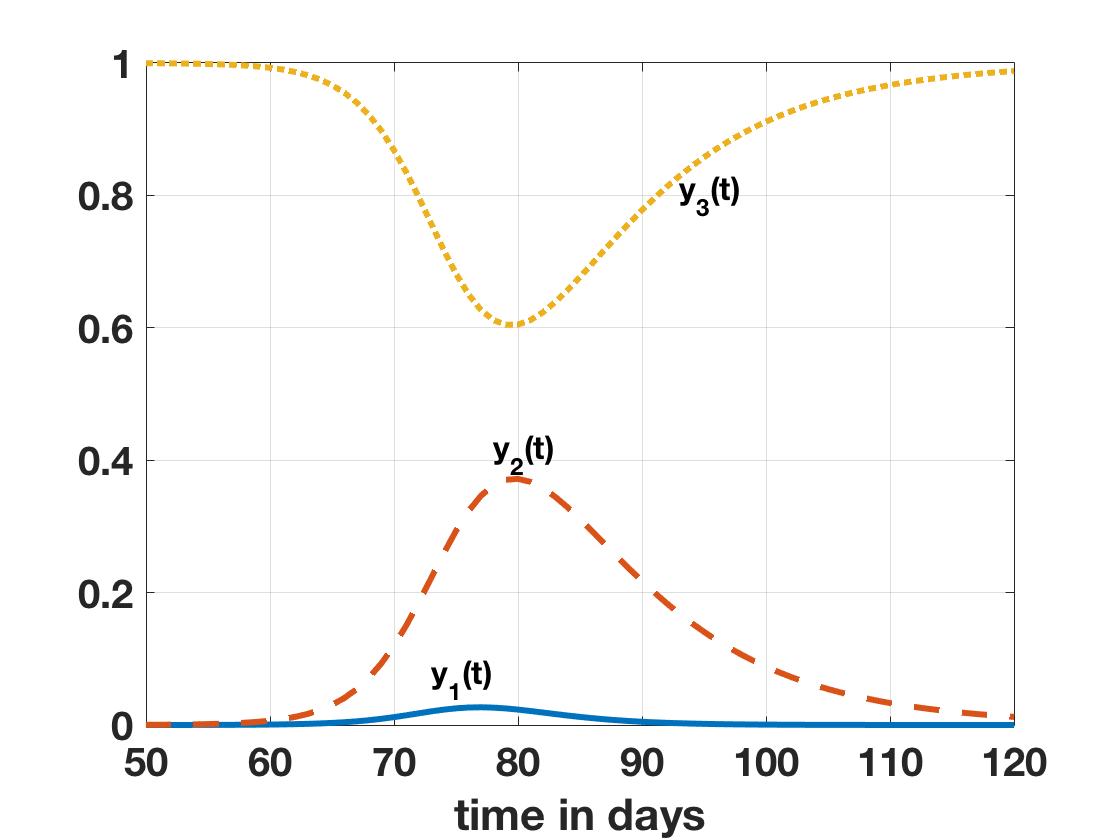}
\caption{The three outputs $y_1(t)=I$ (solid blue line), $y_2(t)=A$ (dashed red line), and $y_3(t)=S+E+R$ (dotted yellow line). The three profiles are independent of the parameter $\tau$, as expected.} \label{FigYCovidBeta1}
\end{center}
\end{figure}

We verified that the three outputs, $y_1(t)=I$, $y_2(t)=A$, and $y_3(t)=S+E+R$ are independent of $\tau$, at any $t$.
This is obtained by proceeding as follows.

We considered several values of $\tau$ in $\mathcal{U}_0$. For each $\tau$, we integrated the differential equation in (\ref{EquationCovidSystem}) with the initial conditions given in (\ref{EquationCovidIndistinguishableInitStates2}) for that $\tau$ and by setting the  unknown time-varying parameter as in (\ref{EquationCovidIndistinguishableUI2}) for the same value of $\tau$ (that is one of the profiles plotted in Figure \ref{FigBeta1}). Both the initial conditions in (\ref{EquationCovidIndistinguishableInitStates2}) and the unknown input significantly depend on $\tau$. However, the three outputs (which are displayed in Figure \ref{FigYCovidBeta1}) are independent, in accordance with our results.

\subsection{$\beta(t)=\cos\left(\frac{\pi}{400}t\right)$}
We proceed exactly as in the previous case. In this case we are allowed to also include smaller $\tau$ and we set:
\[
\mathcal{U}_0:=[-0.10, ~10].
\]

\begin{figure}[htbp]
\begin{center}
\includegraphics[width=.9\columnwidth]{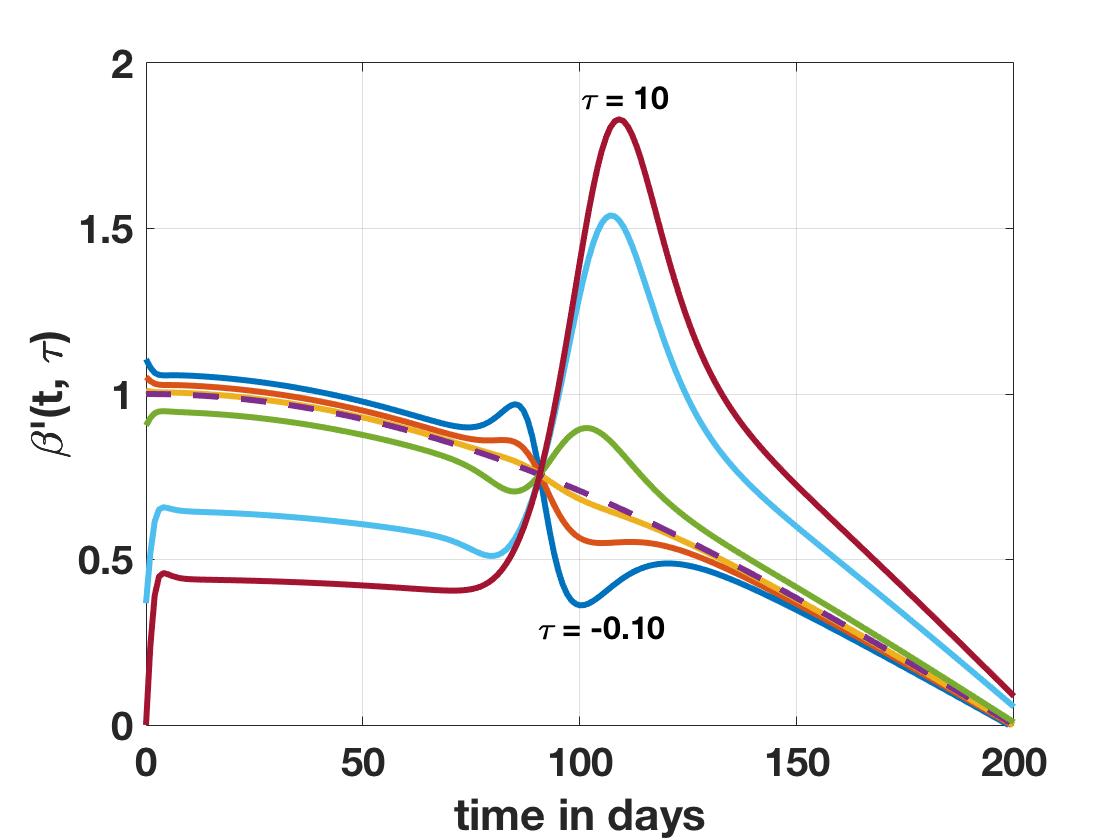}
\caption{Several indistinguishable profiles for the time-varying parameter $\beta'(t,~\tau)$ for several values of the parameter $\tau$ ranging from $-0.1$ up to $10$. The dashed purple line is the profile given by $\beta(t)=\cos\left(\frac{\pi}{400}t\right)$, i.e., $\beta'(t,~\tau)$ for $\tau=0$.} \label{FigBetaV}
\end{center}
\end{figure}

\begin{figure}[htbp]
\begin{center}
\includegraphics[width=.9\columnwidth]{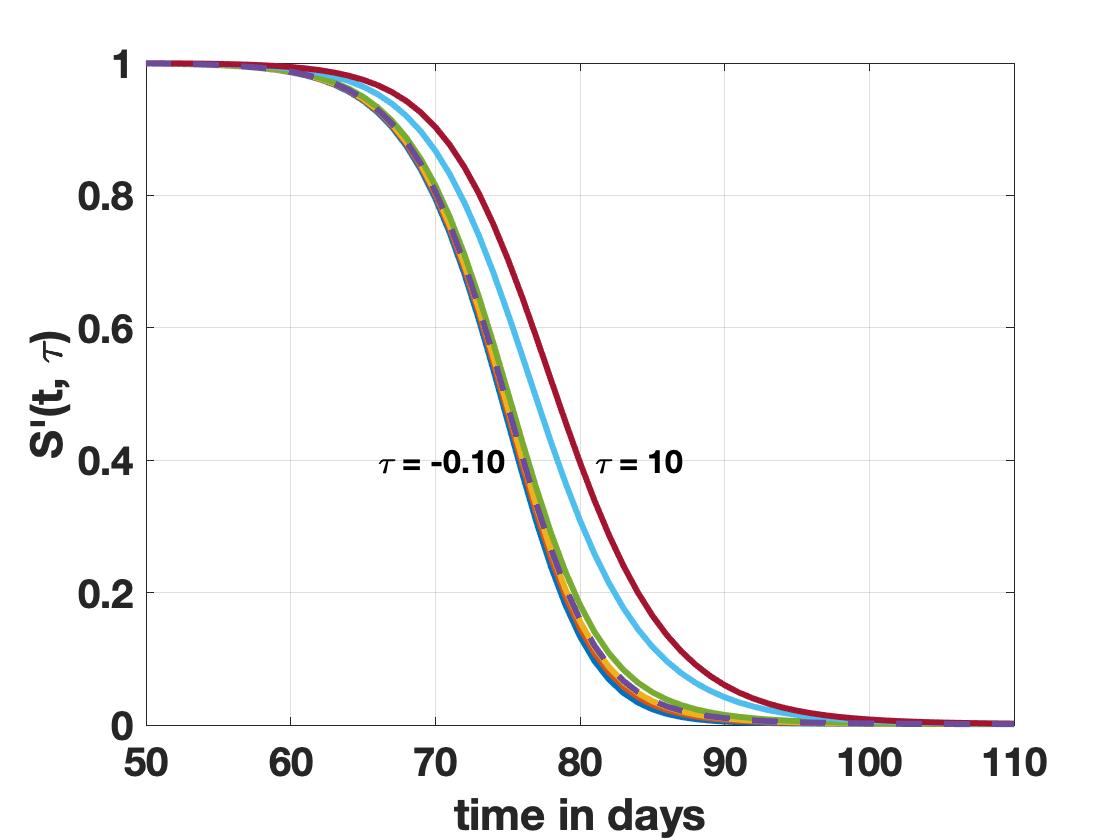}
\caption{The profiles $S'(t,~\tau)$ for several values of the parameter $\tau$ ranging from $-0.1$ up to $10$. The dashed purple line is the profile for $\tau=0$.} \label{FigSbetaV}
\end{center}
\end{figure}

\begin{figure}[htbp]
\begin{center}
\includegraphics[width=.9\columnwidth]{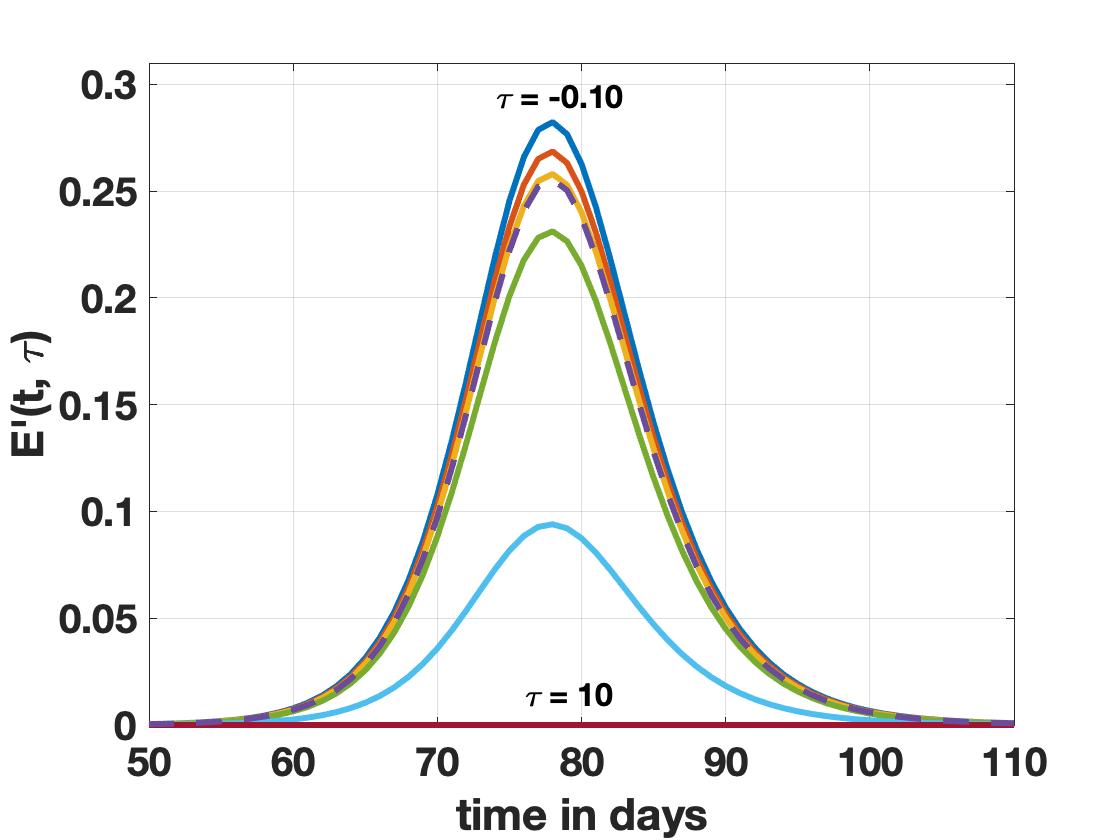}
\caption{The profiles $E'(t,~\tau)$ for several values of the parameter $\tau$ ranging from $-0.1$ up to $10$. The dashed purple line is the profile for $\tau=0$.} \label{FigEbetaV}
\end{center}
\end{figure}

Figure \ref{FigBetaV} displays the profiles of $\beta'(t,~\tau)$. The profile for $\tau=0$ (purple dashed line) is precisely the one set in (\ref{EquationCovidEtaDataSet}). By varying $\tau$, we obtain a significant change of the profile meaning that this parameter is "strongly" unidentifiable.

Figures \ref{FigSbetaV} and \ref{FigEbetaV} display the profiles of $S'(t,~\tau)$ and $E'(t,~\tau)$, respectively.

\begin{figure}[htbp]
\begin{center}
\includegraphics[width=1.\columnwidth]{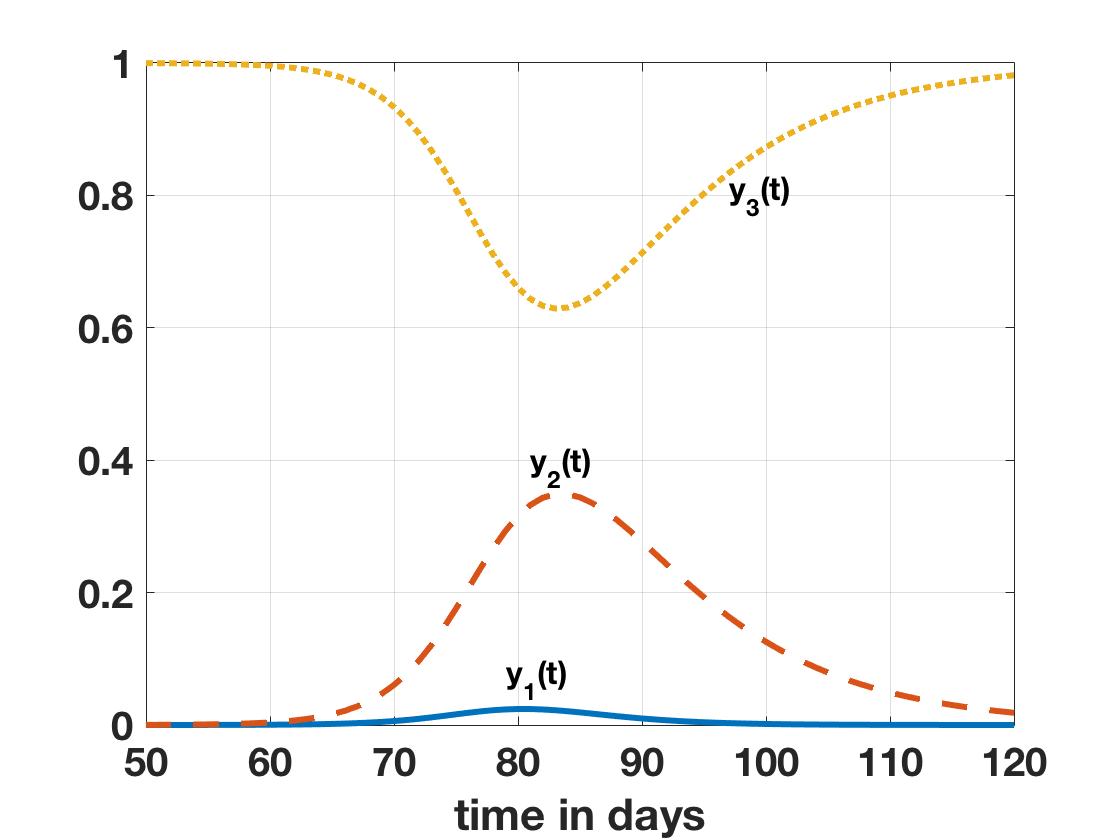}
\caption{The three outputs $y_1(t)=I$ (solid blue line), $y_2(t)=A$ (dashed red line), and $y_3(t)=S+E+R$ (dotted yellow line). The three profiles are independent of the parameter $\tau$, as expected.} \label{FigYCovidBetaV}
\end{center}
\end{figure}

Also in this case we verified that the three outputs, $y_1(t)=I$, $y_2(t)=A$, and $y_3(t)=S+E+R$ are independent of $\tau$, at any $t$.
They are displayed in Figure \ref{FigYCovidBetaV}.

\chapter{Genetic toggle switch}\label{ChapterTS}

This example is provided to illustrate the use of our results when the dynamics of the system are not affine in the inputs.
%
We study the observability and the identifiability properties of the ODE model characterized by the following dynamics and outputs:

\begin{equation}\label{EquationTSSystemNonAffine}
\left\{\begin{array}{ll}
\dot{x}_1 &= k_{01} + \frac{k_1}{1+\left(\frac{x_2}{W_1}\right)^{n_1}} -x_1 \\
\dot{x}_2 &= k_{02} + \frac{k_2}{1+\left(\frac{x_1}{W_2}\right)^{n_2}} -x_2 \\
y &= [x_1, ~x_2], \\
\end{array}\right.
\end{equation}
This model includes two TV-parameters, $W_1$ and $W_2$, and six constant parameters, $k_{01}$, $k_{02}$, $k_1$, $k_2$, $n_1$, $n_2$. All of them are assumed to be unknown.
This model is commonly adopted to characterize the genetic toggle switch. The interested reader is addressed to Section 3.2 of \cite{Villa19b} to find a biological description of this system, which includes the physical meaning of all the above states and parameters. Note that, (\ref{EquationTSSystemNonAffine}) coincides with Eq. (3.5) in \cite{Villa19b} with $n_1=\eta_{TetR}$, $n_2=\eta_{LacI}$, $W_1=1+w_1$, and $W_2=1+w_2$.

The model in (\ref{EquationTSSystemNonAffine}) is 
a special case of (\ref{EquationSystemDefinitionUIOGeneral})
but not directly a special case of
(\ref{EquationSystemDefinitionUIO}).
In accordance with what we mentioned at the end of Section \ref{SectionProblem}, we need to introduce a state that, in addition to the six constant parameters, also includes $W_1$ and $W_2$. In other words:

\begin{equation}\label{EquationTSState}
x=[x_1, ~x_2, ~W_1, ~W_2, ~k_{01}, ~k_1, ~n_1, ~k_{02}, ~k_2, ~n_2]^T
\end{equation}
and the dynamics are:
\begin{equation}\label{EquationTSSystem}
\left\{\begin{array}{ll}
\dot{x}_1 &= k_{01} + \frac{k_1}{1+\left(\frac{x_2}{W_1}\right)^{n_1}} -x_1 \\
\dot{x}_2 &= k_{02} + \frac{k_2}{1+\left(\frac{x_1}{W_2}\right)^{n_2}} -x_2 \\
\dot{W}_1 &= w_1 \\
\dot{W}_2 &= w_2 \\
\dot{k}_{01} &=\dot{k}_1=\dot{n}_1=\dot{k}_{02}=\dot{k}_2=\dot{n}_2=0.  \\
y &= [x_1, ~x_2], \\
\end{array}\right.
\end{equation}

The system is now a special case of (\ref{EquationSystemDefinitionUIO}). In particular, 
$m_u=0$, $m_w=2$, $p=2$, $h_1(x)=x_1$, $h_2(x)=x_2$,

\begin{equation}\label{EquationTSg0g1}
g^0=
\left[
\begin{array}{c}
k_{01} + \frac{k_1}{1+\left(\frac{x_2}{W_1}\right)^{n_1}} -x_1\\
k_{02} + \frac{k_2}{1+\left(\frac{x_1}{W_2}\right)^{n_2}} -x_2 \\
0\\
0\\
0_6\\
\end{array}
\right],~
g^1=
\left[
\begin{array}{c}
0\\
0\\
1\\
0\\
0_6\\
\end{array}
\right],~
g^2=
\left[
\begin{array}{c}
0\\
0\\
0\\
1\\
0_6\\
\end{array}
\right]
\end{equation}

\section{Implementation of Algorithm \ref{AlgoFull}}\label{SectionTSObs}

We run Algorithm \ref{AlgoFull} to obtain the observability codistribution.
Line \ref{AlgoLineOmegaINIT} provides:
 
\[
\Omega=\textnormal{span}\{\nabla h_1,~\nabla h_2\}
=\textnormal{span}\{[1, 0, 0, 0, 0, 0, 0, 0, 0, 0],~[0, 1, 0, 0, 0, 0, 0, 0, 0, 0]\}.
\]
$\uideg \left({\Omega}\right)=0~(<m_w)$, as $\Li_{g^1}h_1=\Li_{g^1}h_2=\Li_{g^2}h_1=\Li_{g^2}h_2=0$ (and the rank of the unknown input reconstructability matrix from $h_1,~h_2$ vanishes).
The system is not in canonical form with respect to its unknown input. 
We have $m=0$ and 
$\mu^0_0=\nu^0_0=1$, obtained from (\ref{EquationTensorMSynchrom}) (these tensors have a single entry as the indices only take the value 0).
 In addition, from (\ref{Equationgalpham}) we obtain: $\widehat{g}^0=\nu^0_0g^0=g^0$, 
 $\giat^1=g^1$, and $\giat^2=g^2$.

 Algorithm \ref{Algo***} provides Finish=False. In addition, from (\ref{Equationgiatdinfty}), we obtain: $\giat=\giat^0=g^0$.
We have:
\[
\Omega=
\left<\left.\left\{ 
\begin{array}{ll}
 \dtv{\Li}_{\giat}& \\
 \giat^1&\giat^2\\
\end{array}
\right\}~\right|\textnormal{span}\left\{\nabla h_1,~\nabla h_2\right\}\right>=
\left<\left.\left\{ 
\begin{array}{ll}
 g^0& \\
 g^1&g^2\\
\end{array}
\right\}~\right|\textnormal{span}\left\{\nabla h_1,~\nabla h_2\right\}\right>.
\]
By a direct computation, we obtain:

\[
\Omega=\textnormal{span}\left\{\nabla h_1,~\nabla h_2,~\nabla\Li_{g^0}h_1,~\nabla\Li_{g^0}h_2\right\}=
\]
\[
\textnormal{span}\left\{\nabla h_1,~\nabla h_2,~\nabla h_3,~\nabla h_4\right\}
\]
where we adopted the following notation:
\[
h_3:= \Li_{g^0}h_1,~~
h_4:= \Li_{g^0}h_2
\]

Now, $\uideg\left(\Omega\right)=2=m_w$. Hence, the execution of Algorithm \ref{AlgoFull} continues with Line \ref{AlgoLineFINITSelection}. The operation at this line ($[\widetilde{h}_1,\ldots, \widetilde{h}_{m_w}]=\mathcal{S}(\Sigma,~\Omega)$) sets 

\begin{equation}\label{EquationTShtilde}
\widetilde{h}_1=h_3=k_{01} + \frac{k_1}{1+\left(\frac{x_2}{W_1}\right)^{n_1}} -x_1,~~~
\widetilde{h}_2=h_4=k_{02} + \frac{k_2}{1+\left(\frac{x_1}{W_2}\right)^{n_2}} -x_2
\end{equation}

Then, Line \ref{AlgoLineFINITMuNuTOBS} provides:

\[
\mu=\left[
\begin{array}{ccc}
1&0&0\\
\Li_{g^0}\tih_1&\Li_{g^1}\tih_1&\Li_{g^2}\tih_1\\
\Li_{g^0}\tih_2&\Li_{g^1}\tih_2&\Li_{g^2}\tih_2\\
\end{array}
\right],
\]
whose expression is obtained by using (\ref{EquationTSg0g1}) and (\ref{EquationTShtilde}). From $\mu$ we obtain $\nu$ (by computing its inverse).
Finally, from $\nu$, we obtain $\widehat{g}^0$, $\widehat{g}^1$, and $\widehat{g}^2$, by using (\ref{Equationgalpha}).

From (\ref{EquationTOBSDef}) we obtain that, when $m_u=0$, the codistribution $\tobs$ vanishes. 
The final step of Algorithm \ref{AlgoFull} is the execution of Line \ref{AlgoLineFINALSTEP}, namely:

\[
\OBS=
\left<
\left.
\widehat{g}^0,~
\widehat{g}^1,~\widehat{g}^2,
\right|
~\textnormal{span}\left\{\nabla h_1,~\nabla h_2,~\nabla h_3,~\nabla h_4\right\}\right>.
\]

The algorithm 
in (\ref{EquationAlgorithmsMinimalCod}) converges at the first step ($\Omega_1=\Omega_0$), and

\[
\OBS=\textnormal{span}\left\{
\nabla h_1,~\nabla h_2,~\nabla h_3,~\nabla h_4
\right\}.
\]

The rank of $\OBS$ is $4$, which is smaller than the dimension of the state in (\ref{EquationTSState}). Hence, this state is not observable. Note that $x_1$ and $x_2$ are observable and we know this even without the implementation of Algorithm \ref{AlgoFull}, as they are the outputs.

\section{Identifiability analysis and indistinguishable states and parameters}\label{SectionTSIdentifiability}

Regarding the constant parameters, it is immediate to check that none is identifiable (it suffices to show that their gradient does not belong to $\OBS$).
This result contradicts the result available in the state of the art (e.g., see Section 3.2 of \cite{Villa19b}).
In Section \ref{SectionTSComparison}, we explicitly prove the validity of our result.

Let us study the identifiability of the time varying parameters.

First of all we remark that the two original TV-parameters, $W_1$ and $W_2$, were included in the state (defined in Eq. (\ref{EquationTSState})). As result, their identifiability is obtained by simply checking if their differential belongs to $\OBS$. It is immediate to check that this is not the case. This result contradicts the result available in Section 3.2 of \cite{Villa19b}. In Section \ref{SectionTSComparison}, we explicitly prove the validity of our result.

Let us check the identifiability of their time derivatives, i.e., $w_1$ and $w_2$.

We execute Algorithm \ref{AlgoFullIDE}.
First, as the system is canonic with respect to its unknown inputs, we do not have symmetries of the unknown input due to the system non canonicity ($^c\chi$). 

Let us compute the distribution orthogonal to the observable codistribution. We have:

\begin{equation}\label{EquationTSSymmetry}
\OBS^\bot=\textnormal{span}\left\{
\xi_1,~\xi_2,~\xi_3,~\xi_4,~\xi_5,~\xi_6
\right\}.
\end{equation}
with:
\[
\xi_1=\left[\begin{array}{c}
0_2\\
-\frac{W_1 ((x_2/W_1)^{n_1} + 1)^2}{n_1 k_1 (x_2/W_1)^{n_1}}\\
0\\
1\\
0_5\\
\end{array}\right],~
\xi_2=\left[\begin{array}{c}
0_2\\
-\frac{W_1 ((x_2/W_1)^{n_1} + 1)}{n_1 k_1 (x_2/W_1)^{n_1}}\\
0_2\\
1\\
0_4\\
\end{array}\right],~
\xi_3=\left[\begin{array}{c}
0_2\\
W_1 \log\left(\frac{x_2}{W_1}\right)\\
0_3\\
n_1\\
0_3\\
\end{array}\right],
\]
\[
\xi_4=\left[\begin{array}{c}
0_3\\
-\frac{W_2 ((x_1/W_2)^{n_2} + 1)^2}{n_2 k_2 (x_1/W_2)^{n_2}}\\
0_3\\
1\\
0_2\\
\end{array}\right],~
\xi_5=\left[\begin{array}{c}
0_3\\
-\frac{W_2 ((x_1/W_2)^{n_2} + 1)}{n_2 k_2 (x_1/W_2)^{n_2}}\\
0_4\\
1\\
0\\
\end{array}\right],~
\xi_6=\left[\begin{array}{c}
0_3\\
W_2 \log\left(\frac{x_1}{W_2}\right)\\
0_4\\
0\\
n_2\\
\end{array}\right]
\]

Hence, 
we have six symmetries of the state. From them, it is immediate to compute the symmetries of the unknown inputs (i.e., by using Eq. (\ref{EquationSymmetryUIObs})). It is easy to check that they do not vanish. We conclude that also $w_1$ and $w_2$ are unidentifiable, even locally.

As for the case study, starting from the generators of $\OBS^\bot$ we compute the indistinguishable states (although the determination of the indistinguishable states is not requested by Algorithm \ref{AlgoFullIDE}).
We need to solve the differential equation in (\ref{EquationDiffEqStates}). 

As we have six generators, we obtain six distinct sets of solutions. We directly provide the results.

Let us denote the true state and parameters as follows:
$x_1(t)$, $x_2(t)$, $W_1(t)$, $W_2(t)$, $w_1(t)$, $w_2(t)$, $k_{01}$, $k_1$, $n_1$, $k_{02}$, $k_2$, and $n_2$.
The new parameters will be denoted by
\[
W_1'(t),~W_2'(t), ~w_1'(t),~w_2'(t), k_{01}', ~k_1', ~n_1', ~k_{02}', ~k_2', ~n_2'.
\]
 They will depend on a given 
parameter, denoted by $\tau$, that can take all the values in a given interval $(-\epsilon,\epsilon)$ (for a given $\epsilon\in\mathbb{R}^+$).

\subsubsection{First set}

\begin{equation}\label{EquationTSFirstSet}
W_1(t)'=x_2\left(\frac{1-n_1\left(1+\left(\frac{x_2}{W_1(t)}\right)^{n_1}\right)\tau}{\left(\frac{x_2}{W_1(t)}\right)^{n_1}+n_1\left(\frac{x_2}{W_1(t)}\right)^{n_1}\tau+n_1\tau}\right)^{\frac{1}{n_1}},
\end{equation}
\[
k_{01}'=k_{01}+n_1k_1\tau,~
\]
and the remaining parameters do not change, i.e.:
$W_2'(t)=W_2(t),~ k_1'=k_1,~ n_1'=n_1,~k_{02}'=k_{02},~k_2'=k_2,~ n_2'=n_2$.

\subsubsection{Second set}

\begin{equation}\label{EquationTSSecondSet}
W_1'(t)=x_2\left(\frac{1}{\left(1+\left(\frac{x_2}{W_1(t)}\right)^{n_1}\right) e^{n_1\tau}-1}\right)^{\frac{1}{n_1}}
\end{equation}
\[
k_1'=k_1 e^{n_1\tau}
\]
and the remaining parameters do not change, i.e.: $W_2'(t)=W_2(t),~ k_{01}'=k_{01},~ n_1'=n_1,~k_{02}'=k_{02},~k_2'=k_2,~ n_2'=n_2$.

\subsubsection{Third set}

\begin{equation}\label{EquationTSThirdSet}
W_1'(t)=x_2 ~e^{\left(
e^{-\tau}\log\frac{W_1(t)}{x_2}
\right)}
\end{equation}
\[
n_1'=n_1e^{\tau}
\]
and the remaining parameters do not change, i.e.: $W_2'(t)=W_2(t),~ k_{01}'=k_{01},~ k_1'=k_1,~k_{02}'=k_{02},~k_2'=k_2,~ n_2'=n_2$.

\subsubsection{Fourth set}
As the first set, with the change $1\leftrightarrow2$. 

\subsubsection{Fifth set}
As the second set, with the change $1\leftrightarrow2$. 

\subsubsection{Sixth set}
As the third set, with the change $1\leftrightarrow2$.

\section{Direct proof}\label{SectionTSComparison}

We provide an unequivocal proof of the validity of our results.
In particular, we directly prove that the values of the parameters in the six sets are indistinguishable from the true values.

The indistinguishability will be proved by showing that, for any $\tau$, the new parameters provide exactly the same dynamics of $x_1$ and $x_2$, obtained with the true parameters, i.e, from (\ref{EquationTSSystemNonAffine}), by proving the following two equalities:

\begin{equation}\label{EquationTSCheck1}
k_{01}' + \frac{k_1'}{1+\left(\frac{x_2(t)}{W_1'(t)}\right)^{n_1'}}=
k_{01} + \frac{k_1}{1+\left(\frac{x_2(t)}{W_1(t)}\right)^{n_1}}
\end{equation}

\begin{equation}\label{EquationTSCheck2}
k_{02}' + \frac{k_2'}{1+\left(\frac{x_1(t)}{W_2'(t)}\right)^{n_2'}}=
k_{02} + \frac{k_2}{1+\left(\frac{x_1(t)}{W_2(t)}\right)^{n_2}}
\end{equation}

\subsection{First set}

By a direct substitution of (\ref{EquationTSFirstSet}) in (\ref{EquationTSCheck1}), we obtain the validity of (\ref{EquationTSCheck1}). The validity of (\ref{EquationTSCheck2}) is trivial. 
The existence of this set of indistinguishable parameters proves the unidentifiability of $W_1(t)$ and $k_{01}$. In addition, 
as $w_1(t)=\dot{W}_1(t)\neq\dot{W}_1'(t)=w_1'(t)$, it also proves the unidentifiability of $w_1(t)$.
Finally, as $\tau$ can take infinitesimal values, and $\lim_{\tau\rightarrow0}W_1(t)'=W_1(t)$, and $\lim_{\tau\rightarrow0}k_{01}'=k_{01}$, the unidentifiability is even local.

\subsection{Second set}
By a direct substitution of (\ref{EquationTSSecondSet}) in (\ref{EquationTSCheck1}), we obtain the validity of the equality in (\ref{EquationTSCheck1}). The validity of (\ref{EquationTSCheck2}) is trivial. 
The existence of this set of indistinguishable parameters proves the unidentifiability of $W_1(t)$, $k_1$, and $w_1(t)$.
~Finally, as $\tau$ can take infinitesimal values, and $\lim_{\tau\rightarrow0}W_1(t)'=W_1(t)$, and $\lim_{\tau\rightarrow0}k_1'=k_1$, the unidentifiability is even local.
\subsection{Third set}

By a direct substitution  of (\ref{EquationTSThirdSet}) in (\ref{EquationTSCheck1}), we obtain the validity of the equality in (\ref{EquationTSCheck1}). The validity of (\ref{EquationTSCheck2}) is trivial. 
The existence of this set of indistinguishable parameters proves the unidentifiability of $W_1(t)$, $n_1$, and $w_1(t)$.
Finally, as $\tau$ can take infinitesimal values, and $\lim_{\tau\rightarrow0}W_1(t)'=W_1(t)$, and $\lim_{\tau\rightarrow0}n_1'=n_1$, the unidentifiability is even local.

\subsection{Fourth set}
In this case the validity of (\ref{EquationTSCheck1}) is trivial while to prove the validity of (\ref{EquationTSCheck2}) we need to use (\ref{EquationTSFirstSet}) with the change $1\leftrightarrow2$.
The existence of this set of indistinguishable parameters proves the unidentifiability of $W_2(t)$, $k_{02}$, and $w_2(t)$.
Finally, as $\tau$ can take infinitesimal values, and $\lim_{\tau\rightarrow0}W_2(t)'=W_2(t)$, and $\lim_{\tau\rightarrow0}k_{02}'=k_{02}$, the unidentifiability is even local.

\subsection{Fifth set}
In this case the validity of (\ref{EquationTSCheck1}) is trivial while to prove the validity of (\ref{EquationTSCheck2}) we need to use (\ref{EquationTSSecondSet}) with the change $1\leftrightarrow2$.
The existence of this set of indistinguishable parameters proves the unidentifiability of $W_2(t)$, $k_2$, and $w_2(t)$.
Finally, as $\tau$ can take infinitesimal values, and $\lim_{\tau\rightarrow0}W_1(t)'=W_2(t)$, and $\lim_{\tau\rightarrow0}k_2'=k_2$, the unidentifiability is even local.

\subsection{Sixth set}
In this case the validity of (\ref{EquationTSCheck1}) is trivial while to prove the validity of (\ref{EquationTSCheck2}) we need to use (\ref{EquationTSThirdSet}) with the change $1\leftrightarrow2$.
The existence of this set of indistinguishable parameters proves the unidentifiability of $W_2(t)$, $n_2$, and $w_2(t)$.
Finally, as $\tau$ can take infinitesimal values, and $\lim_{\tau\rightarrow0}W_2(t)'=W_2(t)$, and $\lim_{\tau\rightarrow0}n_2'=n_2$, the unidentifiability is even local.

\chapter{Conclusion}\label{ChapterConclusion}
In this paper we provided the following four main contributions:

\begin{itemize}

\item {\bf First contribution:} Compact presentation of the systematic procedure introduced in \cite{IF22} to
determine the observability of the state of any nonlinear system in the presence of time-varying parameters (or unknown inputs). In particular, the systematic procedure was presented to make it usable by a non specialist user.

\item {\bf Second contribution:} Introduction of the systematic procedure to determine the identifiability of all the time-varying parameters of any ODE model.

\item {\bf Third contribution:} Introduction of the continuous transformations (one parameter Lie groups) that allow us to build the set of indistinguishable states and indistinguishable unknown inputs 
in the presence of unobservability and unidentifiability.

\item {\bf Fourth contribution:} 
Introduction of new important results about the observability and the identifiability 
of a very popular HIV model, a Covid-19 model, and a genetic toggle switch model. The results on the HIV model and the genetic toggle switch model are in contrast with the current state of the art.

\end{itemize}

\section{First contribution}

The systematic procedure was provided in Chapter \ref{ChapterSolutionUIO} and is Algorithm \ref{AlgoFull}. This algorithm can be executed by a non specialist user. All the functions/operations that appear in the algorithm were defined in Sections \ref{SectionUIOBasicOperations}-\ref{SectionSolutionUIONonCanonic}.
Algorithm \ref{AlgoFull} can be executed without restrictions on the system investigated. It provides the so-called {\it observability codistribution} (denoted by $\OBS$) for any ODE model in the presence of time-varying parameters, independently of its complexity and type of nonlinearity. Note that the system was modelled by Equation (\ref{EquationSystemDefinitionUIO}), and, in Section \ref{SectionProblem}, we showed that any system modelled by (\ref{EquationSystemDefinitionUIOGeneral}) can be easily converted to a system modelled by (\ref{EquationSystemDefinitionUIO}).

Once $\OBS$ is computed, we can easily check whether a given physical quantity, which can be analytically expressed in terms of the state, is observable or not. It suffices to compute its gradient with respect to the state and verify if this gradient belongs to $\OBS$. This holds for any physical quantity that can be analytically expressed in terms of the state. In particular, it holds for the unknown constant parameters, which were included in the state. To check their identifiability it suffices to compute their gradient. This is trivial. The gradient of a given parameter is a row vector with zero everywhere with the exception of the entry that has the same index of the parameter in the state, which is 1.

The theoretical foundation that ensures the general validity of Algorithm \ref{AlgoFull} is available in \cite{IF22}.

\section{Second contribution}
The systematic procedure was provided in Chapter \ref{ChapterUIRec} and is Algorithm \ref{AlgoFullIDE}. 
Also in this case, to obtain the identifiability of all the time-varying parameters, is sufficient to simply follow the steps of this systematic procedure (Algorithm \ref{AlgoFullIDE}). The validity of the algorithm was ensured by two new theoretical results which are Theorems \ref{TheoremIdentifiabilityCan} and \ref{TheoremIdentifiability}.

Algorithm \ref{AlgoFullIDE} determines the identifiability of the unknown time-varying parameters without 
restrictions on the system investigated.
It can handle any system, regardless of its complexity and type of nonlinearity (as for Algorithm \ref{AlgoFull}, the system was modelled by Equation (\ref{EquationSystemDefinitionUIO}), which is equivalent to (\ref{EquationSystemDefinitionUIOGeneral})).

For the special case when the system is characterized by a state that is observable,
Chapter \ref{ChapterUIRec} provided a simpler theoretical result, which was Theorem \ref{TheoremUIRecObservable}. In this special case, the identifiability can be easily verified by checking if the system is canonic with respect to its unknown inputs.
Chapter \ref{ChapterUIRec} showed that the results stated by Theorems \ref{TheoremIdentifiabilityCan} and \ref{TheoremIdentifiability} include the result of Theorem \ref{TheoremUIRecObservable}, as a special case.

\section{Third contribution}

When the state is unobservable and/or a set of parameters (constant and/or time-varying) is unidentifiable, there are more values for all these unobservable and unidentifiable quantities that reproduce exactly the same outputs (and also agree with the same known inputs, when present).
Chapter \ref{ChapterIndistinguishableStatesAndUIsAndMI} introduced the mathematical tool that quantitatively allows us to determine all these indistinguishable values. This tool consists of two sets of first order ordinary differential equations. Specifically,  they are Equations (\ref{EquationDiffEqUICanonic}) and (\ref{EquationDiffEqStatesUI}).
The former is a consequence of the system non-canonicity with respect to its unknown inputs and only regards the time-varying parameters (and not the state). The solution of this first set of differential equations is trivial and can be obtained analytically (and automatically) for any system. Note that this set of equations exists if and only if the system is not canonic with respect to its unknown inputs. In addition, if the system is not canonic with respect to its unknown inputs, we certainly have simple analytical solutions of (\ref{EquationDiffEqUICanonic}). In other words, we can certainly generate values of the time-varying parameters that agree with the same system known inputs and outputs. In particular, this regards the last $m_w-m$ time-varying parameters, where $m_w$ is the total number of them and $m$ the highest unknown input degree of reconstructability (note that the order of the $m_w$ time-varying parameters may have changed after the execution of Algorithm \ref{AlgoFull} and, moreover, after this execution some of these parameters may have been redefined as the time derivatives of the original ones, of a given order).

In contrast, the second set of differential equations (i.e., (\ref{EquationDiffEqStatesUI})) regards the time-varying parameters and the state, simultaneously, and provides their indistinguishable values provided that the condition in (\ref{EquationCONDITIONCommutativitaTempoSimmetria}) is honoured.
This set of equations exists only when the state is unobservable (i.e., when the dimension of $\OBS$ is strictly smaller than the dimension of the state).
The state unobservability  is a necessary but not a sufficient condition for the existence of indistinguishable time-varying parameters of this second type. In other words, there exist cases for which the set of equations in (\ref{EquationDiffEqStatesUI}) only regards the state (i.e., the quantities $^u\chi^i$ in (\ref{EquationDiffEqStatesUI}) vanish for all the values of $i$ (i.e., for $i=1,\ldots,S$). This was for instance the case of the first scenario of our case study (see Section \ref{SubSectionCaseStudyIdentifiability}).

Unfortunately, in contrast with the equations system in (\ref{EquationDiffEqUICanonic}), the system in (\ref{EquationDiffEqStatesUI}) is complex and its solution cannot be obtained automatically. Even worse, in many cases the solution can only be obtained numerically. On the other hand, when it is possible to obtain an analytical solution, we obtain very useful properties about the ODE model under investigation. This was the case of our viral applications in Chapters \ref{ChapterHIV} and \ref{ChapterCovid}. On the other hand, the solution of both the sets of differential equations in (\ref{EquationDiffEqUICanonic}) and (\ref{EquationDiffEqStatesUI}) was not requested to determine the observability of the state and the identifiability of all the time-varying parameters. The determination of them was obtained by simply following the steps of the two systematic procedures (i.e., Algorithm \ref{AlgoFull} and Algorithm \ref{AlgoFullIDE}, respectively). This is always the case and makes possible to obtain the observability and the identifiability of any ODE model, automatically.

\section{Fourth contribution}

Chapters \ref{ChapterHIV}, \ref{ChapterCovid}, and \ref{ChapterTS} provided a detailed study of the observability and identifiability properties of two viral ODE models and a genetic toggle switch model. In particular, the ODE model investigated in Chapter  \ref{ChapterHIV} is very popular and regards the HIV dynamics. It was studied in many previous works (e.g., \cite{Miao11,Villa19b,Per99,Per02,Chen08}).
The second ODE model, studied in Chapter \ref{ChapterCovid}, was a Covid-19 model adopted in
\cite{VillaCovid,SEIARPribylova}. The third model was investigated also in \cite{Villa19b}.

We started by simply executing Algorithm \ref{AlgoFull} and \ref{AlgoFullIDE} to obtain the observability of the state and the identifiability of the time-varying parameters. In the first two models, the state is unobservable and the time-varying parameters unidentifiable. In the third case, all the model parameters are unidentifiable.
In the case of the HIV ODE model, and in the case of the genetic toggle switch model, the results are in contrast with the current state of the art\footnote{The interested reader can find a detailed explanation of the serious error made by the authors of \cite{Miao11} in \cite{arXivErratum}.
Note that Section 6.2 of \cite{Miao11} contains two distinct errors. The former, which is a simple typo, was recently acknowledged by the authors (see \cite{MiaoErratum}). However, in \cite{MiaoErratum}, the authors still concluded that the system is observable/identifiable.
The very serious error committed by the authors is highlighted in Section 3.2 of \cite{arXivErratum} and lies in an incorrect use of the Inverse Function Theorem. In particular, when using this theorem, the authors did not take into account the constraints due to the dynamics of the system, as shown in Section 3.2 of \cite{arXivErratum}.}.

For all these ODE models, it was possible to obtain an analytical solution of the differential equations in (\ref{EquationDiffEqStatesUI}) (the set of equations in (\ref{EquationDiffEqUICanonic}) does not exist because these systems are canonic with respect to their unknown input).
This allowed us to determine indistinguishable unknown inputs that agree with the same outputs of the model (these models do not contain known inputs and, consequently, the only source of information about the state and the model parameters comes from the outputs).

Regarding the HIV ODE model, the differential equations in (\ref{EquationDiffEqStatesUI}) become the system in (\ref{EquationHIVDiffEqStatesUI}). This system is complex and it is not surprising that, in the state of the art, no value of the  time-varying parameter, different from the true one but indistinguishable from it, was detected, and the model was classified {\it identifiable}. In other words, the solution of (\ref{EquationHIVDiffEqStatesUI}), although obtained analytically, is complex and not possible to be determined by following an intuitive reasoning, as for the second scenario of our case study. In this last case, the differential equations in (\ref{EquationDiffEqStatesUI}) become the system in (\ref{EquationCaseStudyDiffEqStateUI}), which is trivial. Its solution is (\ref{EquationCaseStudyIndStateUI}) and could have been obtained with intuitive reasoning (basically, the physical meaning of this solution is that we cannot reconstruct the absolute scale, which was intuitive).

In Section \ref{SectionHIVComparisonSOTA}, we used a data set available in the literature and, by using the solution of (\ref{EquationDiffEqStatesUI}), we generated infinite different unknown inputs that produce the same outputs, starting from the true unknown input and the true states. This unequivocally showed the unidentifiability of the system, in contrast with the results available in the state of the art.
Finally, we determined the minimal external information (external to the knowledge of the outputs) requested to uniquely determine the system parameters.
In Section \ref{SectionCovidNumerical}, we performed a similar analysis for the Covid-19 ODE model here investigated.

Finally, for the first two ODE models, we provided a description based on observable states (Eq. (\ref{EquationHIVLocalDec}) and (\ref{EquationCovidLocalDec}), respectively). This may be useful for future investigations on the HIV and Covid-19 dynamics.

\vskip.6cm

We conclude by emphasizing the generality of the two algorithms (Algorithms \ref{AlgoFull} and \ref{AlgoFullIDE}) introduced in this paper. They provide, automatically, the state observability and the parameters identifiability, of any nonlinear system, independently of its complexity and type of nonlinearity and in the presence of time-varying parameters (or unknown inputs). This generality also regards the two sets of equations in (\ref{EquationDiffEqUICanonic}) and (\ref{EquationDiffEqStatesUI}) here introduced,
which allow us to build the set of indistinguishable states and indistinguishable unknown inputs 
in the presence of unobservability and unidentifiability.
This paper provided a first application of all these theoretical results on two viral models and several important new properties (and also an error in the state of the art) were automatically discovered.
Due to the aforementioned generality, we strongly believe that our theoretical results can be very useful in many other scientific domains, basically to investigate any system that can be described by a an ODE model with time-varying parameters. This occurs in many scientific domains, ranging from the natural and applied sciences up to the social sciences.
In particular, we are interested in studying the identifiability of different astrophysical models in the context of the G-dwarf problem \cite{Caimmi08}. To solve this problem, several models have been proposed since more than half century. Many of them are based on time-varying parameters. For instance, these time-varying parameters characterize the  
star formation rate, the presence of gas radial flows through our galaxy, and even the possibility that the Initial Mass Function (IMF) is time-varying, as we suggested long time ago, in \cite{AA00}. It would be interesting to use the results introduced by this paper to investigate the identifiability of the time-varying parameters that characterize these models in order to determine which measurements are suitable to provide sufficient information that makes these parameters identifiable and, eventually, to discriminate between all the models proposed so far.

\appendix

\chapter{Proof of Theorem \ref{TheoremIdentifiabilityCan}}\label{AppendixTheoremCan}

\proof{
To check the reconstructability of the $m_w$ unknown inputs, we consider the extended state:

\begin{equation}\label{EquationProofTheoremIdCanState}
X:= [x^T,~v_1,\ldots,v_m,~ w_{m+1},\ldots,w_{m_w}]^T
\end{equation}

and we compute its observability codistribution. We denote it by $\OBS^X$.
We remind the reader that the quantities $v_j$ are given in (\ref{Equationvi}).

We denote the dimension of $\OBS$ by $n_o(\le n)$. Algorithm \ref{AlgoFull} builds the $n_o$ generators of $\OBS$. They will be denoted by 
$\theta_1(x),\ldots,\theta_{n_o}(x)$.
We have:

\[
\OBS=\textnormal{span}\left\{
\nabla\theta_1,\ldots,\nabla\theta_{n_o}
\right\}.
\]

As $m$ is the highest unknown input degree of reconstructability, we know that the dimension of $\OBS^X$ is $n_o+m$. Therefore:

\begin{equation}\label{EquationProofTheoremIdCanOBS}
\OBS^X=\textnormal{span}\left\{
\nabla\theta_1,\ldots,\nabla\theta_{n_o}, ~\nabla v_1,\ldots,\nabla v_m
\right\}.
\end{equation}

Now, by definition, any vector field that belongs to the null space of $\OBS^X$ is a simultaneous symmetry of both the state and the unknown input.
Let us study the structure of this null space.
As all the $n_o+m$ functions $\theta_1,\ldots,\theta_{n_o}, v_1,\ldots,v_m$
 are independent of the last $m_w-m$ unknown inputs, any vector with dimension $n+m_w$, with the first $n+m_w-m$ entries equal to zero, certainly belongs to the null space of $\OBS^X$. Therefore, the vectors $^c\chi$ in (\ref{EquationIndwCanonic}) are $m_w-m$ independent symmetries of the unknown input.
$\blacktriangleleft$}

\chapter{Proof of Theorem \ref{TheoremIdentifiability}}\label{AppendixTheoremObs}

\proof{

From (\ref{Equationvi}) we have:

\[
w_k=\sum_{\beta=0}^m\nu_k^\beta \left(v_\beta- \sum_{l=m+1}^{m_w}w_l~\Li_{g^l}\tih_\beta\right).
\]

By using the condition in (\ref{EquationConditionGk}), we obtain:

\[
w_k=\sum_{\beta=0}^m\nu_k^\beta v_\beta,
\]

and:

\[
v_\beta=\sum_{\alpha=0}^m\mu^\alpha_\beta w_\alpha.
\]

Hence:

\[
^u\chi_k=\sum_{\beta=0}^m\Li_\xi\nu_k^\beta v_\beta=
\sum_{\beta=0}^m\Li_\xi\nu_k^\beta \sum_{\gamma=0}^m\mu^\gamma_\beta~w_\gamma=
-\sum_{\beta=0}^m\nu_k^\beta
\sum_{\gamma=0}^m
\left(\Li_\xi\mu^\gamma_\beta\right) w_\gamma=
\]
\[
-\sum_{\beta=0}^m\nu_k^\beta
\left(
\sum_{\gamma=0}^m
\left(\Li_\xi\Li_{g^\gamma}\tih_\beta\right) w_\gamma
\right)=
-\sum_{i=1}^m\nu_k^i
\left(
\left(\Li_\xi\Li_{g^0}\tih_i\right)
+\sum_{j=1}^m
\left(\Li_\xi\Li_{g^j}\tih_i\right) w_j
\right)=
\]
\[
-\sum_{i=1}^m\nu_k^i
\left(
\xi^0_i
+\sum_{j=1}^m
\xi^j_i ~w_j
\right).
\]

$\blacktriangleleft$}

\end{document}